\documentclass{USG}

\usepackage{anyfontsize}
\usepackage{colortbl}
\usepackage{xcolor}
\usepackage{steinmetz}     
\usepackage{siunitx}
\usepackage{booktabs}   
\pdftrailerid{}
\usepackage{tikz}        
\usetikzlibrary{calc}    
\usepackage{afterpage}   
\pgfdeclarelayer{background}  
\pgfsetlayers{background,main}
\usepackage{amsmath,amssymb}       
\usepackage{lipsum}        
\usepackage{bm}
\usepackage{mathtools}
\usepackage{placeins}

\usepackage{eso-pic}
\newcommand{\PreprintMark}{PREPRINT}
\AddToShipoutPictureFG*{%
  \AtPageCenter{%
    \makebox(0,0){%
      \rotatebox{45}{%
        \tikz\node[opacity=0.13, inner sep=0pt,
                   font=\sffamily\bfseries\fontsize{105}{105}\selectfont]
                  {\PreprintMark};}}}}

\graphicspath{{./figures/}{./images/}}

\usepackage[numbers,sort&compress]{NJDnatbib}
\makeatletter
\renewcommand\refname{\fontsize{9}{11.5}\selectfont References}
\renewcommand{\bibfont}{\reset@font\fontfamily{\rmdefault}\fontsize{8.5}{11}\selectfont\baselineskip=10\p@}
\makeatother

\AtBeginDocument{\setlength{\parindent}{1em}}

\makeatletter
\if@twocolumn
  \renewcommand\paragraph{\@startsection{paragraph}{4}{\z@}%
    {-12\@p@t \@plus -2\@p@t \@minus -2\@p@t}{-6\@p@t}%
    {\fontsize{9}{11}\selectfont\bfseries}}%
\else
  \renewcommand\paragraph{\@startsection{paragraph}{4}{\z@}%
    {-12\@p@t \@plus -2\@p@t \@minus -2\@p@t}{-6\@p@t}%
    {\fontsize{10}{12}\selectfont\bfseries}}%
\fi
\makeatother

\makeatletter
\g@addto@macro\normalsize{\belowdisplayshortskip\belowdisplayskip}
\makeatother

\newcommand{\tens}[1]{\mathbf{#1}}          
\newcommand{\vect}[1]{\bm{#1}}          

\newcommand{\bF}{\tens{F}}                      
\newcommand{\bC}{\tens{C}}                      
\newcommand{\bI}{\tens{I}}                      
\newcommand{\isoc}[1]{\overline{#1}}            
\newcommand{\Ione}{\isoc{I}_1}                  
\newcommand{\Itwo}{\isoc{I}_2}                  
\newcommand{\Ithree}{\isoc{I}_3}                

\newcommand{\bP}{\tens{P}}                      
\newcommand{\Wenergy}{\psi}                     
\newcommand{\Wckan}{\tilde{\Wenergy}^{\mathrm{CKAN}}}   

\newcommand{\fibangle}{\zeta}                   
\newcommand{\afib}{\vect{a}}                    
\newcommand{\Mfib}{\tens{M}}                    
\newcommand{\Ifibn}[1]{\isoc{I}_{4#1}}          
\newcommand{\Ifour}{\isoc{I}_4}                 
\newcommand{\Isix}{\isoc{I}_6}                  
\newcommand{\mac}[1]{\langle#1\rangle}          
\newcommand{\KANfib}{\operatorname{KAN}_{\mathrm{fib}}}  
\newcommand{\KANiso}{\operatorname{KAN}_{\mathrm{iso}}}  
\newcommand{\Wckaniso}{\tilde{\Wenergy}^{\mathrm{CKAN}}_{\mathrm{iso}}}  
\newcommand{\Wckanfib}{\tilde{\Wenergy}^{\mathrm{CKAN}}_{\mathrm{fib}}}  
\newcommand{\Wcheck}{\check{\Wenergy}}            
\newcommand{\Gsym}{\mathrm{G}}                  
\newcommand{\pcav}{\hat{p}_{\mathrm{c}}}         
\newcommand{\bFs}{\tens{F}_{\mathrm{s}}}        
\newcommand{\bCs}{\tens{C}_{\mathrm{s}}}        
\newcommand{\bPs}{\tens{P}_{\mathrm{s}}}        
\newcommand{\Grads}{\operatorname{Grad}_{\mathrm{s}}}   

\newcommand{\bu}{\vect{u}}                      
\newcommand{\bX}{\vect{X}}                      
\newcommand{\bx}{\vect{x}}                      
\newcommand{\bR}{\vect{R}}                      
\newcommand{\btheta}{\vect{\theta}}             

\newcommand{\obs}{\mathrm{obs}}                 
\newcommand{\Qu}{\tens{Q}_{u}}                  
\newcommand{\QF}{\tens{Q}_{f}}                  
\newcommand{\Pen}{\mathcal{P}}                   

\newcommand{\tr}{\operatorname{tr}}
\newcommand{\cof}{\operatorname{cof}}
\newcommand{\diag}{\operatorname{diag}}

\newcommand{\GL}{\mathrm{GL}}                   
\newcommand{\SL}{\mathrm{SL}}                   
\newcommand{\SO}{\mathrm{SO}}                   
\newcommand{\Sym}{\mathrm{Sym}}                 

\newcommand{\Grad}{\operatorname{Grad}}         
\newcommand{\Divg}{\operatorname{Div}}          
\newcommand{\argmin}{\operatorname*{arg\,min}}
\newcommand{\dd}{\,\mathrm{d}}
\newcommand{\diff}{\mathrm{d}}                  
\newcommand{\norm}[1]{\left\lVert#1\right\rVert}
\newcommand{\abs}[1]{\left\lvert#1\right\rvert}

\articletype{ORIGINAL ARTICLE}%

\received{Added at production}
\revised{Added at production}
\accepted{Added at production}
\journal{Journal Name}
\volume{0}
\copyyear{2026}
\startpage{1}
\articledoi{}

\begin{document}

\title{Constitutive Kolmogorov--Arnold Networks in Finite Element Model Updating: Closed-Form Hyperelastic Strain Energies from Heterogeneous Experiments}

\author[1]{Kian P. Abdolazizi}
\author[2]{Kevin Linka}
\author[1,3]{Christian J. Cyron}

\authormark{ABDOLAZIZI \textsc{et al.}}
\titlemark{TITLE}

\address[1]{\orgdiv{Institute for Artificial Intelligence and Simulation in Mechanics, }\orgname{Hamburg University of Technology, }%
\orgaddress{\city{Hamburg, }\country{Germany}}}

\address[2]{\orgdiv{Department of Computational Mechanics in Medicine, Institutes of Applied Medical Engineering, }\orgname{RWTH Aachen University, }%
\orgaddress{\city{Aachen, }\country{Germany}}}

\address[3]{\orgdiv{Institute of Material Systems Modeling, }\orgname{Helmholtz-Zentrum Hereon, }%
\orgaddress{\city{Geesthacht, }\country{Germany}}}

\corres{Kian Abdolazizi (\email{kian.abdolazizi@tuhh.de})}




\abstract[ABSTRACT]{
Calibrating a hyperelastic constitutive model for a soft material traditionally poses two problems: an expert must choose its functional form, and several homogeneous tests must determine its parameters. Neural constitutive models solve the first, but either remain black boxes or draw on a library of candidate terms fixed in advance. Full-field measurements, such as those from digital image correlation (DIC), promise to solve the second, replacing those tests by a single heterogeneous one. Two gaps remain: flexibility during training with interpretability afterwards, and a way to train such a model on one heterogeneous test. Here we close both. We embed a constitutive Kolmogorov--Arnold network (CKAN) in finite element model updating (FEMU) and train it by minimizing the mismatch between the predicted and the measured kinematic fields and reaction forces. Pruning and symbolic regression then reduce the trained network to a compact closed-form strain energy that satisfies every constitutive requirement. On perforated natural-rubber membranes, the CKANs reproduce the DIC displacement fields with median errors between one and two percent, and transfer to a hole pattern withheld from training. On a perforated silicone plate, a CKAN trained on one stretch profile at a single load reproduces that profile and the measured reaction force to within a few percent. On an aortic specimen inflated under pressure, an anisotropic CKAN predicts the inflated three-dimensional shape. The discovered closed-form energies match the CKANs they replace on the training data and can be compared directly with established models.
}

\abbr{BVP, boundary-value problem; CANN, constitutive artificial neural network; CKAN, constitutive Kolmogorov--Arnold network; DIC, digital image correlation; EGM, equilibrium gap method; FE, finite element; FEMU, finite element model updating; HGO, Holzapfel--Gasser--Ogden; KAN, Kolmogorov--Arnold network; MT1.0, material testing 1.0; MT2.0, material testing 2.0; PS, pure shear; UT, uniaxial tension; VFM, virtual fields method.}


\copyright{}

\maketitle

\section{Introduction}\label{sec:intro}

Constitutive models provide the closure relations for the governing equations of continuum mechanics and therefore determine the accuracy and predictive capability of a finite element analysis. Soft materials such as natural rubber, silicones, and biological tissue are particularly difficult to describe: they undergo large deformations, are (nearly) incompressible, and their response is strongly nonlinear \cite{Ogden1972,fung2013biomechanics,Humphrey2002}. Which model is adequate for a given material is rarely obvious a priori, and selecting or developing one requires expert knowledge.

Most classical constitutive models are phenomenological \cite{Holzapfel2000a}. They satisfy the constitutive requirements \cite{Truesdell2004}, such as objectivity and material symmetry, but are formulated from experimental observation alone, guided by intuition and experience. Developing such a model is time-consuming, and the result is often limited to one class of materials. Whichever model is chosen, its functional form is fixed a priori, before the data are seen, and only a few material parameters remain to be identified \cite{Marckmann2006}.

\paragraph*{Material testing}
Determining these parameters is an inverse problem. Classically, the data stem from statically determinate tests, above all uniaxial tension, where the stress follows from the measured force and the cross-sectional area and the strain from extensometers or digital image correlation (DIC) \citep{Peters1982,Sutton1986}.
However, such an idealized test probes limited deformation states, and the parameters identified from it generally do not generalize to others \cite{Johlitz2011}. Multiple tests are therefore necessary, such as planar biaxial tension for anisotropic materials \cite{Sommer2013}, or bulge inflation \cite{Vandemaele2025}. \citet{Pierron2021a} call this classical approach material testing 1.0 (MT1.0).

Often, however, material inhomogeneity or a complex specimen geometry and boundary conditions invalidate the assumption of a statically determinate test \citep{Grediac1998}. The test is then statically indeterminate, and the stress state no longer follows from the measured force and the geometry alone. Imaging techniques, most prominently DIC, capture the resulting heterogeneous deformation and yield full-field displacement data \citep{Grediac2012}. Together with the global reaction forces recorded at the same time, these heterogeneous data offer a more informative basis for the parameter identification. \citet{Pierron2021a} call this approach material testing 2.0 (MT2.0). The promise is that a single heterogeneous test suffices to identify parameters that generalize to other loading conditions. Nevertheless, MT1.0 tests remain useful: the maximum stretches in a uniaxial tensile test often far exceed those reached in statically indeterminate tests \citep{Abbasi2026}.

\paragraph*{Identification methods}
Several methods exploit full-field data for the parameter identification of a fixed model \citep{Avril2008}, among them the virtual fields method (VFM) \cite{Pierron2012}, the equilibrium gap method (EGM), and finite element model updating (FEMU) \citep{Kavanagh1971}. More recently, machine learning has extended them by replacing the fixed model: EUCLID builds on the equilibrium gap and discovers the model by sparse regression over a library of candidate terms \citep{Flaschel2021a}, recently also for specimens that combine several materials \citep{Shi2025}. The same gap has since been combined with physics-augmented networks, on planar full-field data \citep{Friedrichs2026} and on the volumetric fields of digital volume correlation \citep{Bourdyot2026}, and the VFM with neural constitutive models \citep{Meng2025}. Physics-informed neural networks \citep{Raissi2019} have been applied to the same task, with the network representing the displacement field \citep{Moon2025} or the spatially varying material properties \citep{Wu2025}. Regardless of the variant, the methods divide into two families. Weak-form methods such as the VFM and EGM insert the measured displacement fields directly into the balance equations: they need no forward simulation, but differentiating the fields amplifies their noise and the fields must cover the complete domain. Methods built on a finite element model of the experiment, above all FEMU, instead compute the fields and compare them with the measurement. They can also accommodate sparse data that cover only part of the specimen, at the price of an iterative and computationally expensive solution.

In FEMU, a finite element model replicates the experiment and computes, for a given constitutive model, the kinematic fields and reaction forces under the boundary conditions of the test. The material parameters are then updated iteratively to minimize an objective function that typically combines the mean squared error of the kinematic fields with that of the reaction forces. Numerous refinements of this classical variant, which we employ here, have been proposed; see \citet{Chen2025}. Stereo DIC and digital volume correlation extend it from thin specimens under plane stress to curved surfaces and volumetric measurements \citep{ShafieiDizaji2018,Bay2008}. The method succeeds only if the numerical model mirrors its physical counterpart: geometry, boundary conditions, and loading must match the test, which in practice requires registering the nominal specimen geometry to the measured one. Two burdens remain: every iteration requires solving a nonlinear forward problem, and deriving the sensitivities of the objective function with respect to the material parameters by hand is laborious and error-prone.

\paragraph*{Neural constitutive models}
In parallel, machine learning has sparked interest in data-driven constitutive modeling. To avoid fixing the functional form of a material model a priori, researchers have begun to replace the constitutive model by neural networks \citep{Ghaboussi1991}. The first neural constitutive models were black boxes, flexible function approximators that disregarded the constitutive requirements. Modern architectures satisfy these requirements by construction. Constitutive artificial neural networks (CANNs) \citep{Linka2021,Linka2022a,Abdolazizi2023} and related physics-augmented approaches \citep{Klein2022,Linden2023,Tepole2025,Geuken2025} encode objectivity, material symmetry, a stress- and energy-free reference configuration, and polyconvexity directly into the architecture without sacrificing flexibility. Even a constrained network, however, remains a black box: its parameters carry no physical meaning, and the learned energy is difficult to interpret or to compare with established models. Existing remedies restore interpretability by selection: sparse regression keeps the few terms of a large library that survive a sparsity penalty \citep{Flaschel2021a,McCulloch2024}, whereas symbolic regression searches the expressions that a fixed set of operations can build \citep{Bahmani2023,Abdusalamov2025}. In both cases, the candidates are fixed a priori.

In nonlinear finite element analysis, stresses and consistent tangent operators must be computed in every Newton iteration, and deriving them analytically is tedious \citep{Rothe2015,Vigliotti2021}, all the more for a neural constitutive model. Automatic differentiation removes this manual derivation, including that of the objective-function sensitivities, and directly supports inverse material identification \citep{Ogawa2025,Wan2026}. Differentiable finite element frameworks implement this throughout: the complete solver is differentiable and executes on graphics processing units, which also mitigates the cost of the repeated forward solves \citep{Xue2023}. Training the network inside a differentiable solver also avoids integrating it into commercial finite element software, a challenge that material subroutines for such models have only recently begun to address \citep{Peirlinck2024,Abdolazizi2026}.

Flexible energies have also been identified by FEMU without a network: \citet{Wiesheier2024} represent the strain energy by a spline-based interpolant  and determine its parameters from DIC displacements and reaction forces, which fits the data but yields no closed form. Recently, two groups independently replaced FEMU's a priori chosen constitutive model by a neural network trained by automatic differentiation through the finite element solve: \citet{Regazzoni2026} on synthetic data with FEniCS, and \citet{Knipper2026} on experimental data with JAX-FEM \citep{Xue2023}. Of the two network variants \citet{Knipper2026} consider, the flexible gray-box CANN \citep{Linka2021} performed best when full-field data were available, whereas the interpretable white-box CANN \citep{Linka2022a}, restricted to a fixed library of terms, proved more reliable when the data were sparse. Flexibility and interpretability thus remain split across two architectures.

\paragraph*{Kolmogorov--Arnold networks}
This split is bridged by Kolmogorov--Arnold networks (KANs) \citep{Liu2024a,Liu2024}, which, motivated by the Kolmogorov--Arnold representation theorem \citep{Kolmogorov1957}, compose multivariate functions from univariate ones. A conventional network learns the weights of fixed activation functions; a KAN instead learns the univariate activation functions themselves, parameterized as splines on the edges. This structure lends itself to pruning and symbolic regression: after training, each remaining univariate function can be matched by a closed-form expression, and a compact formula is read off the trained network a posteriori rather than selected from a fixed library. Constraints such as polyconvexity are equally natural to impose. Constitutive Kolmogorov--Arnold networks (CKANs) build on these properties to represent hyperelastic strain energies \citep{Abdolazizi2025} and inelastic dissipation potentials \cite{Ji2026}. \citet{Thakolkaran2025} replaced EUCLID's library by an input-convex KAN, which as a weak-form method inherits the noise sensitivity and the need for the complete displacement field. To date, KAN-based constitutive models have been trained either directly on stress--strain data from homogeneous tests, or through the weak form on complete full-field data.

\paragraph*{Objectives and contributions}
Two gaps remain. First, no KAN-based constitutive model has been trained through the finite element model updating method. When no stress is measured and the kinematic data cover only part of the specimen, such as a stretch profile along a single path or the displacements of an observed surface, neither existing route applies, and FEMU is what remains. Second, within FEMU the split between flexibility and interpretability persists: no single constitutive model has yet combined flexibility during training with a closed-form expression afterwards that satisfies every constitutive requirement.

We close both gaps by embedding a CKAN into FEMU and training it by automatic differentiation through the finite element model of the experiment, so that any quantity that the experiment measures and the simulation computes can supervise the training. After training, we prune the network and replace its splines by closed-form expressions under the same constraints, which yields a compact strain energy formula with all constitutive requirements intact. We demonstrate the approach on three experimental datasets: full-field displacements of perforated natural-rubber membranes \citep{Abbasi2026}, where we test generalization to a geometry not used for training; a single stretch profile along a path on a perforated silicone plate at one load level \citep{Meunier2008}; and three-dimensional surface displacements of an aortic specimen under measured pressure \citep{Romo2014}, which identify an anisotropic model. The three cases together probe how much constitutive information different kinds of kinematic data carry.

\paragraph*{Outline}
Section~\ref{sec:theory} reviews the background on continuum mechanics, constitutive modeling, and Kolmogorov--Arnold networks, and introduces the CKAN. Section~\ref{sec:femu} embeds it in a FEMU framework for constitutive model discovery. Section~\ref{sec:applications} applies the approach to three experimental datasets: full-field displacements, a single stretch profile along a path, and three-dimensional surface displacements under pressure. Section~\ref{sec:conclusions} summarizes the findings and outlines future work.

\paragraph*{Notation}
Within this manuscript, italic symbols $\psi$ denote zeroth-order tensors (scalars). For first-order tensors (vectors), we use bold italic symbols $\vect{a}$, which also denote one-dimensional arrays of real numbers or of functions, e.g.\ $\vect{z}\in\mathbb{R}^n$. For second-order tensors $\tens{A}$, we use bold Roman symbols, which also denote matrices, i.e., two-dimensional arrays of real numbers or of functions.
Components refer to the fixed Cartesian basis vectors $\vect{e}_i$, through which we identify vectors with $\mathbb{R}^3$ and second-order tensors with $\mathbb{R}^{3\times 3}$, and the Einstein summation convention applies to all repeated indices. The dyadic product of two vectors is denoted by $\otimes$, $\vect{a}\otimes\vect{b}= a_i b_j\,\vect{e}_i\otimes\vect{e}_j$.
Single contractions are written by juxtaposition, $\tens{A}\vect{a}=A_{ij}a_j\vect{e}_i$ and $\tens{A}\tens{B}=A_{ik}B_{kj}\vect{e}_i\otimes\vect{e}_j$, the scalar product of vectors by $\vect{a}\cdot\vect{b}=a_ib_i$ with the norm $\norm{\vect{a}}=\sqrt{\vect{a}\cdot\vect{a}}$, and the double contraction by $\tens{A}:\tens{B}=A_{ij}B_{ij}$.
The transpose and inverse of $\tens{A}$ are denoted by $\tens{A}^\top$ and $\tens{A}^{-1}$, respectively, and $\tens{A}^{-\top}=(\tens{A}^{-1})^\top$.
The trace, determinant, and cofactor of $\tens{A}$ are denoted by $\tr\tens{A}$, $\det\tens{A}$, and $\cof\tens{A}$, respectively, with $\cof\tens{A}=\det(\tens{A})\,\tens{A}^{-\top}$ for invertible $\tens{A}$.
The general linear group is given by $\GL=\{\tens{A}\in\mathbb{R}^{3\times 3}:\det\tens{A}\neq 0\}$, its subgroup of tensors with positive determinant by $\GL^+=\{\tens{A}\in\GL:\det\tens{A}>0\}$, and the special linear group by $\SL=\{\tens{A}\in\GL:\det\tens{A}=1\}$.
The set of symmetric tensors is given by $\Sym=\{\tens{A}\in\mathbb{R}^{3\times 3}:\tens{A}^\top=\tens{A}\}$, and the set of positive definite tensors by $\Sym^+=\{\tens{A}\in\Sym:\vect{a}\cdot\tens{A}\vect{a}>0~\forall\vect{a}\in\mathbb{R}^3\setminus\{\vect{0}\}\}$.
The special orthogonal group is given by $\SO=\{\tens{Q}\in\mathbb{R}^{3\times 3}:\tens{Q}^\top\tens{Q}=\tens{I},~\det\tens{Q}=1\}$, where $\tens{I}=\delta_{ij}\vect{e}_i\otimes\vect{e}_j$ is the second-order identity tensor and $\delta_{ij}$ is the Kronecker delta.
Derivatives with respect to the referential coordinates $\bX=X_i\vect{e}_i$ are written with the material gradient of a scalar and of a vector field and the material divergence of a second-order tensor field, $\Grad w=\frac{\partial w}{\partial X_j}\vect{e}_j$, $\Grad\vect{u}=\frac{\partial\vect{u}}{\partial X_j}\otimes\vect{e}_j=\frac{\partial u_i}{\partial X_j}\vect{e}_i\otimes\vect{e}_j$ and $\Divg\tens{A}=\frac{\partial\tens{A}}{\partial X_j}\,\vect{e}_j=\frac{\partial A_{ij}}{\partial X_j}\vect{e}_i$.
Quantities carrying an overline, $\overline{\cdot}$, refer to the isochoric part of the deformation or are functions of it. A tilde, $\tilde{\cdot}$, marks a function expressed in terms of invariants and a check, $\check{\cdot}$, one in which the fiber invariants enter only through their tension-only form (Sec.~\ref{sec:theory:continuum}); a hat, $\hat{\cdot}$, marks prescribed boundary data. Diagonal matrices with entries $\lambda_i$ are denoted by $\diag(\lambda_1,\ldots,\lambda_n)$. As is common, we do not distinguish notationally between a function and its values: a symbol such as $\bP$ denotes both the stress field and the constitutive function, and its meaning follows from the context. Likewise, a function of a vector argument is written with the components as arguments, e.g.\ $f(a_1,a_2)$ for $f\bigl((a_1,a_2)\bigr)$.


\section{Theoretical background}\label{sec:theory}

\subsection{Continuum mechanical fundamentals}\label{sec:theory:continuum}

\paragraph*{Kinematics}
In its reference configuration, a continuum body occupies the domain $\Omega_0\subset\mathbb{R}^3$ with boundary $\partial\Omega_0$. Under external loads it deforms, at each time instant $t\in\mathcal{T}=[0,T]$, to a current configuration $\Omega\subset\mathbb{R}^3$ with boundary $\partial\Omega$. A smooth, invertible, and orientation-preserving map $\vect{\chi}:\Omega_0\times\mathcal{T}\to\mathbb{R}^3$ carries each material point $\bX\in\Omega_0$ to its current position $\bx=\vect{\chi}(\bX,t)$, so that $\Omega=\vect{\chi}(\Omega_0,t)$.
The displacement field $\bu:\Omega_0\times\mathcal{T}\to\mathbb{R}^3$ is $\bu(\bX,t)=\vect{\chi}(\bX,t)-\bX$. The deformation gradient $\bF = \Grad\vect{\chi} = \bI + \Grad\vect{u}\in\GL^+$ measures the local deformation, its determinant $J=\det\bF$ the local change in volume.
The right Cauchy--Green deformation tensor $\bC=\bF^\top\bF\in\Sym^+$ has the eigenvalues $\lambda_i^2$, $i\in\{1,2,3\}$, where $\lambda_1\geq\lambda_2\geq\lambda_3>0$ are the principal stretches.

Following \citet{Flory1961}, we introduce the isochoric deformation gradient $\isoc{\bF}=J^{-1/3}\bF\in\SL$ and the isochoric right Cauchy--Green deformation tensor $\isoc{\bC}=\isoc{\bF}^\top\isoc{\bF}\in\Sym^+$. The principal invariants of $\isoc{\bC}$ are $\Ione=\tr\isoc{\bC}$, $\Itwo=\tr(\cof\isoc{\bC})$, and $\Ithree=\det\isoc{\bC}=1$.

The anisotropic materials considered here are soft biological tissues reinforced by collagen fibers. The fibers form families; the unit vector $\afib_\kappa$ is the direction of family $\kappa$ in the reference configuration. Of the invariants that $\isoc{\bC}$ and these directions generate \citep{Spencer1984,Boehler1987,Zheng1994}, we use the squared fiber stretches $\Ifibn{\kappa}=\isoc{\bC}:\Mfib_\kappa$, where $\Mfib_\kappa=\afib_\kappa\otimes\afib_\kappa$ are structural tensors. The remaining invariants are hard to identify from mechanical tests, and most soft-tissue models, including that of \citet{Holzapfel2000}, leave them out.\footnote{In a compressible material, fiber invariants of $\isoc{\bC}$ make the fibers respond to changes in volume, which is nonphysical \citep{Sansour2008,Helfenstein2010,Nolan2014}. The material here is incompressible, so that $\bC=\isoc{\bC}$. In the computations this holds pointwise, since the reductions of Appendix~\ref{app:thin} eliminate the constraint through the thickness stretch.}

\paragraph*{Hyperelasticity}
We assume an incompressible, hyperelastic material: its stress derives from a strain energy density per unit reference volume, $\isoc{\Wenergy}:\SL\to\mathbb{R}$, $\isoc{\bF}\mapsto\isoc{\Wenergy}(\isoc{\bF})$.
Incompressibility constrains the admissible deformations to $J=1$ in $\Omega_0$. A Lagrange multiplier field $p:\Omega_0\times\mathcal{T}\to\mathbb{R}$, the hydrostatic pressure, enforces it \citep{Truesdell2004,Holzapfel2000a}. The augmented energy density is
\begin{equation}\label{eq:augmented}
    \Wenergy:\GL^+\times\mathbb{R}\to\mathbb{R},\qquad (\bF,p)\mapsto\Wenergy(\bF,p) = \isoc{\Wenergy}(\isoc{\bF})-p(J-1).
\end{equation}
The first Piola--Kirchhoff stress follows as
\begin{equation}\label{eq:stress}
    \bP(\bF,p) = \frac{\partial \Wenergy}{\partial \bF} = \bP_{\mathrm{iso}}(\bF) - p J \bF^{-\top}, \qquad \bP_{\mathrm{iso}}(\bF) = \frac{\partial \isoc{\Wenergy}(\isoc{\bF})}{\partial \bF}.
\end{equation}
Deriving the stress from an energy ensures \emph{thermodynamic consistency}. The isochoric stress $\bP_{\mathrm{iso}}$ is the constitutive part, while the hydrostatic part $-pJ\bF^{-\top}$ is the reaction to the incompressibility constraint.

\paragraph*{Constitutive requirements}
Beyond thermodynamic consistency, the energy must be invariant under rigid rotations of the current configuration (\emph{objectivity}) and under those rotations of the reference configuration that leave the material response unchanged (\emph{material symmetry}). The two requirements combine into
\begin{equation}\label{eq:objiso}
    \isoc{\Wenergy}(\tens{Q}_1\isoc{\bF}\tens{Q}_2) = \isoc{\Wenergy}(\isoc{\bF}) \qquad\forall\,\isoc{\bF}\in\SL,\,\tens{Q}_1\in\SO,\,\tens{Q}_2\in\Gsym,
\end{equation}
where $\Gsym\subseteq\SO$ is the symmetry group of the material.
For an \emph{isotropic} material, $\Gsym=\SO$. By the representation theorem for isotropic scalar-valued tensor functions \citep{Rivlin1955,Truesdell2004}, Eq.~\eqref{eq:objiso} then holds if and only if $\isoc{\Wenergy}$ depends on $\isoc{\bF}$ through the invariants of $\isoc{\bC}$ alone; as $\Ithree=1$, two remain:
\begin{equation}\label{eq:invariants}
    \isoc{\Wenergy}(\isoc{\bF}) = \tilde{\Wenergy}(\Ione,\Itwo).
\end{equation}
For a fiber-reinforced material, in contrast, $\Gsym$ contains only the rotations that leave its structural tensors unchanged:
\begin{equation}\label{eq:symmetry_group}
  \Gsym=\bigl\{\tens{Q}\in\SO \bigm| \tens{Q}\Mfib_\kappa\tens{Q}^\top=\Mfib_\kappa\bigr\} .
\end{equation}
Such a rotation maps $\isoc{\bC}$ to $\tens{Q}_2^\top\isoc{\bC}\tens{Q}_2$ and leaves $\Ione$, $\Itwo$ and the $\Ifibn{\kappa}$ unchanged, so an energy of the form
\begin{equation}\label{eq:invariants_fiber}
  \isoc{\Wenergy}(\isoc{\bF})=\tilde{\Wenergy}(\Ione,\Itwo,\Ifibn{\kappa})
\end{equation}
satisfies Eq.~\eqref{eq:objiso}. Unlike Eq.~\eqref{eq:invariants}, this form is sufficient but not necessary, since it uses only the squared fiber stretches.
Collagen fibers buckle under compression and carry load only when stretched \citep{Holzapfel2000}. The energy therefore depends on the fiber invariants only through $\mac{\Ifibn{\kappa}-1}^2$, with the positive part $\mac{x}=\max(x,0)$:
\begin{equation}\label{eq:tension_only}
  \tilde{\Wenergy}(\Ione,\Itwo,\Ifibn{\kappa})=\Wcheck\bigl(\Ione,\Itwo,\mac{\Ifibn{\kappa}-1}^2\bigr).
\end{equation}
Each fiber argument vanishes while its family is unstretched, and the material response stays isotropic until some family is stretched.

The reference configuration is free of energy and of isochoric stress (\emph{normalization conditions}):
\begin{equation}\label{eq:normalization}
    \tilde{\Wenergy}(3,3,1) = 0, \qquad \bP_{\mathrm{iso}}(\bI) = \tens{0},
\end{equation}
where $\Ione=\Itwo=3$ and $\Ifibn{\kappa}=1$ at $\bF=\bI$, so that the arguments of $\Wcheck$, Eq.~\eqref{eq:tension_only}, are $(3,3,0)$ there; without fibers the third argument is absent.

Furthermore, the energy must be \emph{polyconvex} \citep{Ball1976}, which implies ellipticity and thus excludes material instabilities, a prerequisite for solving the forward problem reliably. For an incompressible material, polyconvexity requires a representation $\isoc{\Wenergy}(\isoc{\bF})=g(\isoc{\bF},\cof\isoc{\bF})$ on $\SL$ with $g$ jointly convex in its two arguments \citep{Ball1976,Klein2026}. Since $\Ione=\isoc{\bF}:\isoc{\bF}$, $\Itwo=\cof\isoc{\bF}:\cof\isoc{\bF}$ and $\Ifibn{\kappa}=\norm{\isoc{\bF}\afib_\kappa}^2$ are convex functions of the pair $(\isoc{\bF},\cof\isoc{\bF})$, a sufficient condition is
\begin{equation}\label{eq:polyconvex}
    \tilde{\Wenergy}\ \text{ is jointly convex in } (\Ione,\Itwo,\Ifibn{\kappa}) \text{ and non-decreasing in each argument},
\end{equation}
because the composition of a convex, non-decreasing function with convex functions is convex. The same argument covers the tension-only form, Eq.~\eqref{eq:tension_only}: $x\mapsto\mac{x-1}^2$ is convex and non-decreasing, so $\mac{\Ifibn{\kappa}-1}^2$ is convex in $\isoc{\bF}$ as well, and it suffices that
\begin{equation}\label{eq:polyconvex_fiber}
  \Wcheck\ \text{ is jointly convex in its arguments and non-decreasing in each}.
\end{equation}
Lastly, the energy must be non-negative, $\tilde{\Wenergy}\geq 0$, so that the unloaded reference configuration is the state of lowest energy. For an isotropic material this requirement need not be imposed separately: $\Ione,\Itwo\geq 3$ for every $\isoc{\bF}\in\SL$, so it follows from Eqs.~\eqref{eq:normalization} and \eqref{eq:polyconvex}. With fibers it follows in the same way from the tension-only form, whose arguments $\mac{\Ifibn{\kappa}-1}^2$ are non-negative and vanish at $\bF=\bI$.

\subsection{Kolmogorov--Arnold networks}\label{sec:kan}
We represent the strain energy density of Sec.~\ref{sec:theory:continuum} by a Kolmogorov--Arnold network (KAN) \citep{Liu2024a}.

\paragraph*{Architecture} A KAN with $L$ hidden layers maps its input vector $\vect{z}^{(0)}\in\mathbb{R}^{n_0}$ to its output vector $\vect{z}^{(L+1)}\in\mathbb{R}^{n_{L+1}}$, where $n_l$ is the number of nodes in layer $l$.
The output of layer $l$ is the vector $\vect{z}^{(l)}$. Its element $z^{(l)}_{i}$, the output of node $i$ of layer $l$, passes through the univariate activation function $\phi^{(l)}_{ji}$ on the \emph{edge} from node $i$ of layer $l$ to node $j$ of layer $l+1$; the output of node $j$ is the sum of the inputs it receives from layer $l$:
\begin{equation}\label{eq:activation}
    z^{(l+1)}_{j} = \sum_{i=1}^{n_l} \phi^{(l)}_{ji}\bigl(z^{(l)}_{i}\bigr), \qquad j\in\{1,\ldots,n_{l+1}\}.
\end{equation}
Each activation function is the weighted sum of a fixed base function $\eta(z)$ and a spline $s^{(l)}_{ji}(z)$,
\begin{equation}\label{eq:activationfn}
    \phi^{(l)}_{ji}(z) = \alpha^{(l)}_{ji}\,\eta(z) + \beta^{(l)}_{ji}\,s^{(l)}_{ji}(z),
\end{equation}
where $\alpha^{(l)}_{ji}$ and $\beta^{(l)}_{ji}$ are weights. The spline is a linear combination of $n_s$ B-spline basis functions $B^{(l)}_k(z)$ with control points $c^{(l)}_{jik}\in\mathbb{R}$:
\begin{equation}\label{eq:spline}
    s^{(l)}_{ji}(z)=\sum_{k=1}^{n_s} c^{(l)}_{jik}\, B^{(l)}_k(z).
\end{equation}
Every edge $(l,j,i)$ thus carries its own weights and control points; where a statement holds for every edge, these indices are dropped below. The basis functions $B^{(l)}_k$ are B-splines of degree $d$ ($d=3$ in this work) on a uniform knot vector that divides the grid interval $[z^{(l)}_{\min},z^{(l)}_{\max}]$ of layer $l$ into $n_s-d$ subintervals and extends it by $d$ knots on either side. All edges of a layer share this knot vector.\footnote{B-splines of degree $d$ on a knot vector without repeated knots are $d-1$ times continuously differentiable. Hence, $d\geq 2$ yields a continuous stress, Eq.~\eqref{eq:stress}, and $d\geq 3$ a continuous tangent matrix of the forward problem (Sec.~\ref{sec:femu:forward}) within the grid intervals; at the ends of the grid intervals, where the splines pass into their linear continuation (Sec.~\ref{sec:ckan}), they are only once continuously differentiable.} We arrange the activation functions of layer $l$ in a matrix $\bm{\Phi}^{(l)}$:
\begin{equation}
    \vect{z}^{(l+1)} =
    \underbrace{
    \begin{bmatrix}
        \phi^{(l)}_{11}            & \phi^{(l)}_{12}            & \cdots & \phi^{(l)}_{1\,n_l}            \\
        \phi^{(l)}_{21}            & \phi^{(l)}_{22}            & \cdots & \phi^{(l)}_{2\,n_l}            \\
        \vdots                     & \vdots                     & \ddots & \vdots                         \\
        \phi^{(l)}_{n_{l+1}\,1}    & \phi^{(l)}_{n_{l+1}\,2}    & \cdots & \phi^{(l)}_{n_{l+1}\,n_l}      \\
    \end{bmatrix}}_{\eqqcolon\,\bm{\Phi}^{(l)}}
    \vect{z}^{(l)},
\end{equation}
where the matrix--vector product is understood in the sense of Eq.~\eqref{eq:activation}: the entry $\phi^{(l)}_{ji}$ acts on the component $z^{(l)}_i$, and the results are summed over $i$.
The KAN is the composition of its layers,
\begin{equation}\label{eq:KAN}
    \vect{z}^{(L+1)} = \operatorname{KAN}\bigl(\vect{z}^{(0)}\bigr) = \bigl(\bm{\Phi}^{(L)} \circ \bm{\Phi}^{(L-1)} \circ \cdots \circ \bm{\Phi}^{(1)} \circ \bm{\Phi}^{(0)}\bigr)\vect{z}^{(0)}.
\end{equation}

\paragraph*{Pruning and symbolic regression} A trained KAN can be reduced to a closed-form expression in two steps \citep{Liu2024a,Liu2024}. First, edges and nodes with attribution scores below a small threshold are \emph{pruned}. The attribution score of an edge or node measures its share in the output; it is computed layer by layer, passing the scores backward from the output to the input. Second, \emph{symbolic regression} replaces the activation function $\phi$ of each remaining edge by a symbolic function $\omega_1 f(\omega_2 z+\omega_3)+\omega_4$, where $f$ is a candidate from a fixed library. To this end, $\phi$ is sampled at the inputs that the training data produce, and every candidate is fitted to these samples: the inner parameters $(\omega_2,\omega_3)$ are swept over a grid, and the outer ones $(\omega_1,\omega_4)$ follow by least-squares fitting. The fitted candidate with the best score, which rewards the coefficient of determination of the fit and penalizes a complexity assigned to each library function, replaces $\phi$. Since the symbolic functions change the network, the coefficients $\omega_1,\ldots,\omega_4$ of all edges are then re-fitted with the training procedure of Sec.~\ref{sec:femu:training}.

\subsection{Constitutive Kolmogorov--Arnold networks}\label{sec:ckan}
The constitutive Kolmogorov--Arnold network (CKAN) \citep{Abdolazizi2025} meets the requirements of Sec.~\ref{sec:theory:continuum} by construction. It represents the energy by a KAN that takes as input the invariants the material symmetry admits and returns a scalar, $n_{L+1}=1$: for an isotropic material the two of Eq.~\eqref{eq:invariants}, $\vect{z}^{(0)}=(\Ione,\Itwo)$; for a fiber-reinforced material those of Eq.~\eqref{eq:tension_only}, $\vect{z}^{(0)}=\bigl(\Ione,\Itwo,\mac{\Ifibn{\kappa}-1}^2\bigr)$. We write $\vect{z}^{(0)}_{\bI}$ for the input in the reference configuration, Sec.~\ref{sec:theory:continuum}. The energy is $\Wckan$ of Eq.~\eqref{eq:energy_ckan} below, that is, $\tilde{\Wenergy}$ of Eq.~\eqref{eq:invariants} without fibers and $\Wcheck$ of Eq.~\eqref{eq:tension_only} with them, and the stress follows from Eq.~\eqref{eq:stress}, evaluated by automatic differentiation through the network.

\paragraph*{Constitutive requirements}
\emph{Objectivity and material symmetry}, Eq.~\eqref{eq:objiso}, hold identically: the network depends on $\isoc{\bF}$ only through the invariants, so the invariance follows from the choice of inputs and is not learned from data. For an isotropic material this is Eq.~\eqref{eq:objiso} with $\Gsym=\SO$.

The first of the \emph{normalization conditions}, Eq.~\eqref{eq:normalization}, is enforced by shifting the network output by its value in the reference configuration,
\begin{equation}\label{eq:energy_ckan}
\Wckan(\vect{z}^{(0)}) = \operatorname{KAN}(\vect{z}^{(0)}) - \operatorname{KAN}(\vect{z}^{(0)}_{\bI}).
\end{equation}
Since this shift is constant with respect to $\bF$, it does not affect the stress. The second normalization condition, $\bP_{\mathrm{iso}}(\bI)=\tens{0}$, holds without correction: $\partial\Ione/\partial\bF$ and $\partial\Itwo/\partial\bF$ vanish at $\bF=\bI$, and hence so does $\bP_{\mathrm{iso}}$. The fiber inputs need the chain rule, since $\partial\Ifibn{\kappa}/\partial\bF$ does not vanish at $\bF=\bI$, but the tension-only form does:
\begin{equation}\label{eq:fiber_stress}
  \frac{\partial\mac{\Ifibn{\kappa}-1}^2}{\partial\bF}=2\,\mac{\Ifibn{\kappa}-1}\,\frac{\partial\Ifibn{\kappa}}{\partial\bF}
\end{equation}
vanishes at $\Ifibn{\kappa}=1$ and thus at $\bF=\bI$.

\emph{Polyconvexity} is imposed through the architecture of the KAN: each activation function is constructed to be convex and non-decreasing, while the nodes only sum. A sum of convex, non-decreasing functions is convex and non-decreasing, as is the composition of convex, non-decreasing functions. Hence, $\Wckan$ is jointly convex and non-decreasing in its inputs, and the isochoric energy is polyconvex by Eq.~\eqref{eq:polyconvex} for an isotropic material and by Eq.~\eqref{eq:polyconvex_fiber} for a fiber-reinforced one.

\paragraph*{Convex, non-decreasing activation functions}
Three design choices make the activation functions, Eq.~\eqref{eq:activationfn}, convex and non-decreasing. First, the weights $\alpha$ and $\beta$ are constrained to be non-negative. Second, the base function is the softplus, $\eta(z)=\log(1+\exp(z))$, which is positive, convex, and non-decreasing on $\mathbb{R}$. Third, the control-point sequence $(c_k)$ of each spline is constrained to have non-negative and non-decreasing first differences, which, on the uniform knot vector, ensures that the spline is convex and non-decreasing on its grid interval. The constraints on the weights and control points are imposed by a reparameterization in terms of unconstrained parameters (Appendix~\ref{app:param}).

A spline decays to zero outside its grid interval, where it is neither convex nor non-decreasing. Each spline is therefore continued there by its endpoint tangents \citep{Abdolazizi2025}, so that $\operatorname{KAN}$ in Eq.~\eqref{eq:energy_ckan} is convex and non-decreasing on all of $\mathbb{R}$, irrespective of the grid (Appendix~\ref{app:grids}).

Pruning and symbolic regression (Sec.~\ref{sec:kan}) keep the activation functions convex and non-decreasing. Pruning only removes edges and nodes. Symbolic regression uses only candidates $f$ that are convex and non-decreasing on all of $\mathbb{R}$ and keeps $\omega_1$ and $\omega_2$ non-negative, so that every symbolic function $\omega_1 f(\omega_2 z+\omega_3)+\omega_4$ is convex and non-decreasing \citep{Abdolazizi2025}. The resulting closed-form energy therefore satisfies the same requirements as the CKAN it replaces.

\paragraph*{Parameters and implementation}
The unconstrained parameters of all edges are the trainable parameters of the CKAN, which we collect in the vector $\btheta\in\mathbb{R}^{n_\theta}$. The fiber directions are prescribed here, although they could be identified along with the network parameters.\footnote{With $n_0=2$ and $n_{L+1}=1$, and assuming that all $L\geq1$ hidden layers have the same width, $n_1=\ldots=n_L$, there are in total $n_\theta=\bigl(3n_1 + n_1^2(L-1)\bigr)(n_s+2)$ trainable parameters.} We implement the CKAN in JAX \citep{jax2018github}, starting from a port of the implementation of \citet{Liu2024}; JAX provides automatic differentiation and just-in-time compilation for evaluating the network and its derivatives.

\section{Finite element model updating with CKANs}\label{sec:femu}

\subsection{Forward problem}\label{sec:femu:forward}
\paragraph*{Strong form}
The boundary $\partial\Omega_0$ is decomposed into a Dirichlet part $\partial\Omega_{0}^{\bu}$ and a Neumann part $\partial\Omega_{0}^{\vect{t}}$, with $\partial\Omega_0=\partial\Omega_{0}^{\bu}\cup\partial\Omega_{0}^{\vect{t}}$ and $\partial\Omega_{0}^{\bu}\cap\partial\Omega_{0}^{\vect{t}}=\emptyset$.\footnote{Where only some displacement components are prescribed, e.g., on a plane of symmetry, the decomposition and Eqs.~\eqref{eq:bvp_displacement} and \eqref{eq:bvp_traction} apply per component: $u_i=\hat{u}_i$ on $\partial\Omega_0^{u_i}$ and $(\bP\vect{N})_i=\hat{t}_i$ on $\partial\Omega_0^{t_i}=\partial\Omega_0\setminus\partial\Omega_0^{u_i}$, and likewise for the function spaces in Eq.~\eqref{eq:spaces}.} The body is subjected to prescribed displacements $\hat{\bu}:\partial\Omega_{0}^{\bu}\times\mathcal{T}\to\mathbb{R}^3$, prescribed tractions per unit reference area $\hat{\vect{t}}:\partial\Omega_{0}^{\vect{t}}\times\mathcal{T}\to\mathbb{R}^3$, and a body force per unit reference volume $\vect{b}:\Omega_0\times\mathcal{T}\to\mathbb{R}^3$. The nonlinear boundary-value problem (BVP) of incompressible elastostatics is to find, for every $t\in\mathcal{T}$, the displacement field $\bu$ and the pressure field $p$ such that
\begin{subequations}\label{eq:bvp}
\begin{align}
    \Divg\bP(\bF,p;\btheta) + \vect{b} &= \vect{0} \quad\text{in }\Omega_0, \label{eq:bvp_equilibrium}\\
    J - 1 &= 0 \quad\text{in }\Omega_0, \label{eq:bvp_constraint}\\
    \bu &= \hat{\bu} \quad\text{on }\partial\Omega_{0}^{\bu}, \label{eq:bvp_displacement}\\
    \bP(\bF,p;\btheta)\,\vect{N} &= \hat{\vect{t}} \quad\text{on }\partial\Omega_{0}^{\vect{t}}, \label{eq:bvp_traction}
\end{align}
\end{subequations}
where $\bP(\bF,p;\btheta)$ is the stress of Eq.~\eqref{eq:stress}, which depends on the CKAN parameters $\btheta$, and $\vect{N}$ is the outward unit normal vector on $\partial\Omega_0$. Since the BVP cannot be solved analytically in general, we approximate its solution by the finite element method, starting from its weak form.

\paragraph*{Weak form}
The mixed weak form of Eq.~\eqref{eq:bvp} uses the function spaces
\begin{equation}\label{eq:spaces}
    \mathcal{S} = \{\bu \in [H^1(\Omega_0)]^3 \mid \bu = \hat{\bu}
        \text{ on } \partial\Omega_0^{\bu}\}, \qquad
    \mathcal{V} = \{\vect{v} \in [H^1(\Omega_0)]^3 \mid \vect{v} = \vect{0}
        \text{ on } \partial\Omega_0^{\bu}\}, \qquad
    \mathcal{Q} = L^2(\Omega_0)
\end{equation}
$\mathcal{S}$ for the displacements, $\mathcal{V}$ for their test functions, and $\mathcal{Q}$ for the pressure and its test functions. It reads: for every $t\in\mathcal{T}$, find $(\bu, p) \in \mathcal{S} \times \mathcal{Q}$ such that the residuals $r_u$ and $r_p$ vanish,
\begin{subequations}\label{eq:weakform}
\begin{align}
    r_u(\bu, p, \vect{v}; \btheta) &=
    \int_{\Omega_0} \bP(\bF, p; \btheta) : \Grad\vect{v} \dd V
    - \int_{\Omega_0} \vect{b} \cdot \vect{v} \dd V
    - \int_{\partial\Omega_0^{\vect{t}}} \hat{\vect{t}} \cdot \vect{v} \dd A = 0
    && \forall\, \vect{v} \in \mathcal{V}, \label{eq:weakform_u} \\
    r_p(\bu, q) &= -\int_{\Omega_0} q \,(J - 1) \dd V = 0
    && \forall\, q \in \mathcal{Q}. \label{eq:weakform_p}
\end{align}
\end{subequations}
\paragraph*{Discretization} We partition the domain $\Omega_0$ into $n_e$ elements $\Omega_0^e$, $e\in\{1,\ldots,n_e\}$, with $n_a$ nodes at the reference positions $\bX_a$, $a\in\{1,\ldots,n_a\}$. We approximate the displacements by continuous, piecewise trilinear functions on hexahedral elements and the pressure by piecewise constant ones, i.e., in the finite element spaces
\begin{equation}
    \mathcal{X}_h=\Bigl\{\sum_{a=1}^{n_a}N_a\vect{c}_a \Bigm| \vect{c}_a\in\mathbb{R}^3\Bigr\}\subset[H^1(\Omega_0)]^3, \qquad
    \mathcal{Q}_h=\{q\in\mathcal{Q} \mid q \text{ is constant in each } \Omega_0^e\},
\end{equation}
where $N_a$ is the shape function of node $a$, with $N_a(\bX_b)=\delta_{ab}$. The discrete test and trial spaces are
\begin{equation}
    \mathcal{V}_h = \mathcal{X}_h \cap \mathcal{V}, \qquad
    \mathcal{S}_h = \{\bu_h \in \mathcal{X}_h \mid \bu_h = \hat{\bu}_h \text{ on } \partial\Omega_0^{\bu}\},
\end{equation}
where $\hat{\bu}_h$ interpolates the prescribed displacements at the Dirichlet nodes. The Galerkin approximation reads: find $(\bu_h, p_h) \in \mathcal{S}_h \times \mathcal{Q}_h$ such that Eq.~\eqref{eq:weakform} holds for all $\vect{v}_h \in \mathcal{V}_h$ and $q_h \in \mathcal{Q}_h$. Since $q_h$ is constant per element, Eq.~\eqref{eq:weakform_p} enforces incompressibility in an element-averaged sense.

Trial and test functions are expanded in the same bases,
\begin{equation}\label{eq:expansion}
    \bu_h=\sum_{a=1}^{n_a}N_a\vect{d}_a, \quad \vect{v}_h=\sum_{a=1}^{n_a}N_a\,\delta\vect{d}_a, \qquad
    p_h=\sum_{e=1}^{n_e}M_e\,p_e, \quad q_h=\sum_{e=1}^{n_e}M_e\,\delta p_e,
\end{equation}
where the pressure shape function $M_e\in\mathcal{Q}_h$ equals one in $\Omega_0^e$ and zero elsewhere. The nodal values $\vect{d}_a,\delta\vect{d}_a\in\mathbb{R}^3$ and the element values $p_e,\delta p_e\in\mathbb{R}$ are collected in $\vect{d},\delta\vect{d}\in\mathbb{R}^{3n_a}$ and $\vect{p},\delta\vect{p}\in\mathbb{R}^{n_e}$. Inserting Eq.~\eqref{eq:expansion} into Eq.~\eqref{eq:weakform} gives $r_u=\delta\vect{d}\cdot\bR_u(\vect{d},\vect{p};\btheta)$ and $r_p=\delta\vect{p}\cdot\bR_p(\vect{d})$, where $\bR_u$ and $\bR_p$ collect, in the order of $\vect{d}$ and $\vect{p}$, the nodal and element residuals
\begin{equation}\label{eq:global_residual}
\begin{aligned}
    \bR_{u,a}(\vect{d},\vect{p};\btheta) &= \int_{\Omega_0}\bigl(\bP(\bF_h,p_h;\btheta)\,\Grad N_a - N_a\,\vect{b}\bigr)\dd V
      - \int_{\partial\Omega_0^{\vect{t}}} N_a\,\hat{\vect{t}}\dd A, & \quad a\in\{1,\ldots,n_a\},\\
    R_{p,e}(\vect{d}) &= -\int_{\Omega_0^e} (J_h-1)\dd V,  & \quad e\in\{1,\ldots,n_e\},
\end{aligned}
\end{equation}
with $\bF_h=\bI+\Grad\bu_h$ and $J_h=\det\bF_h$. The integrals are evaluated element by element and assembled into $\bR_u$ through the element connectivity.

The displacement degrees of freedom, i.e., the entries of $\vect{d}$, are split into the set $\mathcal{D}\subset\{1,\ldots,3n_a\}$ of those prescribed on $\partial\Omega_0^{\bu}$, known from $\hat{\bu}_h$, and the set $\mathcal{F}=\{1,\ldots,3n_a\}\setminus\mathcal{D}$ of the free ones. A subscript $\mathcal{D}$ or $\mathcal{F}$ selects the corresponding entries of an array; after reordering, $\vect{d}=(\vect{d}_\mathcal{F},\vect{d}_\mathcal{D})$ and $\bR_u=(\bR_{u,\mathcal{F}},\bR_{u,\mathcal{D}})$. Since the mesh conforms to $\partial\Omega_0^{\bu}$, $\vect{v}_h\in\mathcal{V}_h$ exactly when $\delta\vect{d}_\mathcal{D}=\vect{0}$, so the Galerkin approximation is equivalent to
\begin{equation}\label{eq:discrete_residual}
    \bR(\vect{d}, \vect{p}; \btheta) =
    \begin{bmatrix}
        \bR_{u,\mathcal{F}}(\vect{d}, \vect{p}; \btheta) \\
        \bR_p(\vect{d})
    \end{bmatrix} = \vect{0}\in\mathbb{R}^{|\mathcal{F}|+n_e}
\end{equation}
for the unknowns $\vect{d}_\mathcal{F}$ and $\vect{p}$. The residuals $\bR_{u,\mathcal{D}}$ at the prescribed displacement degrees of freedom are the reaction forces (Sec.~\ref{sec:femu:objective}). All three specimens of Sec.~\ref{sec:applications} are thin bodies, and each is computed with the corresponding reduction of this formulation (Appendix~\ref{app:thin}): the sheets of Cases~1 and~2 in plane stress on bilinear quadrilaterals, the arterial wall of Case~3 as a membrane on linear triangles. In both reductions the incompressibility constraint is eliminated pointwise through the thickness stretch, so that no pressure unknown remains and the tangent is symmetric; the objective function of Sec.~\ref{sec:femu:objective} and the sensitivities of Sec.~\ref{sec:femu:sensitivities} apply unchanged. The mixed formulation is the general statement from which both reductions follow, and the three-dimensional model against which the plane-stress results are checked (Appendix~\ref{app:thin}).

\paragraph*{Newton--Raphson solution} At each $t\in\mathcal{T}$, with $\vect{d}_\mathcal{D}$ fixed by $\hat{\bu}_h$, we solve the nonlinear system of Eq.~\eqref{eq:discrete_residual} for the unknowns $\vect{y}=(\vect{d}_\mathcal{F},\vect{p})\in\mathbb{R}^{|\mathcal{F}|+n_e}$ by the Newton--Raphson method \citep{Bonet2008}. We write $\bR(\vect{y};\btheta)$ for $\bR(\vect{d},\vect{p};\btheta)$, and likewise $\bR_{u,\mathcal{D}}(\vect{y};\btheta)$ for $\bR_{u,\mathcal{D}}(\vect{d},\vect{p};\btheta)$, with $\vect{d}_\mathcal{D}$ and the loads $\vect{b}$, $\hat{\vect{t}}$ taken at the current instant. Starting from an initial guess $\vect{y}_0$, each iteration solves the linearized system for the increment $\Delta\vect{y}_i$ and updates the iterate,
\begin{equation}\label{eq:newton}
    \tens{K}(\vect{y}_i;\btheta)\,\Delta\vect{y}_i = -\bR(\vect{y}_i;\btheta), \qquad
    \vect{y}_{i+1} = \vect{y}_i + \Delta\vect{y}_i, \qquad i=0,1,\ldots,
\end{equation}
with the tangent matrix
\begin{equation}\label{eq:tangent}
    \tens{K}=\frac{\partial\bR}{\partial\vect{y}}=
    \begin{bmatrix}
        \partial\bR_{u,\mathcal{F}}/\partial\vect{d}_\mathcal{F} & \partial\bR_{u,\mathcal{F}}/\partial\vect{p}\\
        \partial\bR_p/\partial\vect{d}_\mathcal{F} & \tens{0}
    \end{bmatrix}.
\end{equation}
The iteration stops once $\norm{\bR(\vect{y}_i;\btheta)}\leq\varepsilon_{\mathrm{abs}}$ or $\norm{\bR(\vect{y}_i;\btheta)}\leq\varepsilon_{\mathrm{rel}}\norm{\bR(\vect{y}_0;\btheta)}$, with $\varepsilon_{\mathrm{abs}}=10^{-8}$ and $\varepsilon_{\mathrm{rel}}=10^{-10}$. We compute $\tens{K}$ by automatic differentiation.

We implement the forward problem in JAX-FEM \citep{Xue2023}, a differentiable finite element solver written in JAX \citep{jax2018github}, which allows us to seamlessly integrate the CKANs into the FE model and to differentiate the residual assembly with respect to the unknowns and the CKAN parameters by automatic differentiation, as the adjoint method of Sec.~\ref{sec:femu:sensitivities} requires.

\subsection{Objective function}\label{sec:femu:objective}

For given CKAN parameters $\btheta$, the forward problem of Sec.~\ref{sec:femu:forward} yields the FE solution $\vect{y}(\btheta)=(\vect{d}_\mathcal{F}(\btheta),\vect{p}(\btheta))$. With it follow the nodal displacements $\vect{d}(\btheta)=(\vect{d}_\mathcal{F}(\btheta),\vect{d}_\mathcal{D})$, where $\vect{d}_\mathcal{D}$ is prescribed, and the reaction forces $\bR_{u,\mathcal{D}}(\btheta)=\bR_{u,\mathcal{D}}(\vect{d}(\btheta),\vect{p}(\btheta);\btheta)$ at the prescribed displacement degrees of freedom. We write the argument $\btheta$ where the dependence on the parameters matters; without it, $\vect{d}$, $\vect{p}$, and $\vect{y}$ denote variables, e.g., the arguments of the residuals, or values at the FE solution. In general, these simulated quantities differ from the measured ones. To quantify the mismatch, we define two \emph{losses}, which, together with a sparsity-promoting penalty, form the \emph{objective function}. The inverse problem is to identify the CKAN parameters $\btheta^{\star}$ that minimize the objective function, so that the FE model reproduces the measurements as closely as possible.

\paragraph*{Measurements} From an experiment, we obtain measurements at $n_T$ time instants ${}^mt\in\mathcal{T}$, $m\in\{1,\ldots,n_T\}$, where a left superscript $m$ marks quantities at ${}^mt$:
\begin{itemize}
  \item \emph{displacements} ${}^m\bu^{\obs}_j\in\mathbb{R}^3$ at $n_u$ material points with reference positions $\bX^{\obs}_j\in\Omega_0\cup\partial\Omega_0^{\vect{t}}$, $j\in\{1,\ldots,n_u\}$, e.g., from digital image correlation (DIC); points on $\partial\Omega_0^{\bu}$ are excluded, since their displacement is prescribed;
  \item \emph{reaction forces} ${}^mf^{\obs}_k\in\mathbb{R}$, $k\in\{1,\ldots,n_f\}$, i.e., the components along unit vectors $\vect{s}_k$ of the resultant forces exerted on the body through parts $\Gamma_k\subset\partial\Omega_0^{\bu}$ of the Dirichlet boundary, e.g., from force transducers, which measure one component per axis; for a multi-axis transducer, several $\Gamma_k$ coincide.
\end{itemize}
Per time instant, these measurements are collected in ${}^m\bu^{\obs}=({}^m\bu^{\obs}_1,\ldots,{}^m\bu^{\obs}_{n_u})\in\mathbb{R}^{3n_u}$ and ${}^m\vect{f}^{\obs}=({}^mf^{\obs}_1,\ldots,{}^mf^{\obs}_{n_f})\in\mathbb{R}^{n_f}$.

The losses compare these measurements with the corresponding quantities of the FE model, but the two are not defined at the same places: the displacements are measured at the points $\bX^{\obs}_j$, which in general do not coincide with the FE nodes, and the measured forces are components of resultants over the boundary parts $\Gamma_k$, whereas the FE model provides nodal reaction forces. Two \emph{observation operators}, one for displacements and one for forces, map the FE quantities to the quantities that are measured.

\paragraph*{Displacement loss} The displacement observation operator $\Qu\in\mathbb{R}^{3n_u\times 3n_a}$ evaluates the FE displacement field, Eq.~\eqref{eq:expansion}, at the measurement points,
\begin{equation}\label{eq:Qu}
  \Qu\,\vect{d}=\bigl(\bu_h(\bX^{\obs}_1),\ldots,\bu_h(\bX^{\obs}_{n_u})\bigr), \qquad
  \bu_h(\bX^{\obs}_j)=\sum_{a=1}^{n_a}N_a(\bX^{\obs}_j)\,\vect{d}_a,
\end{equation}
ordered like the measurements ${}^m\bu^{\obs}$; only the nodes of the element containing $\bX^{\obs}_j$ contribute to the sum. The entries of $\Qu$ are the shape functions evaluated at the measurement points, $N_a(\bX^{\obs}_j)$, so $\Qu$ is a constant matrix.\footnote{In practice, often only some displacement components are measured; e.g., planar (2D) DIC observes only the in-plane ones. $\Qu$ then keeps only the rows of the observed components, which restricts the displacement loss to them.} The displacement loss thus reads
\begin{equation}\label{eq:displacement_loss}
  \mathcal{L}_{u}(\btheta) = \sum_{m=1}^{n_T} \norm{\Qu\,{}^m\vect{d}(\btheta) - {}^m\bu^{\obs}}^2.
\end{equation}

\paragraph*{Force loss} The nodal reaction forces $\bR_{u,\mathcal{D}}$ are the forces that the supports exert on the body; they are the residuals at the prescribed displacement degrees of freedom, which Eq.~\eqref{eq:discrete_residual} excludes. The force observation operator $\QF\in\mathbb{R}^{n_f\times|\mathcal{D}|}$ sums the components of these reactions along $\vect{s}_k$ over the nodes on each measured boundary part,\footnote{The products $\vect{s}_k\cdot\bR_{u,a}$ in Eq.~\eqref{eq:QF} involve only entries of $\bR_{u,\mathcal{D}}$ if the nodes on $\Gamma_k\subset\partial\Omega_0^{\bu}$ are prescribed in every displacement component in which $\vect{s}_k$ is nonzero, e.g., only in the $i$-th one for $\vect{s}_k=\vect{e}_i$; their other components may be free.}
\begin{equation}\label{eq:QF}
  \QF\,\bR_{u,\mathcal{D}}=\Bigl(\vect{s}_1\cdot\sum_{a:\,\bX_a\in\Gamma_1}\bR_{u,a},\ldots,\vect{s}_{n_f}\cdot\sum_{a:\,\bX_a\in\Gamma_{n_f}}\bR_{u,a}\Bigr),
\end{equation}
ordered like the measurements ${}^m\vect{f}^{\obs}$. Like $\Qu$, $\QF$ is a constant matrix; its entries are the components of $\vect{s}_k$ at the nodes on $\Gamma_k$ and zero elsewhere. The force loss reads
\begin{equation}\label{eq:force_loss}
  \mathcal{L}_{f}(\btheta) = \sum_{m=1}^{n_T} \norm{\QF\,{}^m\bR_{u,\mathcal{D}}(\btheta) - {}^m\vect{f}^{\obs}}^2.
\end{equation}


\paragraph*{Requirements on the FE model} The two losses compare FE quantities with measurements, which constrains the FE model beyond the forward problem of Sec.~\ref{sec:femu:forward}. The FE domain must contain the measurement points, $\bX^{\obs}_j\in\Omega_0\cup\partial\Omega_0^{\vect{t}}$, where $\Qu$ evaluates the FE displacement field; FE model and measurement must therefore share one coordinate frame. Each boundary part $\Gamma_k$ must carry the resultant that the corresponding force transducer measures. The prescribed displacements, finally, must reproduce those of the experiment, on which FEMU depends critically \citep[Sec.~5.2]{Chen2025}. The three cases of Sec.~\ref{sec:applications} meet these requirements in different ways; each states its FE domain, its boundary conditions, and where their values come from.

\paragraph*{Identifiability of the energy scale} If the FE model carries no loads besides the prescribed displacements, $\vect{b}=\vect{0}$ and $\hat{\vect{t}}=\vect{0}$, its displacements do not change when the energy is rescaled, $\tilde{\Wenergy}\mapsto c\,\tilde{\Wenergy}$ with $c>0$: scaling the hydrostatic pressure by $c$ as well scales $\bR_{u,\mathcal{F}}$ by $c$ and leaves $\bR_p$ untouched, so the same displacements satisfy Eq.~\eqref{eq:discrete_residual}, while the reactions $\bR_{u,\mathcal{D}}$ scale by $c$. The displacement loss, and equally a loss on any other kinematic observation, is then blind to the scale of the energy, which the force loss alone determines. A load that does not scale with the energy, in Sec.~\ref{sec:applications} the cavity pressure of Case~3, removes the invariance and lets the kinematic measurements fix the scale as well, so that no force has to be measured there.

\paragraph*{Sparsity penalty} A sparsity-promoting penalty $\Pen(\btheta)$ on the network complements the two losses. It drives the magnitudes of redundant activation functions toward zero, so that pruning (Sec.~\ref{sec:kan}) can remove them, and thus prepares the extraction of a closed-form energy from the trained CKAN (Appendix~\ref{app:penalty}).

\paragraph*{Minimization problem} The CKAN parameters are identified by minimizing the weighted sum of the losses and the penalty under the constraint that, at every measured time instant, the unknowns ${}^m\vect{y}=({}^m\vect{d}_\mathcal{F},{}^m\vect{p})$ satisfy the discretized forward problem, Eq.~\eqref{eq:discrete_residual}, with ${}^m\vect{d}_\mathcal{D}$ prescribed by $\hat{\bu}_h$ at ${}^mt$,
\begin{equation}\label{eq:objective}
\begin{gathered}
  \btheta^\star = \argmin_{\btheta\in\mathbb{R}^{n_\theta}} \mathcal{J}(\btheta)
  \quad\text{subject to}\quad \bR\bigl({}^m\vect{y}(\btheta);\btheta\bigr)=\vect{0},\quad m=1,\ldots,n_T,\\
  \mathcal{J}(\btheta) = \mathcal{L}_{u}(\btheta) + \Lambda\,\mathcal{L}_{f}(\btheta) + \gamma\,\Pen(\btheta),
\end{gathered}
\end{equation}
with the objective function $\mathcal{J}$, the weighting factor $\Lambda>0$ of the force loss, and the penalty weight $\gamma\geq0$. Minimizing $\mathcal{J}$ requires its gradient, which the next section derives.



\subsection{Sensitivities by the discrete adjoint method}\label{sec:femu:sensitivities}

We compute the gradient of the objective, Eq.~\eqref{eq:objective}, with the discrete adjoint method, a discretize-then-optimize approach \citep[Sec.~5.1]{Kidger2022}: the adjoint equations follow from implicit differentiation of the discretized forward problem, Eq.~\eqref{eq:discrete_residual}, at its solution, and thus avoid backpropagation through the Newton--Raphson iterations, Eq.~\eqref{eq:newton}.

\paragraph*{Decomposition of the objective} The objective, Eq.~\eqref{eq:objective}, consists of two parts that depend on $\btheta$ differently. The penalty $\Pen$ depends on $\btheta$ only explicitly, and its gradient follows by reverse-mode automatic differentiation of Eq.~\eqref{eq:penalty}. The losses depend on $\btheta$ also implicitly, through the FE solution $\vect{y}(\btheta)$; we derive their gradient below. Both losses are sums over the time instants, so we consider one time instant and drop the superscript $m$. The contribution of this time instant to $\mathcal{L}_u+\Lambda\mathcal{L}_f$, as a function of the displacements and the reaction forces, is
\begin{equation}\label{eq:instant_loss}
  \ell(\vect{d},\bR_{u,\mathcal{D}})=\norm{\Qu\vect{d}-\bu^{\obs}}^2+\Lambda\norm{\QF\bR_{u,\mathcal{D}}-\vect{f}^{\obs}}^2 .
\end{equation}

\paragraph*{Chain rule} The loss $\ell$ depends on $\btheta$ through its two arguments, the displacements $\vect{d}(\btheta)$ and the reactions $\bR_{u,\mathcal{D}}(\btheta)$. By the chain rule, its gradient is
\begin{equation}\label{eq:loss_total}
  \frac{\diff\ell}{\diff\btheta}
  =\Bigl(\frac{\diff\vect{d}}{\diff\btheta}\Bigr)^{\!\top}\vect{g}_u+\Bigl(\frac{\diff\bR_{u,\mathcal{D}}}{\diff\btheta}\Bigr)^{\!\top}\vect{g}_f,
\end{equation}
with the partial derivatives of $\ell$ with respect to its arguments, evaluated at the FE solution,
\begin{equation}\label{eq:loss_partials}
  \vect{g}_u=\frac{\partial\ell}{\partial\vect{d}}=2\,\Qu^\top\bigl(\Qu\vect{d}-\bu^{\obs}\bigr)\in\mathbb{R}^{3n_a}, \qquad
  \vect{g}_f=\frac{\partial\ell}{\partial\bR_{u,\mathcal{D}}}=2\Lambda\,\QF^\top\bigl(\QF\bR_{u,\mathcal{D}}-\vect{f}^{\obs}\bigr)\in\mathbb{R}^{|\mathcal{D}|}.
\end{equation}

The displacements $\vect{d}=(\vect{d}_\mathcal{F},\vect{d}_\mathcal{D})$ depend on $\btheta$ only through the unknowns $\vect{y}=(\vect{d}_\mathcal{F},\vect{p})$, since the prescribed displacements $\vect{d}_\mathcal{D}$ are fixed. With $\diff\vect{d}/\diff\btheta=(\partial\vect{d}/\partial\vect{y})\,\diff\vect{y}/\diff\btheta$, the first term of Eq.~\eqref{eq:loss_total} can therefore be written in terms of $\vect{y}$,
\begin{equation}\label{eq:gy}
  \Bigl(\frac{\diff\vect{d}}{\diff\btheta}\Bigr)^{\!\top}\vect{g}_u
  =\Bigl(\frac{\diff\vect{y}}{\diff\btheta}\Bigr)^{\!\top}\vect{g}_y, \qquad
  \vect{g}_y=\Bigl(\frac{\partial\vect{d}}{\partial\vect{y}}\Bigr)^{\!\top}\vect{g}_u=(\vect{g}_{u,\mathcal{F}},\vect{0})\in\mathbb{R}^{|\mathcal{F}|+n_e}.
\end{equation}
The vector $\vect{g}_y$, partitioned like $\vect{y}$, is thus the derivative of $\ell$ with respect to the unknowns through the displacements. Since $\partial\vect{d}/\partial\vect{y}$ consists of an identity block for $\vect{d}_\mathcal{F}$ and zeros otherwise, $\vect{g}_y$ contains the free entries $\vect{g}_{u,\mathcal{F}}$ of $\vect{g}_u=(\vect{g}_{u,\mathcal{F}},\vect{g}_{u,\mathcal{D}})$ and zeros in place of the pressure; $\vect{g}_{u,\mathcal{D}}$ drops out because $\vect{d}_\mathcal{D}$ does not depend on $\vect{y}$. In the reductions of Appendix~\ref{app:thin}, which the three cases use, there is no pressure unknown: $\vect{y}=\vect{d}_\mathcal{F}$, the pressure block of $\vect{g}_y$ and of the adjoint variable below is absent, and the derivation holds unchanged.

The reactions $\bR_{u,\mathcal{D}}(\btheta)=\bR_{u,\mathcal{D}}(\vect{y}(\btheta);\btheta)$ depend on $\btheta$ in two ways: explicitly, through the stress $\bP(\bF,p;\btheta)$, and implicitly, through the FE solution $\vect{y}(\btheta)$. Accordingly, their total derivative splits into two terms,
\begin{equation}\label{eq:reaction_derivative}
  \frac{\diff\bR_{u,\mathcal{D}}}{\diff\btheta}=\frac{\partial\bR_{u,\mathcal{D}}}{\partial\vect{y}}\,\frac{\diff\vect{y}}{\diff\btheta}+\frac{\partial\bR_{u,\mathcal{D}}}{\partial\btheta},
\end{equation}
where, here and in the following, $\partial/\partial\btheta$ captures the explicit dependence only.

Inserting Eqs.~\eqref{eq:gy} and \eqref{eq:reaction_derivative} into Eq.~\eqref{eq:loss_total} and collecting the terms with $\diff\vect{y}/\diff\btheta$ gives
\begin{equation}\label{eq:chain}
  \frac{\diff\ell}{\diff\btheta}
  =\Bigl(\frac{\diff\vect{y}}{\diff\btheta}\Bigr)^{\!\top}\underbrace{\Bigl[\vect{g}_y+\Bigl(\frac{\partial\bR_{u,\mathcal{D}}}{\partial\vect{y}}\Bigr)^{\!\top}\vect{g}_f\Bigr]}_{\eqqcolon\,\vect{h}\,\in\,\mathbb{R}^{|\mathcal{F}|+n_e}}
  +\Bigl(\frac{\partial\bR_{u,\mathcal{D}}}{\partial\btheta}\Bigr)^{\!\top}\vect{g}_f .
\end{equation}
The vector $\vect{h}$ is the derivative of $\ell$ with respect to the unknowns through both arguments, the displacements and the reaction forces. The first term contains the sensitivity $\diff\vect{y}/\diff\btheta$ of the FE solution, which is not known; the second contains the explicit dependence of the reactions.

\paragraph*{Tangent linear system} The FE solution satisfies $\bR(\vect{y}(\btheta);\btheta)=\vect{0}$, Eq.~\eqref{eq:discrete_residual}, for every $\btheta$. Differentiating this identity with respect to $\btheta$ gives, with $\tens{K}=\partial\bR/\partial\vect{y}$ of Eq.~\eqref{eq:tangent},
\begin{equation}\label{eq:implicit}
  \tens{K}\,\frac{\diff\vect{y}}{\diff\btheta}=-\frac{\partial\bR}{\partial\btheta},
\end{equation}
the tangent linear system for the sensitivity $\diff\vect{y}/\diff\btheta\in\mathbb{R}^{(|\mathcal{F}|+n_e)\times n_\theta}$. The direct way to evaluate Eq.~\eqref{eq:chain} solves Eq.~\eqref{eq:implicit} for $\diff\vect{y}/\diff\btheta$ and inserts the result into the first term of Eq.~\eqref{eq:chain}. This requires one linear solve with $\tens{K}$ per parameter, with the $n_\theta$ columns of $\partial\bR/\partial\btheta$ as right-hand sides, so its cost grows with the number of parameters.

\paragraph*{Adjoint system} Equation~\eqref{eq:chain}, however, does not require $\diff\vect{y}/\diff\btheta$ itself, only its product with the single vector $\vect{h}$. Inserting $\diff\vect{y}/\diff\btheta=-\tens{K}^{-1}\,\partial\bR/\partial\btheta$ from Eq.~\eqref{eq:implicit} into this product gives
\begin{equation}\label{eq:substitution}
  \Bigl(\frac{\diff\vect{y}}{\diff\btheta}\Bigr)^{\!\top}\vect{h}
  =-\overbrace{\Bigl(\frac{\partial\bR}{\partial\btheta}\Bigr)^{\!\top}
     \mathrlap{\underbrace{\phantom{\tens{K}^{-\top}\vect{h}\,}}_{\mathclap{\eqqcolon\,\vect{\xi}:\ \text{one solve}}}}
     \tens{K}^{-\top}}^{\mathclap{n_\theta\ \text{solves}}}\vect{h}.
\end{equation}
Grouping the first two factors is the direct way; grouping the last two requires a single solve for the adjoint variable $\vect{\xi}$, independent of the number of parameters, followed by its product with $(\partial\bR/\partial\btheta)^\top$. With $\vect{\xi}=(\vect{\xi}_\mathcal{F},\vect{\xi}_p)\in\mathbb{R}^{|\mathcal{F}|+n_e}$, partitioned like $\vect{y}$, the adjoint system and the gradient of the loss read
\begin{equation}\label{eq:adjoint}
  \tens{K}^\top\vect{\xi}=\vect{h}, \qquad
  \frac{\diff\ell}{\diff\btheta}=\Bigl(\frac{\partial\bR_{u,\mathcal{D}}}{\partial\btheta}\Bigr)^{\!\top}\vect{g}_f-\Bigl(\frac{\partial\bR}{\partial\btheta}\Bigr)^{\!\top}\vect{\xi}.
\end{equation}

\paragraph*{Evaluation from the forward solve} All quantities in Eq.~\eqref{eq:adjoint} are obtained at the converged Newton--Raphson iterate, without assembling an additional matrix (Appendix~\ref{app:sensitivities}). Per time instant, the gradient thus costs, independent of $n_\theta$, one linear solve with the tangent $\tens{K}$ of that iterate and one vector--Jacobian product, evaluated by reverse-mode automatic differentiation of the residual assembly.

\paragraph*{Gradient of the objective} Summing over the time instants and adding the penalty contribution gives the gradient of the objective,
\begin{equation}\label{eq:total_gradient}
  \frac{\diff\mathcal{J}}{\diff\btheta}=\sum_{m=1}^{n_T}\frac{\diff\,{}^m\ell}{\diff\btheta}+\gamma\,\frac{\partial\Pen}{\partial\btheta}.
\end{equation}


\subsection{Training}\label{sec:femu:training}


\paragraph*{Initialization} 
The forward problem must be solvable from the first epoch, so we start the CKAN from a physically meaningful guess \citep[Sec.~6.7]{Chen2025}. 
We fit the energy of the CKAN and its derivatives to those of a classical model at invariant pairs of isochoric deformations, adjusting only the weights $\alpha$ of the base functions, Eq.~\eqref{eq:activationfn}, while the control points of the splines keep their random initial values, convex by construction (Sec.~\ref{sec:ckan}).
For isotropic materials we use the Neo-Hookean model $\tfrac{\mu}{2}(\Ione-3)$; for anisotropic materials we use the Holzapfel--Gasser--Ogden model \citep{Holzapfel2000} with its fiber term linearized.
We train each set from several random initializations and take the run with the lowest loss, without the penalty, as the identified CKAN.

\paragraph*{Training with the FE model} We train the CKAN by minimizing $\mathcal{J}$ of Eq.~\eqref{eq:objective}. A first pass solves every time instant once at the initial parameters, in order, each from the solution at the preceding instant, so that every instant has a starting iterate. The first instant starts from the reference configuration. Should a solve fail, the prescribed displacements are applied in increments from the reference configuration, whose number is doubled until it converges. One epoch then sweeps the $n_T$ time instants in a fixed order. After each instant, Adam updates $\btheta$ with the gradient of ${}^m\ell$ from Sec.~\ref{sec:femu:sensitivities} plus that of $(\gamma/n_T)\,\Pen$, as in \citet[Alg.~1]{Knipper2026}; over an epoch, these terms sum to $\mathcal{J}$. Each solve starts from the solution at the same time instant for the previous parameters and converges within a few Newton iterations. A time instant whose solve or gradient fails is retried from the starting configuration of the initial solves and otherwise skipped for that epoch. The learning rate is constant or follows a cyclic schedule, and training runs for a fixed number of epochs. Since $\btheta$ changes within an epoch, the sum of the losses, $\mathcal{L}_u+\Lambda\mathcal{L}_f$, is re-evaluated at fixed $\btheta$ every few epochs, and we keep the parameters with the lowest re-evaluated sum.

\paragraph*{Training on homogeneous tests} Homogeneous tests, such as uniaxial tension and pure shear, provide stress--stretch curves directly. Here the stress of the CKAN is fitted directly to the measured stress, unlike in the training with the FE model, where no stress is measured and the CKAN is fitted only through the displacements and forces its stress produces. When training on these curves, the loss is the mean squared error of the nominal stress $P_{11}$ in the loading direction over all measured points, and Adam minimizes it with the gradient over all points, a cyclic learning rate and the same sparsity penalty.



\section{Applications to experimental data}\label{sec:applications}

In this section, we demonstrate the applicability of FEMU with CKANs using three different experimental datasets, which differ in the quantity and type of kinematic information they provide: full-field displacements, a one-dimensional stretch profile, and three-dimensional displacements under pressure. For each dataset, we describe the specimens and measurements and the finite element model used to simulate the experiments, and then present the results.

\subsection{Case~1 --- Full-field displacement data}\label{sec:apps:case1}

\begin{figure*}[!h]
  \centering
  \input{figures/fe_geometry_panel}%
  \tikzset{gaugewedge/.style={draw=black!55, line width=0.4pt,
                              dash pattern=on 2pt off 1.5pt}}
\def\specimenScaleNote{}%
  \tikzset{
    specimenbody/.style={fill=black!22, draw=black!70, line width=0.3pt},
    specimenhole/.style={fill=white, draw=black!70, line width=0.3pt},
    specimenfe/.style={fill={rgb,255:red,243;green,233;blue,216}},
    specimencut/.style={draw={rgb,255:red,238;green,102;blue,119}, line width=0.9pt},
  }
  \setlength{\tabcolsep}{0pt}
  \begin{tabular*}{\textwidth}{@{\extracolsep{\fill}}cccc@{}}
    \toprule
    Uniaxial tension (UT) & Pure shear (PS) & Tensile test 1 (TT1) & Tensile test 2 (TT2) \\
    \midrule
    \addlinespace[1.0ex]
  \begin{tikzpicture}[x=0.11000mm, y=0.11000mm, baseline=7.7000mm]
    \filldraw[specimenbody] (0,0) -- (77,0) -- (77,30) arc[start angle=0.0000, end angle=58.5024, radius=5] -- (55.8877,45.7367) arc[start angle=238.5024, end angle=180.0000, radius=5] -- (53.5,90) arc[start angle=180.0000, end angle=121.4976, radius=5] -- (74.6123,105.7367) arc[start angle=301.4976, end angle=360.0000, radius=5] -- (77,140) -- (0,140) -- (0,110) arc[start angle=180.0000, end angle=238.5024, radius=5] -- (21.1123,94.2633) arc[start angle=418.5024, end angle=360.0000, radius=5] -- (23.5,50) arc[start angle=0.0000, end angle=-58.5024, radius=5] -- (2.3877,34.2633) arc[start angle=121.4976, end angle=180.0000, radius=5] -- cycle;
  \end{tikzpicture}
  &
  \begin{tikzpicture}[x=0.11000mm, y=0.11000mm, baseline=5.5000mm]
    \filldraw[specimenbody] (0,0) rectangle (210,100);
  \end{tikzpicture}
  &
  \begin{tikzpicture}[x=0.11000mm, y=0.11000mm, baseline=9.2180mm, remember picture]
    \filldraw[specimenbody] (0,0) -- (210,0) -- (210,38.8) arc[start angle=0.0000, end angle=66.6430, radius=7] -- (189.2248,52.3736) arc[start angle=246.6430, end angle=180.0000, radius=7] -- (185,108.8) arc[start angle=180.0000, end angle=113.3570, radius=7] -- (205.7752,122.3736) arc[start angle=293.3570, end angle=360.0000, radius=7] -- (210,167.6) -- (0,167.6) -- (0,128.8) arc[start angle=180.0000, end angle=246.6430, radius=7] -- (20.7752,115.2264) arc[start angle=426.6430, end angle=360.0000, radius=7] -- (25,58.8) arc[start angle=0.0000, end angle=-66.6430, radius=7] -- (4.2248,45.2264) arc[start angle=113.3570, end angle=180.0000, radius=7] -- cycle;
    \fill[specimenfe] (24.518,57.977) -- (184.518,58.026) -- (184.518,108.594) -- (24.518,109.108) -- cycle;
    \coordinate (specimenFEtl) at (24.518,109.108);
    \coordinate (specimenFEtr) at (184.518,108.594);
    \coordinate (specimenFEbl) at (24.518,57.977);
    \coordinate (specimenFEbr) at (184.518,58.026);
    \filldraw[specimenhole] (45,95) ellipse [x radius=10, y radius=5];
    \filldraw[specimenhole][rotate around={45:(80,90)}] (80,90) ellipse [x radius=14.15, y radius=7.2];
    \filldraw[specimenhole][rotate around={-27:(50,70)}] (50,70) ellipse [x radius=11.2, y radius=4.6];
    \filldraw[specimenhole] (95,70) ellipse [x radius=15, y radius=5];
    \filldraw[specimenhole][rotate around={-24:(129,83.8)}] (129,83.8) ellipse [x radius=20.8, y radius=11.1];
    \filldraw[specimenhole][rotate around={27:(165,95)}] (165,95) ellipse [x radius=11.15, y radius=6.85];
    \draw[specimencut] (24.518,57.977) -- (184.518,58.026);
    \draw[specimencut] (184.518,108.594) -- (24.518,109.108);
  \end{tikzpicture}
  &
  \begin{tikzpicture}[x=0.11000mm, y=0.11000mm, baseline=9.2180mm]
    \filldraw[specimenbody] (0,0) -- (210,0) -- (210,38.8) arc[start angle=0.0000, end angle=66.6430, radius=7] -- (189.2248,52.3736) arc[start angle=246.6430, end angle=180.0000, radius=7] -- (185,108.8) arc[start angle=180.0000, end angle=113.3570, radius=7] -- (205.7752,122.3736) arc[start angle=293.3570, end angle=360.0000, radius=7] -- (210,167.6) -- (0,167.6) -- (0,128.8) arc[start angle=180.0000, end angle=246.6430, radius=7] -- (20.7752,115.2264) arc[start angle=426.6430, end angle=360.0000, radius=7] -- (25,58.8) arc[start angle=0.0000, end angle=-66.6430, radius=7] -- (4.2248,45.2264) arc[start angle=113.3570, end angle=180.0000, radius=7] -- cycle;
    \fill[specimenfe] (24.891,56.843) -- (184.891,56.133) -- (184.891,107.269) -- (24.891,108.342) -- cycle;
    \filldraw[specimenhole][rotate around={-46:(50,80)}] (50,80) ellipse [x radius=12, y radius=5.3];
    \filldraw[specimenhole] (70,96) ellipse [x radius=16.4, y radius=3.65];
    \filldraw[specimenhole][rotate around={29:(105,83.8)}] (105,83.8) ellipse [x radius=28.5, y radius=7.85];
    \draw[specimencut] (24.891,56.843) -- (184.891,56.133);
    \draw[specimencut] (184.891,107.269) -- (24.891,108.342);
  \end{tikzpicture}
    \\
    \addlinespace[1.0ex]
    \bottomrule
  \end{tabular*}

  \vspace{9.0ex}

  \begin{tikzpicture}[remember picture]
    \node[inner sep=0pt] (fegeom) {\includegraphics{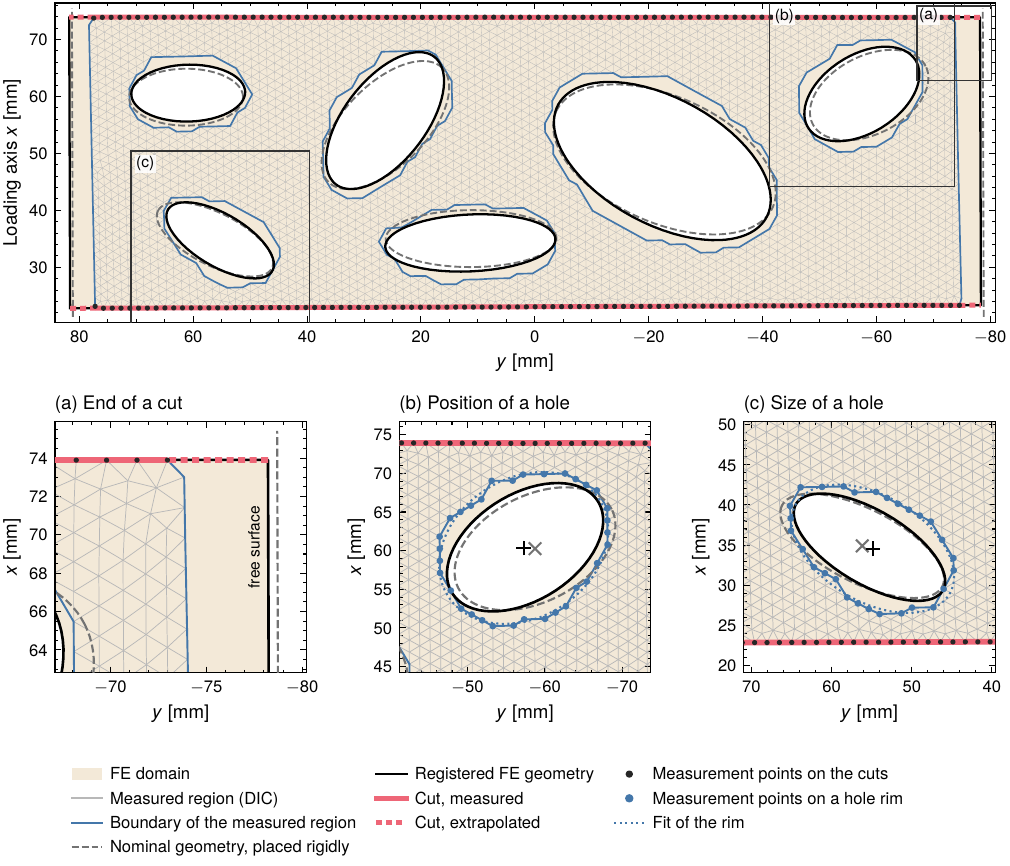}};
    \coordinate (fgL) at ($(fegeom.south west)!\fegeomPanelLeft!(fegeom.south east)$);
    \coordinate (fgR) at ($(fegeom.south west)!\fegeomPanelRight!(fegeom.south east)$);
    \coordinate (fgT) at ($(fegeom.south west)!\fegeomPanelTop!(fegeom.north west)$);
    \coordinate (fgB) at ($(fegeom.south west)!\fegeomPanelBottom!(fegeom.north west)$);
    \coordinate (fegeomNW) at (fgL |- fgT);
    \coordinate (fegeomNE) at (fgR |- fgT);
    \coordinate (fegeomSW) at (fgL |- fgB);
    \coordinate (fegeomSE) at (fgR |- fgB);
    \begin{pgfonlayer}{background}
      \draw[gaugewedge, overlay] (specimenFEbl) -- (fegeomSW);
      \draw[gaugewedge, overlay] (specimenFEbr) -- (fegeomSE);
    \end{pgfonlayer}
  \end{tikzpicture}%
  \begin{tikzpicture}[remember picture, overlay]
    \draw[gaugewedge] (specimenFEtl) -- (fegeomNW);
    \draw[gaugewedge] (specimenFEtr) -- (fegeomNE);
  \end{tikzpicture}%
  \caption{The specimens and the FE model of TT1. \emph{Top:} the NR-40 specimens of the dataset \citep{Abbasi2026} used here: the specimens of the homogeneous uniaxial-tension (UT) and pure-shear (PS) tests, and the perforated specimens TT1 and TT2, all in their nominal geometry\specimenScaleNote. Along the vertical loading direction they measure from \qty{100}{\mm} (PS) to \qty{167.6}{\mm} (TT1, TT2). On TT1 and TT2 the FE domain is shaded and the two cuts bounding it are drawn in red; TT2 is shown the same way in Fig.~\ref{fig:registration}. \emph{Middle:} the shaded area of TT1 (the dataset's specimen TTf), enlarged, viewed normal to the sheet; the dashed lines mark it on the specimen above. The measured region, the gray DIC triangulation inside the blue boundary, does not reach the edge of the specimen. The nominal geometry, placed rigidly in the frame of the measurement, misses the measured holes; the registered FE geometry takes the positions of the holes from the measurement and their sizes from the nominal geometry, and bounds the FE domain. The measured displacements are prescribed on the two cuts: both in-plane components over their measured segment, and both continued with the end slope of the fitted spline over their unmeasured ends. \emph{Bottom:} (a)~End of a cut at the edge of the specimen, extrapolated over \qty{5.2}{\mm} of the \qty{160}{\mm} cut. (b)~\emph{Position} of a hole: its nominal center ($\times$) lies \qty{1.5}{\mm} from the measured one ($+$), the center fitted to its rim. (c)~\emph{Size} of a hole: where the nominal hole would contain measurement points, it is shrunk, here by \qty{5}{\percent}.}
  \label{fig:fe_geometry}
\end{figure*}

\paragraph*{Experimental data} \citet{Abbasi2026} tested natural rubber specimens, cut from sheets of \qty{2.5}{\mm} thickness, in quasi-static uniaxial tensile tests (Fig.~\ref{fig:fe_geometry}). We use the perforated specimens TT1 and TT2, whose holes make the test statically indeterminate and the deformation locally multiaxial, together with the classical uniaxial tension (UT) and pure shear (PS) specimens.
Per specimen and time instant, the dataset provides the reference positions $\bX^{\obs}_j$ and the in-plane displacements ${}^m\bu^{\obs}_j$ at the DIC measurement points, and the reaction force of each gripped edge. These points form the measured region, here well inside the specimen (solid blue boundary in Fig.~\ref{fig:fe_geometry}).
For UT and PS, the dataset also provides the stress--stretch curves of the homogeneous deformation in the central gauge section, the nominal stress $P_{11}$ against the stretch $\lambda$, which enter through the deformation gradients $\bF_\mathrm{UT} = \diag(\lambda, \lambda^{-1/2}, \lambda^{-1/2})$ and $\bF_\mathrm{PS} = \diag(\lambda, 1, \lambda^{-1})$.

\paragraph*{Finite element model -- Geometry} Ideally, the FE domain $\Omega_0$ would follow the actual specimen geometry, but only the nominal geometry is available, and overlaying it on the measured region leaves measurement points outside the material (Fig.~\ref{fig:fe_geometry}). Nor can the DIC triangulation serve as the FE domain in Cases~1 and~2: DIC loses data near the edges of the specimen, and a model based on the measured region omits the material up to the edges. Its load-bearing cross-section is smaller, and at the measured displacements the model identification would compensate for that with a stiffer material.

We therefore register the nominal geometry to the measurement: we choose the nominal width for the specimens outer contour in the gauge section, and the position of each hole is determined from a fit to its measured rim and its size from the nominal geometry, shrunk by the least amount that leaves every measurement point outside it (Fig.~\ref{fig:fe_geometry}b,c and Appendix~\ref{app:registration}). Every measurement point then satisfies $\bX^{\obs}_j\in\Omega_0\cup\partial\Omega_0^{\vect{t}}$, as Sec.~\ref{sec:femu:objective} requires. The model and the measurement now share one coordinate frame, in which $\Qu$ evaluates the FE displacement field at the measurement points.
In the loading direction, the FE domain ends not at the gripped edges but at a cut through the specimen at each end of the measured region (solid red lines in Fig.~\ref{fig:fe_geometry}), where the displacements are known.
The sheets are thin and loaded in their plane, so we compute them in plane stress (Appendix~\ref{app:thin}): the unknowns are the two in-plane displacement components on the midplane, the thickness stretch follows pointwise from incompressibility, and the weak form carries the sheet thickness $H_0=\qty{2.5}{\mm}$. We mesh the registered midplane with bilinear quadrilaterals at an average element size of $h = \qty{2.5}{\mm}$: \num{2616} elements for TT1 and \num{2236} for TT2. 

\paragraph*{Finite element model -- Boundary conditions}

On the two cuts normal to the loading direction we prescribe the measured displacements, or some of their components. The remaining boundary, the lateral edges and the hole rims, is traction-free, $\hat{\vect{t}}=\vect{0}$ on $\partial A_0^{\vect{t}}$; and no body force acts, $\vect{b}=\vect{0}$. The measured force is not applied as a traction but enters the objective function, because its distribution over the grips is unknown. Prescribing displacements on cuts also keeps the grips and the clamping outside the FE domain, and with them uncertainties such as machine compliance, grip slip, and rigid-body motion of the specimen.

The load cell at the upper, moving grip lies outside the FE domain; the reaction force on the cut facing that grip is equal to the measured force, because the material in between is in equilibrium and otherwise traction-free. That cut takes the role of the boundary part $\Gamma_1$ of Sec.~\ref{sec:femu:objective}, with $\vect{s}_1$ the loading direction and $n_f = 1$. Equality requires the cut to span the whole cross-section, with the component along $\vect{s}_1$ prescribed along all of it, a second reason why the FE domain must not end at the measured region.

Because the measured displacements are noisy, we smooth them first: along each cut, we fit both in-plane components by a least-squares cubic B-spline at each time instant. They are also incomplete. Being surface measurements, they are those of the midplane, since the in-plane displacement varies little through the thickness of a sheet loaded in its plane (Appendix~\ref{app:thin}). They stop short of the ends of a cut at the edges of the specimen. Over these short ends (Fig.~\ref{fig:fe_geometry}a) we prescribe both components, continued with the end slope of the fitted spline.

\subsubsection{Results}\label{sec:apps:case1:results}
Obtaining accurate full-field displacement measurements is an art of its own: the camera system and lens are expensive; the measurement quality depends on the lighting, the speckle pattern, the orientation and distance between camera and specimen, the calibration, and the correlation algorithms; and at large stretches the speckle pattern itself degrades. By contrast, statically determinate tests are usually cheap and require far less expertise.

We therefore also investigate whether the additional effort is worthwhile for this material, that is, how the training data influence the identified model and its predictive capabilities. To this end, we train the CKANs on three training sets: (i) UT, (ii) UT \& PS, and (iii) TT1. We treat UT and PS as statically determinate tests and train on their homogeneous stress--stretch curves; only the third is a full-field test.

\begin{figure*}[t]
  \centering
  \includegraphics{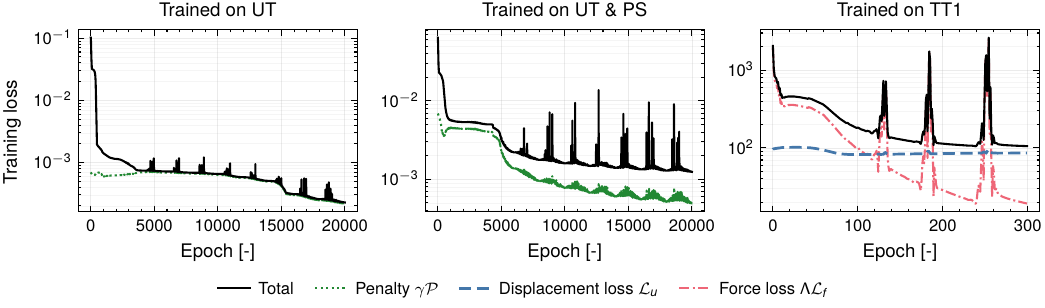}
  \caption{Training loss of the best-performing run of each training set.
    For UT and UT \& PS, the loss is the mean squared error of the nominal stress $P_{11}$ plus the sparsity penalty $\gamma\Pen$ (Sec.~\ref{sec:femu:training}); these fits are full batch, one Adam update per epoch. For TT1, it is the objective $\mathcal{J}$ of Eq.~\eqref{eq:objective}, drawn with its terms, the displacement loss $\mathcal{L}_u$ and the weighted force loss $\Lambda\mathcal{L}_f$; TT1 trains without the penalty
    ($\gamma = 0$).}
  \label{fig:case1_loss}
\end{figure*}

\paragraph*{Training and model discovery}
We train ten CKANs per training set, each starting from the near-Neo-Hookean initialization of Sec.~\ref{sec:femu:training}, with $\mu = \qty{0.67}{\mega\pascal}$ from a fit to the uniaxial test over $\lambda \leq 1.1$. On UT and UT \& PS, Adam runs for \num{20000} epochs at a cyclic learning rate, one full-batch update per epoch. On TT1, training runs for 300 epochs over $n_T = 11$ time instants, spaced evenly in the measured reaction force. On TT1, we switch the penalty off ($\gamma = 0$): a sweep of the penalty weight left the pruned structure unchanged and only degraded the fit. Fig.~\ref{fig:case1_loss} shows the training losses of the best-performing CKANs. On TT1, the displacement loss $\mathcal{L}_u$ hardly decreases after about 100 epochs, while the force loss $\Lambda\mathcal{L}_f$ continues to drop.
Tab.~\ref{tab:case1_hyper} in Appendix~\ref{app:hyperparameters} collects the hyperparameters.

After training, we reduce each CKAN to a closed-form expression by pruning and symbolic regression (Sec.~\ref{sec:ckan}); the coefficients are then re-fitted for another \num{20000} epochs on UT and UT \& PS and another 300 on TT1. Tab.~\ref{tab:case1_symbolic} lists the resulting energies. All three share the same two-term structure: a linear $\Itwo$ term plus an exponential in $\Ione$. Only the coefficients differ; the UT-trained model has the largest $\Itwo$ coefficient and the stiffest $\Ione$ exponent. We compare each model before and after symbolic regression, in particular how the two forms extrapolate. To assess how the trained CKANs generalize, we simulate the withheld specimen TT2 and compare the predicted displacement fields and reaction forces with the measurements, again for both forms.

\begin{table*}[h]
  \centering
  \caption{Closed-form strain energy $\tilde{\Wenergy}(\Ione, \Itwo)$ identified from each training set by pruning and symbolic regression, with re-fitted coefficients; $\Ione$ and $\Itwo$ are the isochoric invariants, and the strain energy is in \unit{\mega\pascal}.}
  \label{tab:case1_symbolic}
  \begin{tabular}{ll}
    \toprule
    Training set & Identified $\tilde{\Wenergy}(\Ione, \Itwo)$ \\
    \midrule
    UT & $0.2538\,\Itwo + 0.5298\,e^{0.11449\,\Ione} - 1.50832$ \\
    UT \& PS & $0.12063\,\Itwo + 1.79323\,e^{0.068\,\Ione} - 2.56093$ \\
    TT1 & $0.10606\,\Itwo + 3.26282\,e^{0.04738\,\Ione} - 4.0793$ \\
    \bottomrule
  \end{tabular}
\end{table*}

\begin{figure*}[t]
  \centering
  \includegraphics{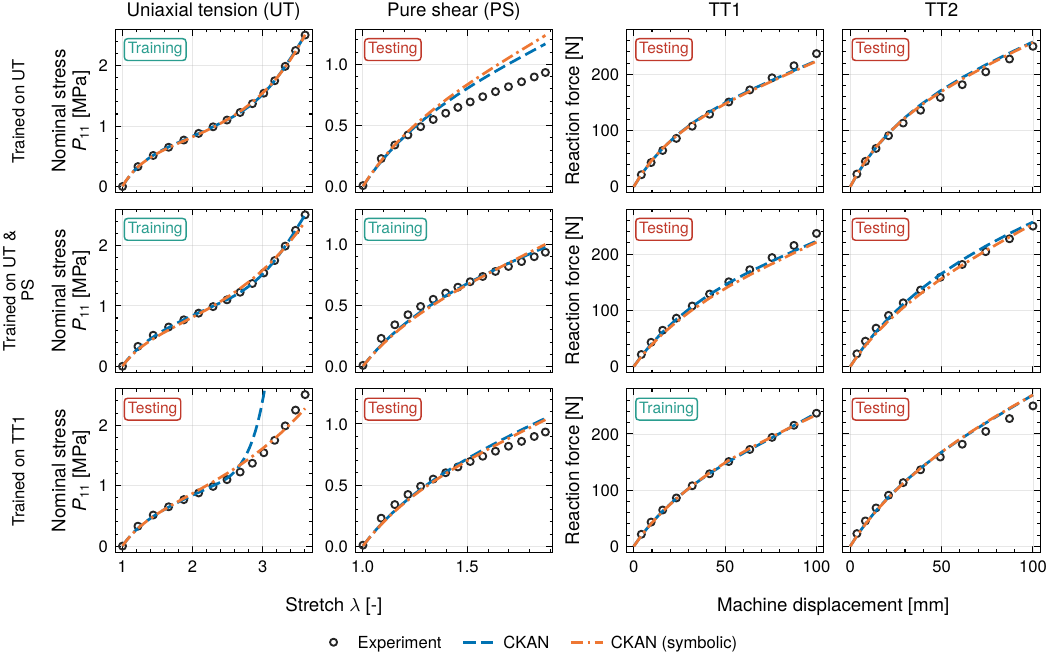}
  \caption{Training--testing matrix: one row per training set, one column
    per test, the badge marking
    whether that subplot's data trained the model. For UT and PS, the nominal
    stress $P_{11}$ over the stretch $\lambda$; for the perforated
    specimens TT1 and TT2, the reaction force
    over the machine displacement. TT2 is never trained on. The UT and PS
    tests are recorded at several hundred stretches; their markers show a
    uniform subsample of the record, for legibility, whereas the reaction
    force is drawn at all of its time instants.}
  \label{fig:case1_matrix}
\end{figure*}

\paragraph*{Global response} Fig.~\ref{fig:case1_matrix} compares the trained CKANs with the measurements on all four tests; the results are similar to those reported by \citet{Knipper2026}. The models trained on the homogeneous tests predict the reaction force of both perforated specimens well, before and after symbolic regression. The TT1-trained model likewise predicts the reaction force of the withheld TT2. However, the homogeneous tests separate the models more sharply. Trained on UT alone, the CKAN predicts PS well only up to $\lambda \approx 1.3$. Trained on TT1, it follows PS over the whole measured range but UT only up to $\lambda \approx 2.6$; there the prediction deteriorates, reaching \qty{8.4}{\mega\pascal} at $\lambda = 3.6$, where \qty{2.5}{\mega\pascal} are measured. The closed form suppresses this extrapolation failure: it stays with the measurement, at almost no cost in training accuracy.

\begin{figure*}[t]
  \centering
  \includegraphics{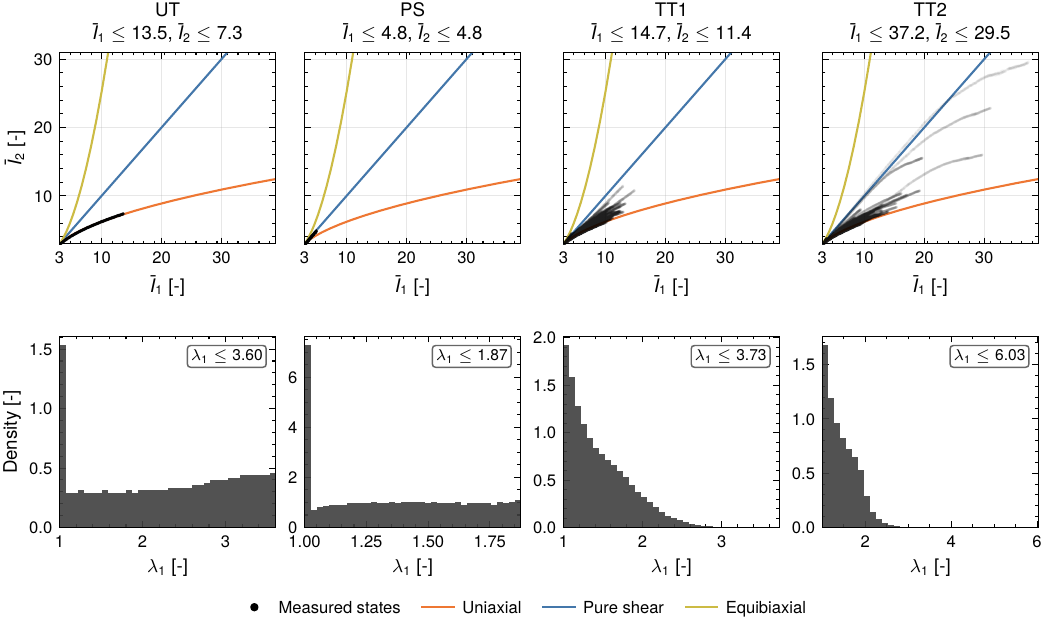}
  \caption{Measured deformation states of the four tests. Top: the states in the plane of the isochoric invariants $(\Ione, \Itwo)$, with the loci of uniaxial, pure-shear, and equibiaxial deformation; the titles state each test's largest invariants. Bottom: density histograms of the maximum principal stretch $\lambda_1$, the badge stating its largest value. For UT and PS, the states follow from the measured stretches of the homogeneous deformation; for TT1 and TT2, from every measured in-plane deformation gradient at all recorded time instants, with the out-of-plane stretch from incompressibility. The few trajectories reaching large invariants belong to a handful of DIC triangles at the hole rims.}  \label{fig:case1_invariants}
\end{figure*}

The invariant plane and the stretch histograms (Fig.~\ref{fig:case1_invariants}) explain this deterioration. 
For an incompressible material, the kinematically admissible deformation states, characterized by the invariant pair $(\Ione, \Itwo)$, fill the region between the loci of uniaxial and equibiaxial tension.
Although TT1 spans almost the whole invariant range of the UT test, its coverage is thin where the UT response stiffens: \qty{99}{\percent} of the \num{57277} training states stay below a maximum principal stretch of $\lambda_1 = 2.6$, and only \qty{0.1}{\percent}, a few DIC triangles at the hole rims, reach beyond $\lambda_1 = 3$. The spline CKAN follows the UT measurement as far as the training states are dense; beyond $\lambda \approx 2.6$ the few rim triangles are too sparse to constrain it, and its response is the generic extrapolation of a convex and non-decreasing network. Symbolic regression replaces the splines by an exponential in $\Ione$ and a linear term in $\Itwo$ (Tab.~\ref{tab:case1_symbolic}); their few coefficients are fitted to all training states at once, and the closed form extrapolates stably.

\begin{figure*}[t]
  \centering
  \includegraphics{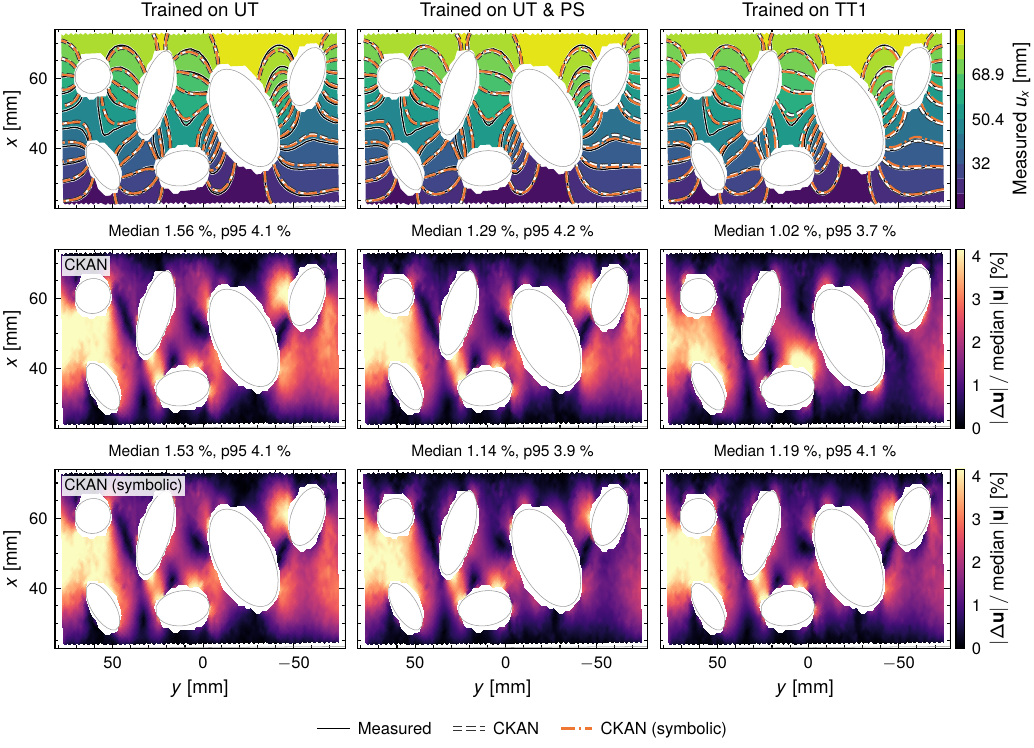}
  \caption{Displacement field on TT1 at the final time instant, one column per training set. Top: measured $u_x$ (filled, black contours) with the predictions of the CKAN (blue dashed) and its symbolic form (orange dash-dotted). Middle and bottom: pointwise error $\norm{\Delta\bu}$ relative to the median $\norm{\bu}$ of the step, for the CKAN and its symbolic form; the subplot titles give the median and the 95th percentile.}
  \label{fig:case1_field_u_TT1}
\end{figure*}

\begin{figure*}[t]
  \centering
  \includegraphics{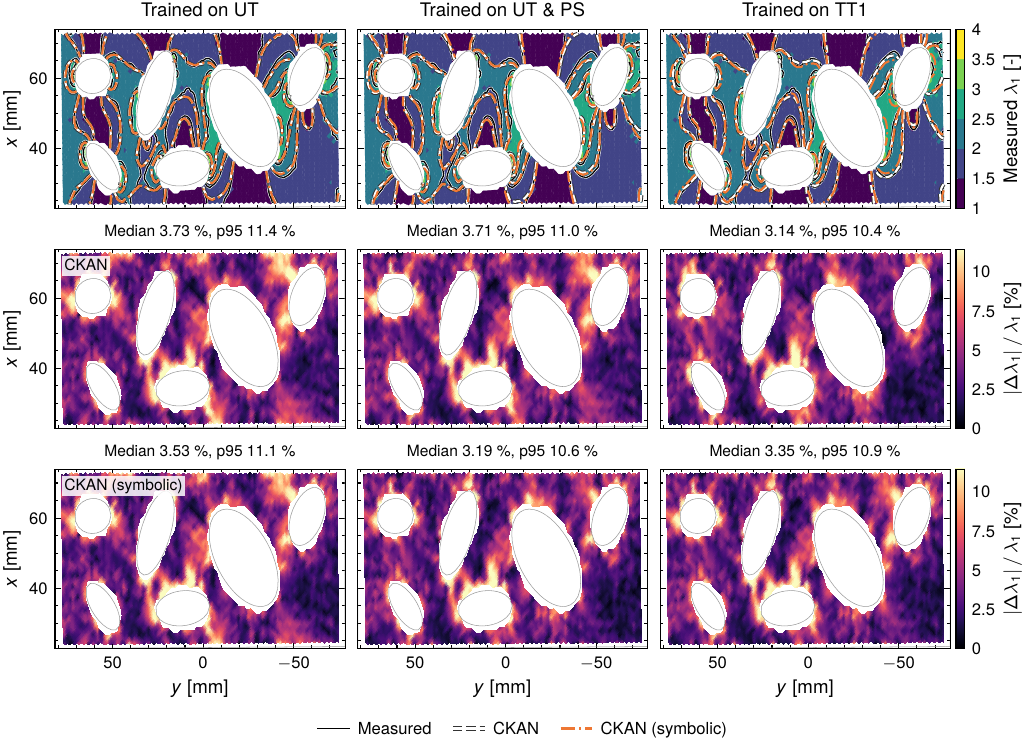}
  \caption{Maximum principal stretch on TT1 at the final time instant, one
    column per training set. Top: measured $\lambda_1$ at the DIC-triangle
    centroids ($\bF$ from the dataset, filled, black contours) with the CKAN
    (blue dashed) and its symbolic form (orange dash-dotted), FE quadrature
    values interpolated cubically. Middle and bottom: pointwise relative
    error $|\Delta\lambda_1| / \lambda_1$ for the two forms; the subplot
    titles give the median and the 95th percentile.}
  \label{fig:case1_stretch_TT1}
\end{figure*}

\paragraph*{Full-field accuracy on TT1} Figs.~\ref{fig:case1_field_u_TT1} and~\ref{fig:case1_stretch_TT1} compare the measured and predicted displacement and stretch fields on TT1 at the final time instant. For every training set, the predicted contours follow the measured ones closely, and the CKAN and its closed form are almost indistinguishable at the contour level. The pointwise displacement error stays near one percent over most of the specimen and peaks locally near hole rims and the edge of the measured region, where DIC is least reliable. The pointwise stretch error is speckled rather than smooth and rises toward the hole rims.

\begin{figure*}[t]
  \centering
  \includegraphics{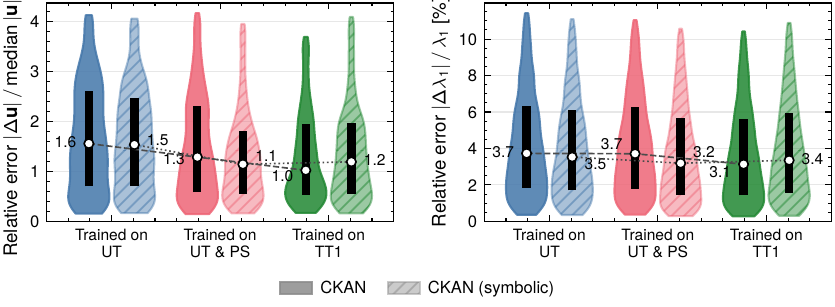}
  \caption{Distributions of the pointwise relative errors on TT1 at the final time instant, for the CKAN and its symbolic form trained on each training set: $\norm{\Delta\bu}$ relative to the median $\norm{\bu}$ of the step (left) and the error of the maximum principal stretch, $|\Delta\lambda_1| / \lambda_1$ (right). The violins span the central \qty{90}{\percent} of the data; the black bar marks the central \qty{50}{\percent} and the white dot the median of the full data, with its value printed beside it. The medians equal those in the subplot titles of Figs.~\ref{fig:case1_field_u_TT1} and~\ref{fig:case1_stretch_TT1}.}
  \label{fig:case1_errors_TT1}
\end{figure*}

Fig.~\ref{fig:case1_errors_TT1} summarizes both errors as distributions. The UT-trained models show the largest displacement errors. The models trained on UT \& PS and on TT1 are nearly tied, with all medians between one and two percent. The stretch errors differ little between the training sets and run two to three times higher than the displacement errors: the stretch is a derived quantity that amplifies the measurement noise, and the interpolation from the FE quadrature points to the DIC triangles adds its own error. The closed form barely changes either distribution. Because the stretch field varies more steeply across the specimen than the displacement field (Fig.~\ref{fig:case1_stretch_TT1}), it is potentially the more informative observable; Case~2 (Sec.~\ref{sec:apps:case2}) uses it as the training target.

\begin{figure*}[t]
  \centering
  \includegraphics{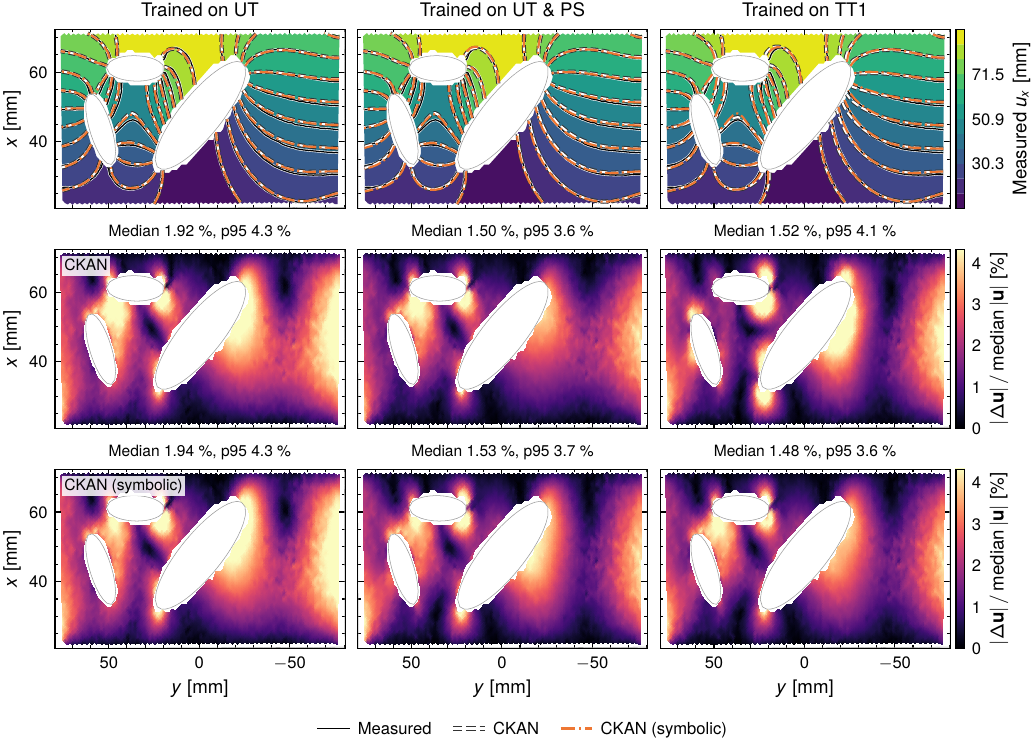}
  \caption{TT2, final time instant: $u_x$ measured (filled, black contours) vs
    the CKAN (blue dashed) and its symbolic form (orange dash-dotted); error
    $\norm{\Delta\bu}$ relative to the median $\norm{\bu}$ of the step.}
  \label{fig:case1_field_u_TT2}
\end{figure*}

\paragraph*{Full-field accuracy on TT2} Fig.~\ref{fig:case1_field_u_TT2} repeats this comparison on TT2, which no model saw during training. The predicted contours follow the measured ones as closely as on TT1, and the CKAN and its closed form remain almost indistinguishable at the contour level. The pointwise error is slightly larger, and it is no longer confined to the hole rims but spreads over a broad region of the specimen. Its peaks sit at the same locations for every training set, which points to the FE model of the experiment rather than to the identified material: to its registered geometry and its prescribed boundary displacements. The ranking of the training sets changes as well: the advantage that training on TT1 held on its own specimen disappears, training on UT \& PS does equally well, and training on UT alone remains the least accurate. Appendix~\ref{app:tt2} shows the stretch field on TT2 and the distributions of both errors. The stretch error is nearly the same for every training set and concentrates in the ligaments between the holes. The training sets thus differ less on the withheld specimen than on TT1, and all of them transfer to the new hole pattern.

\paragraph*{Summary} All three CKANs reproduce the reaction forces and the displacement fields of TT1 and of the withheld TT2 to within a few percent. Uniaxial data alone overestimate pure shear, whereas the heterogeneous test TT1 predicts it well, because its states fill a region of the invariant plane instead of a single curve. Conversely, TT1 leaves the uniaxial response at large stretch unconstrained, where its states are sparse. One heterogeneous test thus delivers what the two homogeneous tests deliver together, the promise of MT2.0, at the price of the full-field measurement and of an FE model that reproduces the experiment. Symbolic regression turns each CKAN into a two-term formula (Tab.~\ref{tab:case1_symbolic}) at almost no cost in accuracy, and it removes the unconstrained extrapolation of the spline model.

\FloatBarrier

\subsection{Case~2 --- One-dimensional stretch profile}\label{sec:apps:case2}

\begin{figure}[t]
  \centering
  \includegraphics{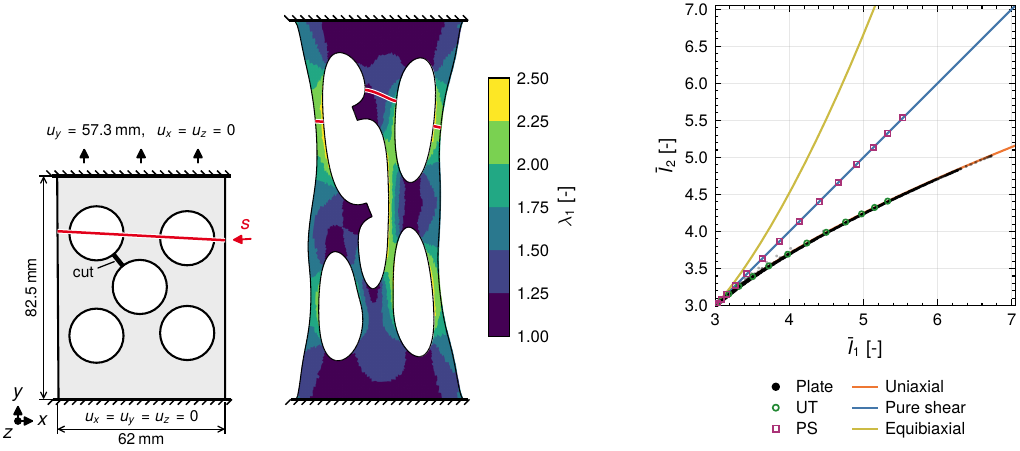}
  \caption{The perforated plate of Case~2; the left two panels are at the same scale. \emph{Left:} the reference configuration, with the corner and hole coordinates tabulated by \citet{Knipper2026}. The plate has five holes of \qty{20}{\mm} diameter, and the ligament between the upper-left and the center hole is cut. The measured path (red) runs through the centers of the two upper holes, which divide it into three material segments; its coordinate $s$ counts from the right edge. The hatched lines mark the gripped edges: the bottom edge is clamped, and the top edge is displaced by \qty{57.3}{\mm} along the load axis $y$, with the other components suppressed. \emph{Middle:} the plate at that displacement, deformed by the CKAN trained on the measured profile (Sec.~\ref{sec:apps:case2:results}) and colored by the maximum principal stretch $\lambda_1$. The red line is the measured path, carried along by the material. \emph{Right:} the deformation states of the training data in the plane of the isochoric invariants, with the loci of uniaxial, pure-shear and equibiaxial deformation. UT and PS are the homogeneous states at the digitized stretches; the plate's are the states at the element centers of the middle panel. Almost all of them lie on the uniaxial locus, which is why the plate alone cannot separate $\Ione$ from $\Itwo$.}
  \label{fig:plate_geometry}
\end{figure}

\paragraph*{Experimental data}
\citet{Meunier2008} tested unfilled RTV~141 silicone in six specimen geometries under quasi-static loading.
One test is statically indeterminate: a perforated plate, as in Case~1. The plate is \qty{82.5}{\mm} tall, \qty{62}{\mm} wide, and \qty{1.75}{\mm} thick, with five circular holes of \qty{20}{\mm} diameter; the narrow ligament (\qty{5.46}{\mm}) between two of the holes has been cut (Fig.~\ref{fig:plate_geometry}). The bottom edge is clamped ($u_x=u_y=u_z=0$), and the top edge is gripped ($u_x=u_z=0$) and displaced along the load axis $y$; at a load of \qty{20}{\N} its displacement is $u_y=\qty{57.3}{\mm}$.
This single equilibrium state is all that is measured on the plate, so $n_T = 1$. Instead of full-field displacements, the kinematic measurement is the maximum principal stretch $\lambda_1$ along the straight path through the centers of the two upper holes, defined in the reference configuration and divided by them into three material segments; its arc-length coordinate $s$ counts from the right edge (Fig.~\ref{fig:plate_geometry}). The stretch profile has 29 points over the \qty{62}{\mm} width: two orders of magnitude fewer kinematic observations than a single time instant of Case~1 carries, and three orders fewer than its eleven steps together.

The other five tests are statically determinate: UT, PS, simple compression, plane-strain compression, and equibiaxial tension (realized as a bulge inflation test). For each, the dataset contains the nominal stress--stretch curve of the homogeneous deformation. As in Case~1, we use the perforated plate together with the UT and PS tests for training. The original data files are no longer available, so we digitized both the stretch profile and the stress--stretch curves from the published figures.

\paragraph*{Finite element model}
Without full-field measurements, there is nothing to register the geometry to and no measured displacements to prescribe, so we use the geometry and the boundary conditions as published. The FE domain follows the corner and hole coordinates tabulated by \citet{Knipper2026}, with the cut modeled as a slit \qty{0.1}{\mm} wide, which joins the two holes it connects into a single boundary loop. The bottom edge is clamped, and the top edge is displaced by \qty{57.3}{\mm} with the transverse component suppressed (a grip). As in Case~1, the plate is thin and loaded in its plane, so we compute it in plane stress (Appendix~\ref{app:thin}) with the plate thickness $H_0=\qty{1.75}{\mm}$ in the weak form, which makes the computed reaction force that on the whole cross-section. We mesh the midplane with bilinear quadrilaterals at an average element size of \qty{1}{\mm}: \num{8829} elements. The prescribed displacement is that of the loading device. Such a displacement rarely matches the actual displacements near the grips; its error matters little, however, where the observations lie far from them, as the path does here.

\paragraph*{Observation operator}
Because the kinematic measurement is a stretch profile, a stretch loss replaces the displacement loss, Eq.~\eqref{eq:displacement_loss}. The observation operator samples $\lambda_1$ at $n_\lambda = 200$ evaluation points on the path, spaced uniformly in arc length within each material segment. At the point at arc length $s_i$ with reference position $\bX_i$, the deformation gradient follows from the shape functions of the element containing it, $\bF_h(\bX_i)=\bI+\sum_{a}\vect{d}_a\otimes\Grad N_a(\bX_i)$, and $\lambda_1(s_i;\btheta)$ is its maximum principal stretch. With $n_T=1$, the stretch loss reads
\begin{equation}\label{eq:stretch_loss}
  \mathcal{L}_{\lambda}(\btheta)=\sum_{i=1}^{n_\lambda}\bigl(\lambda_1(s_i;\btheta)-\lambda^{\obs}_1(s_i)\bigr)^2
\end{equation}
and enters the objective, Eq.~\eqref{eq:objective}, in place of $\mathcal{L}_u$. Here $\lambda^{\obs}_1(s_i)$ is the measured profile at the point, smoothed per material segment by a least-squares cubic polynomial. Since the points are spread over the three material segments in proportion to their length, each segment enters the sum with its length. Unlike $\Qu$, the map from the nodal displacements to the profile is nonlinear, but the sensitivities of Sec.~\ref{sec:femu:sensitivities} involve only its derivative, which automatic differentiation provides. The force loss, Eq.~\eqref{eq:force_loss}, is unchanged, with $\Gamma_1$ the top edge and $n_f=1$.

\subsubsection{Results}\label{sec:apps:case2:results}

\paragraph*{Training and model discovery}
We train two CKANs per training set, each starting from the near-Neo-Hookean initialization of Sec.~\ref{sec:femu:training}, with $\mu = \qty{0.32}{\mega\pascal}$ from a fit to the uniaxial test over $\lambda \leq 1.1$. On UT and UT \& PS, Adam runs for \num{20000} epochs, one full-batch update per epoch; on the plate, where $n_T=1$ makes an epoch a single update, for 800 epochs without the penalty ($\gamma = 0$). Both use a cyclic learning rate. Tab.~\ref{tab:case2_hyper} in Appendix~\ref{app:hyperparameters} collects the hyperparameters.

\begin{figure*}[h]
  \centering
  \includegraphics{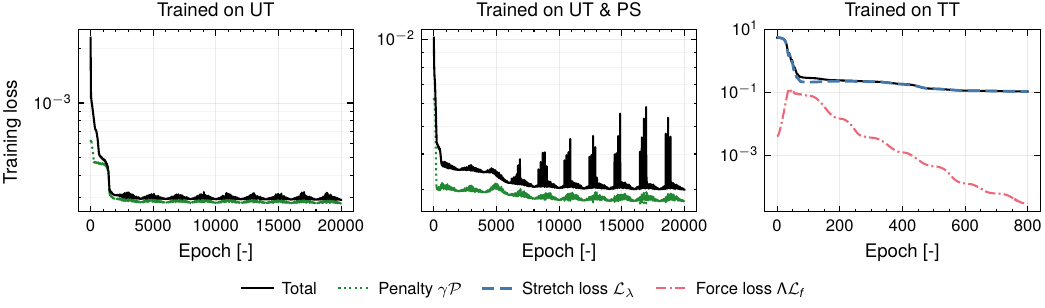}
  \caption{Training loss of the best-performing run of each training set.
    For UT and UT \& PS, the loss is the mean squared error of the nominal stress $P_{11}$ plus the sparsity penalty $\gamma\Pen$ (Sec.~\ref{sec:femu:training}); these fits are full batch. For TT, it is the objective $\mathcal{J}$ of Eq.~\eqref{eq:objective}, drawn with its terms, the stretch loss $\mathcal{L}_\lambda$ and the weighted force loss $\Lambda\mathcal{L}_f$; TT trains without the penalty ($\gamma = 0$), with one update per epoch ($n_T = 1$).}
  \label{fig:case2_loss}
\end{figure*}

All loads on the plate are prescribed displacements, so the measured force alone determines the scale of the energy (Sec.~\ref{sec:femu:objective}). Fig.~\ref{fig:case2_loss} shows how the two terms divide that task. The stretch loss $\mathcal{L}_\lambda$ dominates the objective at every epoch and completes most of its descent in the first fifty. The weighted force loss $\Lambda\mathcal{L}_f$ instead rises at first, while the predicted reaction force drops to \qty{4.7}{\N}, and falls only afterwards, by three and a half decades, not yet converged when training ends; the reaction closes to \qty{19.8}{\N} against the measured \qty{20}{\N}. The CKAN thus fits the profile first, at the cost of the energy scale, and recovers it over the remaining epochs.

After training, we reduce each CKAN to a closed-form expression by pruning and symbolic regression (Sec.~\ref{sec:ckan}); the coefficients are then re-fitted for another \num{20000} epochs on UT and UT \& PS and another 100 on the plate.
Tab.~\ref{tab:case2_symbolic} lists the resulting energies. Both homogeneous training sets prune to a single hidden node and return classical forms: a Neo-Hookean energy from UT, and a Mooney--Rivlin energy from UT \& PS whose $\Itwo$ term carries a fifth of the initial shear modulus. The plate keeps all three hidden nodes, but only one of its terms carries the energy: the square of an exponential in $\Ione$. Over the plate's deformation states the linear $\Ione$ term contributes \qty{1.8}{\percent} of the energy and the $\Itwo$ term inside the square \qty{0.13}{\percent} of the bracket, which stays positive over the whole range, so the positive part never binds. The plate therefore leaves its own $\Itwo$ dependence at zero. We compare each model with its closed form on the three tests, UT, PS and the plate, and on the deformation modes that none of them saw: simple compression, plane-strain compression, and equibiaxial tension.

\begin{figure*}[h]
  \centering
  \includegraphics{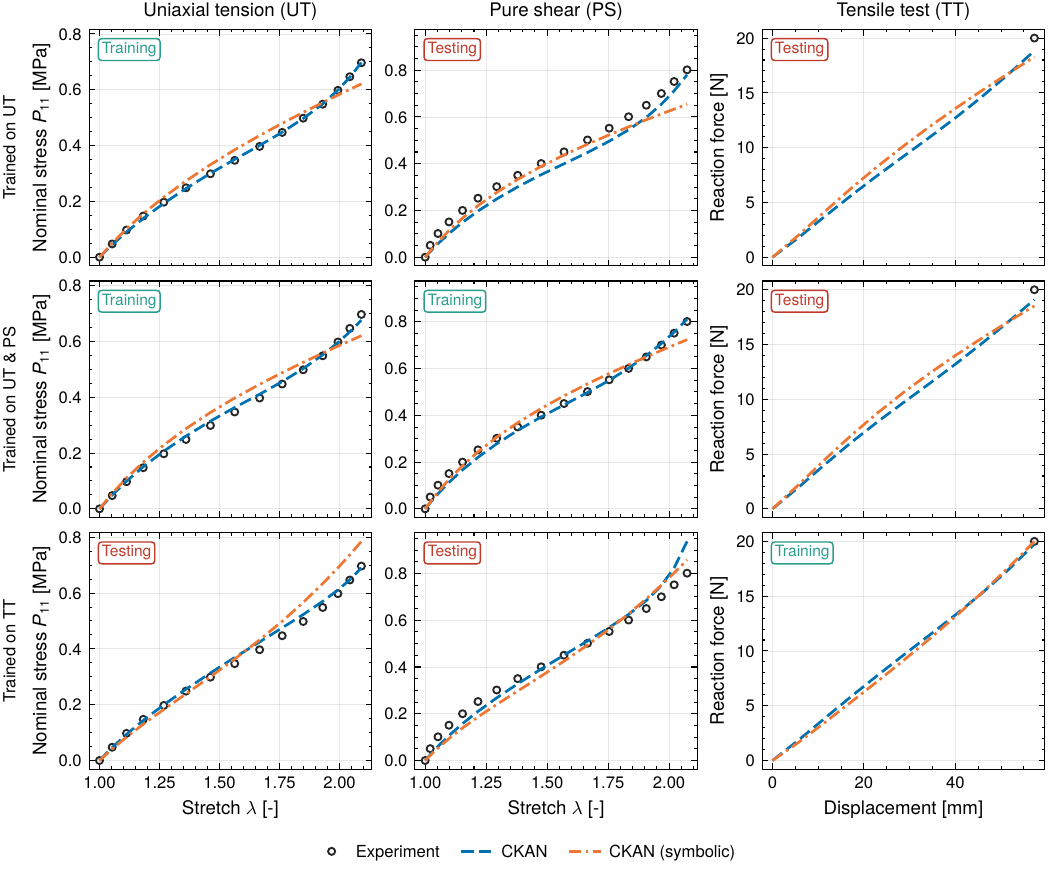}
  \caption{Training--testing matrix of Case~2: one row per training set, one
    column per test, the badge marking
    whether that subplot's data trained the model. For UT and PS, the nominal
    stress $P_{11}$ over the stretch $\lambda$; for the plate (TT), the
    reaction force over the imposed top-edge displacement, where the marker is
    the single measured point, \qty{20}{\N} at \qty{57.3}{\mm}.}
  \label{fig:case2_matrix}
\end{figure*}

\paragraph*{Global response}
Fig.~\ref{fig:case2_matrix} compares the identified models with the measurements on the three tests. All three CKANs reproduce the one measured reaction force, \qty{20}{\N}, to within a few percent, the plate-trained model closest. The stress--stretch curves separate them more sharply. Trained on UT alone, the CKAN underestimates pure shear over the whole measured range; trained on the plate alone, it predicts uniaxial tension well but stiffens beyond $\lambda \approx 1.5$ and misses pure shear by a similar margin. Only the two homogeneous tests together balance the two modes. The measured curves end near $\lambda = 2.1$, so this case does not probe the large-stretch extrapolation that separated the models in Case~1.

Symbolic regression costs the two homogeneous training sets a few percentage points on every stress--stretch curve. On the plate it costs nothing on its own test: its closed form reproduces the measured reaction force to better than a tenth of a percent, closer than the CKAN it replaces, and predicts pure shear as well. It pays for that on uniaxial tension, where the error rises from four to thirteen percent.

\begin{table*}[h]
  \centering
  \caption{Closed-form strain energy $\tilde{\Wenergy}(\Ione, \Itwo)$ identified from each training set of Case~2 by pruning and symbolic regression, with re-fitted coefficients; $\Ione$ and $\Itwo$ are the isochoric invariants, $\mac{\cdot}$ is the positive part, and the strain energy is in \unit{\mega\pascal}.}
  \label{tab:case2_symbolic}
  \begin{tabular}{ll}
    \toprule
    Training set & Identified $\tilde{\Wenergy}(\Ione, \Itwo)$ \\
    \midrule
    UT & $0.16677\,\Ione - 0.50032$ \\
    UT \& PS & $0.15022\,\Ione + 0.03427\,\Itwo - 0.55346$ \\
    TT & $0.0026\,\Ione + 0.00918\,\mac{0.0028\,\Itwo + 19.73803\,e^{0.04736\,\Ione} - 16.00339}^2 - 0.42712$ \\
    \bottomrule
  \end{tabular}
\end{table*}

\begin{figure*}[h]
  \centering
  \includegraphics{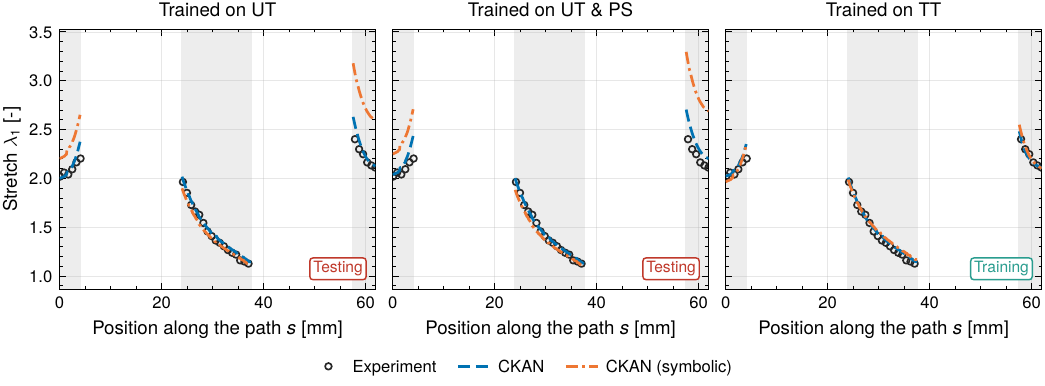}
  \caption{Maximum principal stretch $\lambda_1$ along the measured path at
    the imposed top-edge displacement, one panel per training set, the badge
    marking whether the profile trained the model. Circles: the digitized
    measurement; curves: the CKAN and its symbolic form, drawn over the three
    material segments (shaded); the gaps are the holes.}
  \label{fig:case2_path}
\end{figure*}

\paragraph*{Stretch path}
Fig.~\ref{fig:case2_path} compares the same models with the measured stretch profile. Trained on the plate, the CKAN reproduces it to within about two percent, and as closely at the ends of the path as in its middle. The two homogeneous models follow it too, the UT-trained one about equally well everywhere: UT \& PS fits the long middle segment the best of all three, but overshoots the two short segments at the ends, which leaves it behind the UT-trained model overall. Symbolic regression keeps the profile only for the plate, whose closed form stays near the CKAN it replaces; the closed forms of the two homogeneous models miss those short segments by a wide margin, while still holding the middle one. All of these differences are well above the change in the profile between the training mesh and one twice as fine, so the comparison measures the models, not the mesh.

\begin{figure*}[h]
  \centering
  \includegraphics{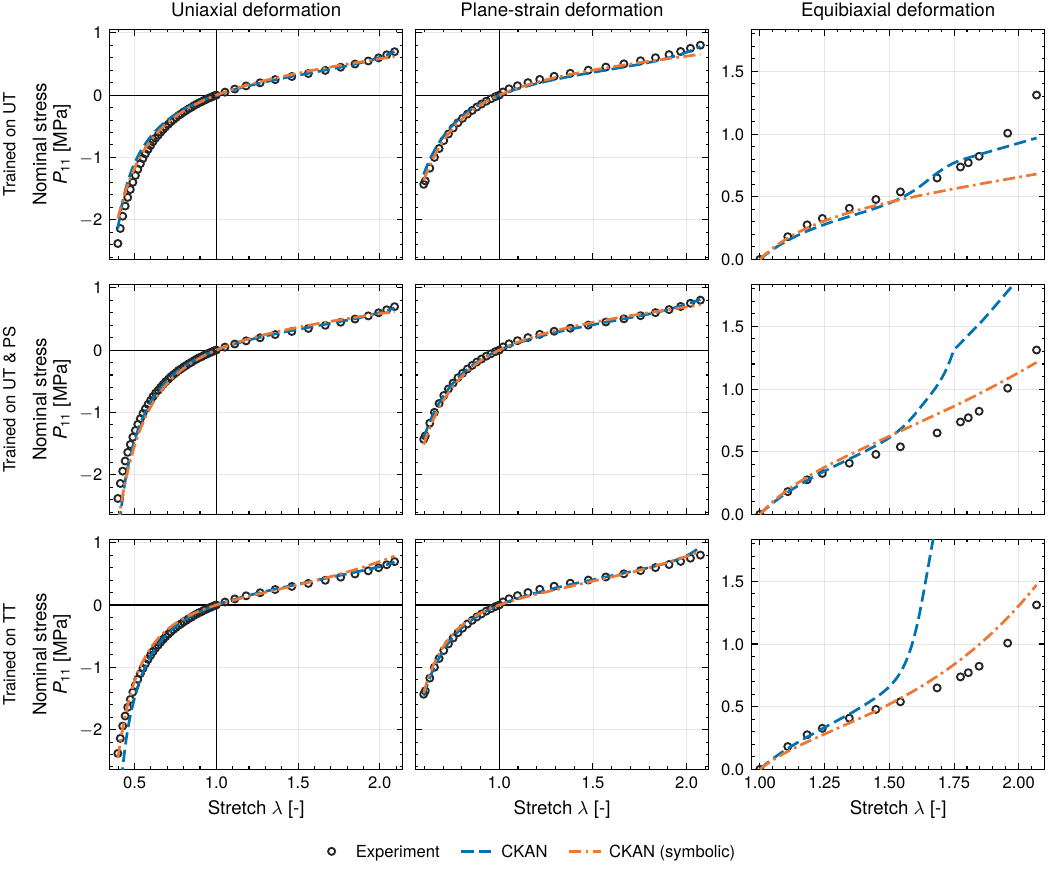}
  \caption{Full material response of the identified models on the statically
    determinate tests, one row per training set, evaluated analytically on
    the incompressible deformation paths: uniaxial tension and compression
    (left, $\bF_\mathrm{UT}$ of Sec.~\ref{sec:apps:case1}), plane-strain
    tension and compression (middle, $\bF_\mathrm{PS}$; the tension branch
    is the PS test), and equibiaxial tension
    (right, $\bF_\mathrm{EB} = \diag(\lambda, \lambda, \lambda^{-2})$;
    realized as the bulge test). Circles: the digitized measurements, which also set the
    frames; a model that extrapolates beyond them leaves the frame. The thin
    lines mark the undeformed state.}
  \label{fig:case2_full_response}
\end{figure*}

\paragraph*{Full material response}
Fig.~\ref{fig:case2_full_response} compares the identified models with the measurements on all five statically determinate tests. The tension branches of the first two columns are the UT and PS tests of Fig.~\ref{fig:case2_matrix}; the compression branches and the bulge test are deformation modes that no model was trained on. All six curves follow simple compression and plane-strain compression over most of their measured range, and they depart from the measurement only at the largest compressive strain, where most of them are too stiff. In the bulge test they agree with the measurement and with one another up to $\lambda \approx 1.6$. Beyond it they fan out: trained on UT, both curves bend below the measurement; trained on UT \& PS, the CKAN leaves the frame while its closed form stays with the measurement; trained on the plate, the CKAN leaves the frame while its closed form stays in view, ending somewhat above the measurement.

The invariant plane explains where the curves separate. Simple compression and equibiaxial tension are the same deformation: the principal stretches of $\bF_\mathrm{UT}$ at $\lambda < 1$ are those of $\bF_\mathrm{EB}$ at the stretch $\lambda^{-1/2}$. The two columns therefore trace one locus in the invariant plane, and the smallest measured compressive stretch, $\lambda = 0.40$, is equibiaxial tension at $\lambda = 1.59$. Plane-strain compression likewise repeats the invariants of the PS test. Of these three deformation modes, only the bulge test beyond that stretch reaches invariants that no other test of the dataset covers, and that is where the predictions separate. There the curves are the generic extrapolation of a convex and non-decreasing energy, as in Case~1 beyond $\lambda \approx 2.6$: they stay admissible, but the data no longer decide between them. The closed form from UT \& PS does follow the bulge measurement to the end, and the plate's stays near it, but no equibiaxial data stand behind either agreement.

\paragraph*{Summary} All three CKANs reproduce the one measured reaction force and the stretch profile to within a few percent. On the homogeneous tests they separate: uniaxial data alone underestimate pure shear, the plate alone predicts uniaxial tension but misses pure shear by a similar margin, and only the two together balance both. The deformation states in the plate are almost everywhere those of uniaxial tension, so they follow a single curve in the invariant plane instead of filling a region of it. All three still predict the two compression tests, and the bulge test as far as $\lambda \approx 1.6$, where the deformation leaves the invariant range of the dataset. Symbolic regression costs the homogeneous sets a few percentage points and returns classical forms, Neo-Hookean from UT and Mooney--Rivlin from UT \& PS; from the plate it returns a squared exponential in $\Ione$ that reproduces the reaction force more closely than the CKAN it replaces and leaves $\Itwo$ at zero.

\clearpage

\subsection{Case~3 --- Three-dimensional displacements under pressure}\label{sec:apps:case3}

Case~3 differs from Cases~1 and~2 in three ways: (i) fibers reinforce the material; (ii) the arterial wall is loaded normal to its surface and modeled as a membrane; (iii) the measured load is a pressure, which is applied to the FE model instead of being compared with it.

\paragraph*{Experimental data}
\citet{Romo2014} inflated a square piece of arterial wall, about \qty{45}{\mm} wide and \qty{0.85}{\mm} thick, cut from a human ascending thoracic aorta; we use their test in the form published with \citet{Meng2025}.
A ring of diameter \qty{30}{\mm} clamps the piece over a cavity, into which saline is pumped at a constant rate, so that the wall bulges out of its plane. The pressure in the cavity is recorded with every image and is the only measured load. Stereo DIC measures the three-dimensional positions of \num{644} points over a central disc of radius \qty{5}{\mm}, triangulated into \num{1203} triangles.
The clamp lies far outside this measured region, whose boundary therefore cuts through the material, as in Case~1.

The dataset holds 43 instants. Over the first seven, the wall creeps at a nearly constant pressure of about \qty{2}{\kilo\pascal}; from the eighth on, the pressure strictly increases, and we use only these 36 instants.

Collagen fibers reinforce the arterial wall, winding helically around the vessel. We idealize them as two families, inclined to either side of the circumferential direction by the same angle $\fibangle=\ang{41}$, the value \citet{Meng2025} assume; we fix it rather than identify it along with the CKAN. On the rig of \citet{Romo2014}, the circumferential direction coincides with the horizontal axis of the clamp, which we take as the $x$-axis of the DIC. We model the wall as a single homogeneous layer, with the families aligned; neither source states whether the sample contains all layers of the wall.

\begin{figure*}[t]
  \centering
  \includegraphics{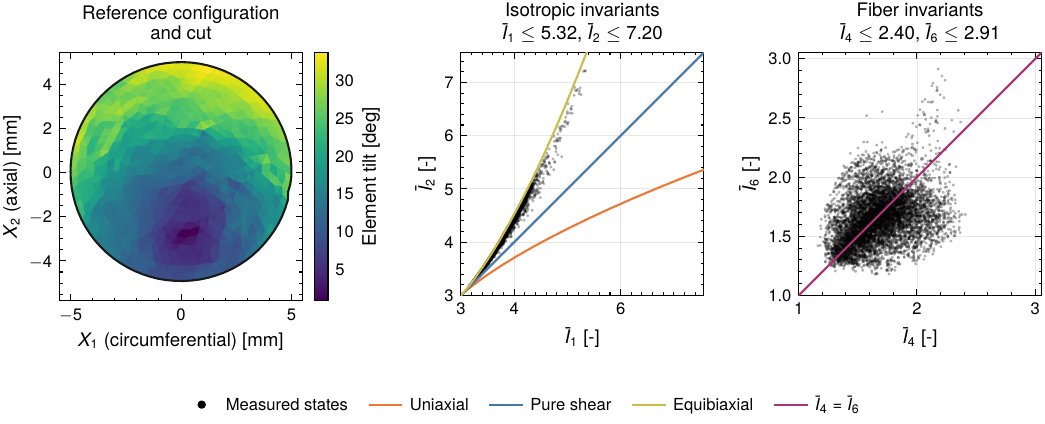}
  \caption{\emph{Left:} the reference configuration of Case~3, the surface measured at the first time instant, colored by the tilt of each triangle against the image plane of the DIC. A flat reference would misrepresent the geometry. The black outline is the boundary of the measured region, a cut through the material at which the measured displacements are prescribed. \emph{Middle and right:} the measured deformation states of the training instants in the two invariant planes. In the plane of the isochoric invariants $(\Ione, \Itwo)$, the states crowd the equibiaxial locus. The squared stretches $\Ifour$ and $\Isix$ of the two fiber families are equal on the diagonal; they differ wherever the deformation shears against the bisector of the families, by \qty{9}{\percent} at the median state and by up to \qty{60}{\percent}.}
  \label{fig:case3_geometry}
\end{figure*}

\paragraph*{Fibers in the arterial wall}
The two fiber families of Sec.~\ref{sec:theory:continuum} lie in the tangent plane of the wall, at the angles $\pm\fibangle$ to its circumferential direction \citep{Holzapfel2000}. With the unit vectors $\vect{E}_1$ and $\vect{E}_2$ in the circumferential and axial directions of the reference configuration, which follow the curved wall, the fiber directions are
\begin{equation}\label{eq:fiber_directions}
  \afib_{\pm}=\cos\fibangle\,\vect{E}_1\pm\sin\fibangle\,\vect{E}_2.
\end{equation}
For their squared stretches we use the names of \citet{Holzapfel2000}: $\Ifour=\isoc{\bC}:\Mfib_{+}$ and $\Isix=\isoc{\bC}:\Mfib_{-}$ are the $\Ifibn{\kappa}$ of Sec.~\ref{sec:theory:continuum} with $\kappa=\pm$, and $\Mfib_{\pm}=\afib_{\pm}\otimes\afib_{\pm}$ are the structural tensors. Both families consist of the same collagen and differ only in the sign of their angle. They are therefore mechanically equivalent, and we require the energy of Eq.~\eqref{eq:invariants_fiber} to be symmetric in their stretches,
\begin{equation}\label{eq:exchange_symmetry}
  \tilde{\Wenergy}(\Ione,\Itwo,\Ifour,\Isix)=\tilde{\Wenergy}(\Ione,\Itwo,\Isix,\Ifour).
\end{equation}

We follow \citet{Holzapfel2000} and split the energy additively, into one KAN over the isotropic invariants and one that each fiber input enters separately. With the CKAN input $\vect{z}^{(0)}=\bigl(\Ione,\Itwo,\mac{\Ifour-1}^2,\mac{\Isix-1}^2\bigr)$ of Sec.~\ref{sec:ckan}, Eq.~\eqref{eq:energy_ckan} reads
\begin{equation}\label{eq:energy_fiber}
  \Wckan(\vect{z}^{(0)})=\Wckaniso(\Ione,\Itwo)+\Wckanfib(\Ifour)+\Wckanfib(\Isix),
\end{equation}
with the isotropic part and the part of one fiber family
\begin{equation}\label{eq:energy_fiber_parts}
  \Wckaniso(\Ione,\Itwo)=\KANiso(\Ione,\Itwo)-\KANiso(3,3), \qquad
  \Wckanfib(I)=\KANfib\bigl(\mac{I-1}^2\bigr)-\KANfib(0).
\end{equation}
Because both families share the single network $\KANfib$, the energy is symmetric in $\Ifour$ and $\Isix$, as Eq.~\eqref{eq:exchange_symmetry} requires.

\paragraph*{Classical counterpart}
With the same kinematics, the model of \citet{Holzapfel2000} (HGO) reads
\begin{equation}\label{eq:hgo}
  \tilde{\Wenergy}=\frac{\mu}{2}(\Ione-3)+\frac{k_1}{2k_2}\sum_{I\in\{\Ifour,\Isix\}}\Bigl[\exp\bigl(k_2\mac{I-1}^2\bigr)-1\Bigr],
\end{equation}
where $\mu$ is the shear modulus of the isotropic part and $k_1>0$ and $k_2>0$ are the fiber parameters. It has the form of Eq.~\eqref{eq:energy_fiber}, with $\KANiso(\Ione,\Itwo)=\tfrac{\mu}{2}\Ione$ and $\KANfib(z)=\tfrac{k_1}{2k_2}\exp(k_2z)$. We use HGO as the reference model of Case~3 and to initialize $\KANfib$ (Sec.~\ref{sec:femu:training}).

\paragraph*{Finite element model}
The arterial wall is thin compared with its radii of curvature, and both sources of the data model it as an incompressible membrane \citep{Romo2014,Meng2025}; so do we. Appendix~\ref{app:thin} reduces the forward problem of Sec.~\ref{sec:femu:forward} to a membrane: the midsurface of the wall deforms in three dimensions and the cavity pressure follows its current normal. We use the DIC triangulation as the FE mesh, its triangles as the elements and its measured points as the FE nodes. The geometry therefore needs no registration to the measurement, and $\Qu$ of Eq.~\eqref{eq:Qu} reduces to a selection of rows. We define the surface measured at the first instant as the midsurface $A_0$ of the reference configuration (Fig.~\ref{fig:case3_geometry}); although it carries a pressure of about \qty{2}{\kilo\pascal}, we treat it as stress-free, as both sources treat theirs. At the \num{83} nodes on the cut that bounds the measured region we prescribe all three measured displacement components, and we apply the measured cavity pressure as a traction boundary condition, uniform over the wall. Since $n_f=0$ here, the objective of Eq.~\eqref{eq:objective} comprises only the displacement loss and the sparsity penalty. At the \num{561} nodes off the cut it compares all three measured components with the displacements of the midsurface.

\subsubsection{Results}\label{sec:apps:case3:results}

\paragraph*{Training and model discovery}
We train on two nested ladders of pressure instants, both beginning at \qty{4.5}{\kilo\pascal}: the \qty{20}{\kilo\pascal} ladder, nine instants to \qty{21.0}{\kilo\pascal}, and the \qty{50}{\kilo\pascal} ladder, thirteen to \qty{49.6}{\kilo\pascal}. The instants a ladder omits are used for testing, up to \qty{72.3}{\kilo\pascal}.

\begin{figure*}[h]
  \centering
  \includegraphics{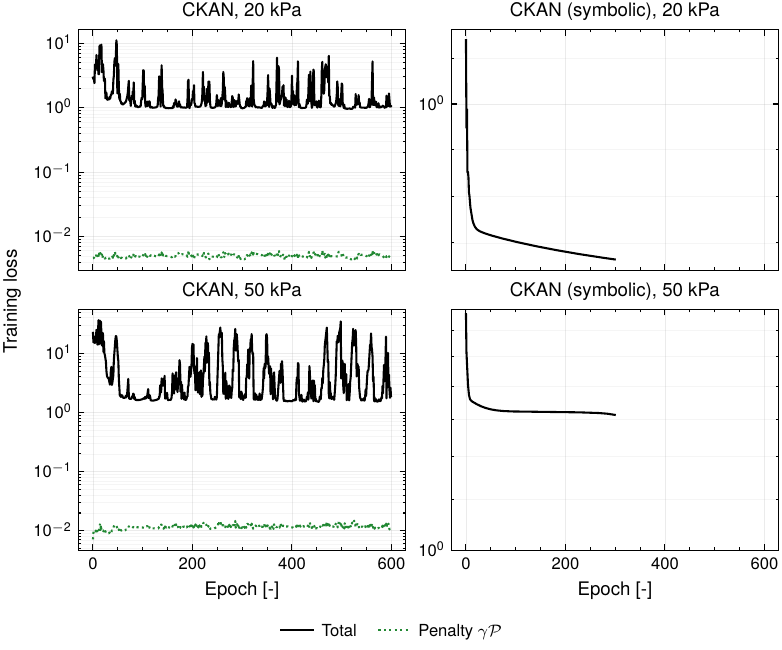}
  \caption{Training loss per epoch of the two models of
    Tab.~\ref{tab:case3_models}: one row per ladder, left the FEMU training
    of the CKAN, right the re-fit of its closed-form coefficients. With no
    measured force ($n_f=0$), the objective of Eq.~\eqref{eq:objective} is
    the displacement loss and the sparsity penalty; the penalty is drawn as
    its own series and enters the FEMU training alone. Each epoch applies
    one update per training instant, and the spikes follow the peaks of the
    cyclic learning rate, while the re-fit runs at a constant rate. The two
    ladders sum over different instants, so their losses do not compare.}
  \label{fig:case3_loss}
\end{figure*}

For each ladder we fit the HGO model, Eq.~\eqref{eq:hgo}, by FEMU at the fixed fiber angle. That fit is the ladder's reference model, and it seeds the ladder's CKANs over $\lambda \leq 1.1$ (Sec.~\ref{sec:femu:training}), as in Cases~1 and~2. From each seed we train five CKANs for 600 epochs at a cyclic learning rate and with the sparsity penalty on both networks. Tab.~\ref{tab:case3_hyper} in Appendix~\ref{app:hyperparameters} collects the hyperparameters, and Fig.~\ref{fig:case3_loss} shows the two trainings reported here.

After training, we reduce each CKAN to a closed-form expression by pruning and symbolic regression (Sec.~\ref{sec:ckan}); the coefficients are then re-fitted for another 300 epochs. Per ladder we report the best performing CKAN. Tab.~\ref{tab:case3_models} gives the two models, Tab.~\ref{tab:case3_formulas} the two parts of each energy. Both keep an exponential in $\Itwo$; the \qty{50}{\kilo\pascal} model also keeps a Neo-Hookean term in $\Ione$, which the \qty{20}{\kilo\pascal} model prunes away. Their fiber terms differ in form: on the \qty{20}{\kilo\pascal} ladder one that stiffens exponentially up to $\mac{\Ifour-1}^2\approx1$ and linearly above it, on the \qty{50}{\kilo\pascal} ladder a square that engages only above $\Ifour=1.50$.

\begin{table*}[h]
  \centering
  \caption{The discovered models of Case~3: one per training ladder, the run
    whose closed form reaches the lowest loss. Tab.~\ref{tab:case3_formulas}
    gives the two parts of each energy. The loss is the objective of
    Eq.~\eqref{eq:objective} with its displacement term alone, summed over
    the ladder's training instants, so the losses compare within a column
    only; each percentage is relative to the HGO reference of that column,
    fitted by FEMU on the same instants. The two ladders share their
    hyperparameters up to the grid intervals (Tab.~\ref{tab:case3_hyper}).
    The last row is the mean displacement error over the measured points at
    \qty{72.3}{\kilo\pascal}, the last instant of
    Fig.~\ref{fig:case3_apex}.}
  \label{tab:case3_models}
  \begin{tabular}{lcc}
    \toprule
     & \qty{20}{\kilo\pascal} ladder & \qty{50}{\kilo\pascal} ladder\\
    \midrule
    Training pressures & \qtyrange{4.5}{21.0}{\kilo\pascal} (9 instants)
                       & \qtyrange{4.5}{49.6}{\kilo\pascal} (13 instants)\\
    Loss, HGO reference & \num{1.1042} & \num{1.7632}\\
    Loss, CKAN & \num{0.9441} & \num{1.5228}\\
    Loss, closed form & \num{0.9640} (\qty{-12.7}{\percent})
                      & \num{1.6102} (\qty{-8.7}{\percent})\\
    Error at \qty{72.3}{\kilo\pascal}, CKAN / closed form / HGO
      & \qty{1.43}{\mm} / \qty{0.54}{\mm} / \qty{0.20}{\mm}
      & \qty{0.17}{\mm} / \qty{0.16}{\mm} / \qty{0.19}{\mm}\\
    \bottomrule
  \end{tabular}
\end{table*}

\begin{table*}[h]
  \centering
  \caption{The two parts of each discovered energy of
    Tab.~\ref{tab:case3_models}, in \unit{\mega\pascal}: the isotropic part
    $\Wckaniso$ and the part of one fiber family $\Wckanfib$, as Eq.
    \eqref{eq:energy_fiber} splits them and Eq.~\eqref{eq:energy_fiber_parts}
    defines them. Both families share $\Wckanfib$, so the energy is
    $\Wckaniso(\Ione,\Itwo)+\Wckanfib(\Ifour)+\Wckanfib(\Isix)$, and
    $\Wckaniso(3,3)=\Wckanfib(1)=0$ by construction. The isotropic part of
    the \qty{50}{\kilo\pascal} model carries a Neo-Hookean term in $\Ione$;
    the fiber part of that model engages only above
    $\Ifour=\num{1.50}$, where its bracket turns positive.}
  \label{tab:case3_formulas}
  \begin{tabular}{lll}
    \toprule
    & $\Wckaniso(\Ione,\Itwo)$ & $\Wckanfib(I)$ \\
    \midrule
    \qty{20}{\kilo\pascal} ladder
      & $0.00407\,e^{0.48412\,\Itwo} - 0.01741$
      & $0.00348\,\ln\!\bigl(1 + 0.01592\,e^{4.16899\,\mac{I-1}^{2}}\bigr) - 5.0\times 10^{-5}$ \\[3pt]
    \qty{50}{\kilo\pascal} ladder
      & $0.01088\,\Ione + 6.0\times 10^{-5}\,e^{1.1397\,\Itwo} - 0.03448$
      & $0.0043\,\mac{0.96005\,\mac{I-1}^{2} - 0.2411}^{2}$ \\
    \bottomrule
  \end{tabular}
\end{table*}

\begin{figure*}[h]
  \centering
  \includegraphics{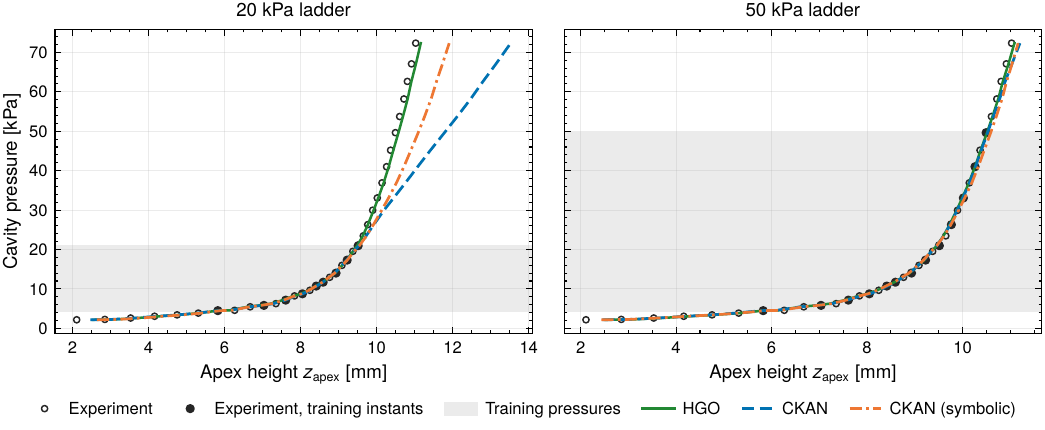}
  \caption{Cavity pressure against the height of the bulge apex over all 36
    instants of increasing pressure, for the two models of
    Tab.~\ref{tab:case3_models}: left the \qty{20}{\kilo\pascal} ladder,
    right the \qty{50}{\kilo\pascal} one. Each model is identified on the
    shaded training pressures alone and then followed to
    \qty{72.3}{\kilo\pascal}. Trained to \qty{21}{\kilo\pascal}, the CKAN
    turns too compliant above its training range, and its closed form
    recovers much of that drift, while HGO stays with the
    measurement. Trained to \qty{49.6}{\kilo\pascal}, all three stay with
    the measurement, and at \qty{72.3}{\kilo\pascal} the network's and
    its closed form's displacement errors, \qty{0.17}{\mm} and
    \qty{0.16}{\mm}, are below HGO's \qty{0.19}{\mm}
    (Tab.~\ref{tab:case3_models}).}
  \label{fig:case3_apex}
\end{figure*}

\begin{figure*}[h]
  \centering
  \includegraphics{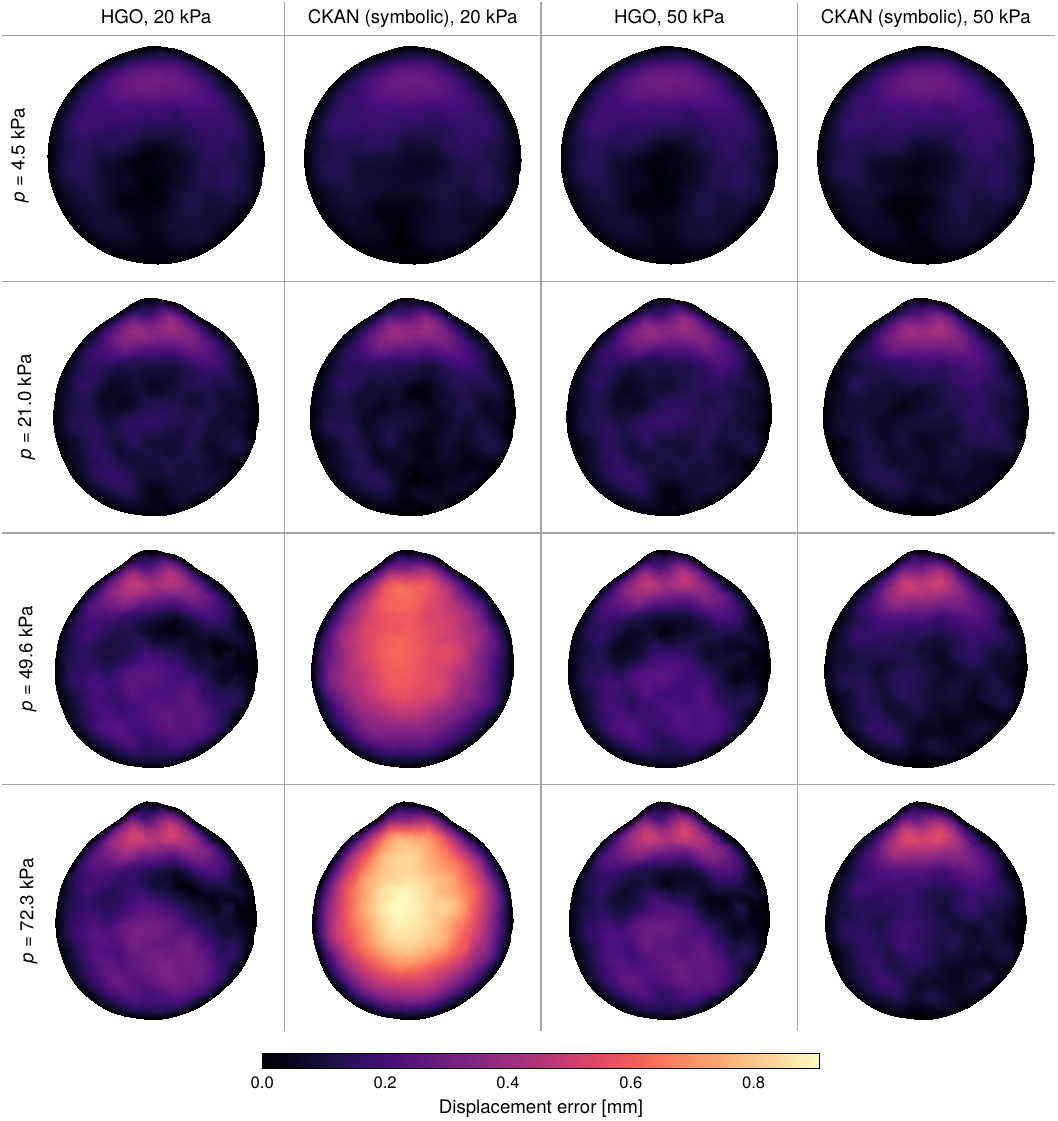}
  \caption{Displacement error over the measured region, drawn on the
    measured deformed surface; one color scale for all panels. Columns: each
    ladder's closed form of Tab.~\ref{tab:case3_models} beside the HGO
    reference of the same ladder, fitted by FEMU on the same instants. Rows:
    the first training pressure (\qty{4.5}{\kilo\pascal}), the last
    training pressure of each ladder (\qty{21.0}{\kilo\pascal} and
    \qty{49.6}{\kilo\pascal}), and the final test pressure
    (\qty{72.3}{\kilo\pascal}). Only the \qty{20}{\kilo\pascal} model
    extrapolates in the last two rows, and its error there is dominated by
    the compliance drift of Fig.~\ref{fig:case3_apex}; the
    \qty{50}{\kilo\pascal} model stays with the measurement throughout and
    keeps a lower error than its own reference at
    \qty{72.3}{\kilo\pascal}.}
  \label{fig:case3_maps}
\end{figure*}

\paragraph*{Extrapolation}
Fig.~\ref{fig:case3_apex} follows both models past their training pressures, to \qty{72.3}{\kilo\pascal}. Within its own training range each reproduces the measured apex height, as does HGO. Past that range the \qty{20}{\kilo\pascal} model turns too compliant: at the last instant its network inflates the bulge to \qty{13.5}{\mm}, where \qty{11.0}{\mm} are measured. Its closed form recovers most of that drift: mean displacement error \qty{0.54}{\mm} against the network's \qty{1.43}{\mm} (Tab.~\ref{tab:case3_models}). The \qty{50}{\kilo\pascal} model stays with the measurement to the end, network and closed form alike, both just under HGO's \qty{0.19}{\mm}. The error maps of Fig.~\ref{fig:case3_maps} show the same over the measured region.

The fiber input explains the difference. Training drove $\mac{\Ifour-1}^2$ to 2.7 on the \qty{20}{\kilo\pascal} ladder and to 3.8 on the \qty{50}{\kilo\pascal} one, while the solutions at \qty{72.3}{\kilo\pascal} ask for values up to 7.7 and 5.0. Three quarters of the \qty{20}{\kilo\pascal} model's fiber states therefore lie beyond the grid interval of its spline, where the energy grows at most linearly in its input (Appendix~\ref{app:grids}): the fiber stiffness stops rising and the wall inflates too far. At the last instant almost every state of the \qty{50}{\kilo\pascal} model still lies inside its grid interval, where network and closed form agree. The closed forms have no grid interval and keep stiffening wherever they are evaluated, so they drift less than the networks they replace, as in Cases~1 and~2. They add no information: beyond the states a ladder visited, nothing in the data decides the fiber energy. The agreement at \qty{72.3}{\kilo\pascal} says how far the \qty{50}{\kilo\pascal} ladder reaches, not that the identified law holds past it.

\paragraph*{Summary} Both CKANs reproduce their training instants more closely than the HGO reference fitted on the same data, and their closed forms keep most of that margin. How far a ladder reaches decides how far its model extrapolates: the \qty{20}{\kilo\pascal} model turns too compliant above its training range, while the \qty{50}{\kilo\pascal} model stays with the measurement to the last instant at \qty{72.3}{\kilo\pascal}. The two ladders also return different fiber terms, because nothing in the data decides the fiber energy beyond the states a ladder visited. The test leaves the split between $\Ione$ and $\Itwo$ open: its states crowd the equibiaxial locus, as Case~2's plate crowds the uniaxial one. It needs no measured force, because the prescribed pressure fixes the scale of the energy. Symbolic regression again returns a compact energy, an isotropic part plus one fiber term (Tab.~\ref{tab:case3_formulas}), at a cost of a few percent inside the training range and with less drift outside it.

\FloatBarrier
\section{Conclusions}\label{sec:conclusions}

Constitutive Kolmogorov--Arnold networks (CKANs) model hyperelastic material behavior without prescribing a functional form in advance. Herein, we embed a CKAN in finite element model updating, so that it can be identified from the kinematics of one heterogeneous experiment, and a reaction force where one is measured, instead of from a series of homogeneous tests. A key strength of this approach is that it leaves the experiment free: specimen geometry and boundary conditions are arbitrary, and the measurement need not cover the whole specimen, as it must for methods that insert the measured fields directly into the balance equations. Automatic differentiation through the finite element solver supplies stresses, tangents, and sensitivities, and no model-specific derivation is needed. The network satisfies all constitutive requirements by construction. After training, we prune it and translate it into a closed-form strain energy. We validated the approach on perforated rubber membranes with full-field digital image correlation data, on a silicone plate with a single stretch profile along one line, and on a human aortic specimen inflated under pressure. In all examples, the closed-form energies matched the networks they replaced on the training data, and several extrapolated beyond the measured states more reliably. The identified energies include the classical Neo-Hookean and Mooney--Rivlin models. We also found that what can be identified, and how complex the resulting expression is, depend on the coverage of the deformation space rather than on the number of measurements: full-field data are not automatically rich in information. The coverage can come from a well-designed heterogeneous specimen or, where full-field measurements are not available, from a combination of homogeneous tests.

Overall, this study provides a route from one mechanical experiment to an interpretable constitutive law ready for use in simulation. Our approach removes both the choice of the functional form and the need for a series of homogeneous tests, and we anticipate that it will make constitutive modeling of soft materials faster and less dependent on expert judgment.


\bmsubsection*{AI Disclosure}
The authors used Claude (Anthropic), through the Claude Code interface, as a coding and writing assistant: to implement and test the software behind the results, to run and evaluate the studies, and revise the text. The authors directed this work, checked its output against the code and the data, and take full responsibility for the content of this article.

\bmsubsection*{Author Contributions}

\bmsubsection*{Acknowledgments}
K.~Linka gratefully acknowledges financial support from Deutsche
Forschungsgemeinschaft (DFG, German Research Foundation) -- 533187597;
517243167. C.~J. Cyron gratefully acknowledges support of the European Research
Council (ERC) under the European Union's Horizon Europe research and innovation
programme (grant agreement No 101167207/MechVivo).
We gratefully acknowledge Stéphane Avril, Ecole des Mines de Saint-Etienne, for providing the arterial wall dataset.

\bmsubsection*{Conflicts of Interest}

The authors declare no conflicts of interest.

\bmsubsection*{Data Availability Statement}

The configuration files of the published runs are part of the repository and pin the options with which they were made, among them the reduction of Appendix~\ref{app:thin} and the conditions on the cut ends.

\bibliography{references}

\appendix

\bmsection{Details of the CKAN\label{app:ckan}}

\bmsubsection{Reparameterization of the constraints\label{app:param}}
The constraints of Sec.~\ref{sec:ckan} on the weights and control points of every edge are imposed by a reparameterization in terms of unconstrained parameters $\theta_\alpha,\theta_\beta,\theta_{c,k}\in\mathbb{R}$,
\begin{equation}\label{eq:reparam}
\begin{aligned}
    \alpha &= |\theta_\alpha|, \qquad \beta = |\theta_\beta|,\\
    c_1 &= \theta_{c,1}, \qquad c_k-c_{k-1} = \sum_{m=2}^{k}\eta(\theta_{c,m}), \quad k=2,\ldots,n_s,
\end{aligned}
\end{equation}
with the softplus $\eta$. The weights are thus non-negative and the first differences of the control points positive and increasing, in particular non-negative and non-decreasing, for every $\btheta\in\mathbb{R}^{n_\theta}$, so that $\btheta$ can be optimized without constraints.

\bmsubsection{Knot grids\label{app:grids}}
The grid intervals $[z^{(l)}_{\min},z^{(l)}_{\max}]$ of the splines, Eq.~\eqref{eq:spline}, are fixed when the network is initialized and are not adapted during training. The grid interval of the input layer is chosen according to the inputs expected in the data and contains the reference state: $z^{(0)}_{\min}\leq 3$ for an invariant $\Ione$ or $\Itwo$, and $z^{(0)}_{\min}=0$ for a fiber input $\mac{\Ifibn{\kappa}-1}^2$, which vanishes in the reference configuration and can leave its grid interval only upward. The inputs of a hidden layer are activations, whose range is not known in advance. Since all activation functions are non-decreasing, the components of $\vect{z}^{(l+1)}$ for $\vect{z}^{(l)}$ in the box $[z^{(l)}_{\min},z^{(l)}_{\max}]^{n_l}$ lie between the smallest of them at the lower corner of the box and the largest at its upper corner. The grid interval of layer $l+1$ is the interval spanned by these two values in the randomly initialized network, widened by \qty{60}{\percent} about its midpoint. Inputs that leave their grid interval during training are evaluated on the linear continuation of Sec.~\ref{sec:ckan}; the constraints hold irrespective of the grid, and only the resolution of the spline is lost there, the energy growing at most linearly in each input beyond the grid interval, for a fiber input quadratically in $\Ifibn{\kappa}-1$. Future investigation could focus on using grids that adapt during training to the range of incoming activations.

\bmsubsection{Sparsity penalty\label{app:penalty}}
The penalty $\Pen$ of Eq.~\eqref{eq:objective} follows \citet{Liu2024a}. It acts on the magnitudes of the edges: for layer $l$, the matrix $\tens{S}^{(l)}\in\mathbb{R}^{n_{l+1}\times n_l}$ has the entries
\begin{equation}\label{eq:edge_magnitude}
  S^{(l)}_{ji}=\frac{\operatorname{std}\bigl[\phi^{(l)}_{ji}(z^{(l)}_i)\bigr]}{\operatorname{std}\bigl[z^{(l)}_i\bigr]+0.1},
\end{equation}
where $\operatorname{std}[\cdot]$ is the standard deviation over a sample of network inputs $\vect{z}^{(0)}$, isotropic invariants and fiber inputs alike; where the energy is split, Eq.~\eqref{eq:energy_fiber}, the penalty runs over the edges of both networks. \citet{Liu2024a} take this sample from the current mini-batch. The FE loop has no such batch, and the inputs at the quadrature points move with the solution, so the sample is fixed once to the deformation states of the training data: the inputs of the measured element deformations at the training instants for a specimen, and those along the analytic deformation paths for the homogeneous tests.
The penalty reads
\begin{equation}\label{eq:penalty}
  \Pen(\btheta)=\sum_{l=0}^{L}\Bigl[\sum_{j,i}S^{(l)}_{ji}
    -\frac{2}{n_{l+1}}\sum_{j}\sum_{i}\rho^{(l)}_{ji}\log_2\rho^{(l)}_{ji}
    -\frac{2}{n_{l}}\sum_{i}\sum_{j}\varrho^{(l)}_{ji}\log_2\varrho^{(l)}_{ji}\Bigr],
  \qquad
  \rho^{(l)}_{ji}=\frac{S^{(l)}_{ji}}{1+\sum_{i'}S^{(l)}_{ji'}}, \quad
  \varrho^{(l)}_{ji}=\frac{S^{(l)}_{ji}}{1+\sum_{j'}S^{(l)}_{j'i}}.
\end{equation}

\bmsection{Plane-stress and membrane reduction\label{app:thin}}
All three specimens of Sec.~\ref{sec:applications} are thin, and each is computed with the reduction derived here, which turns the mixed problem of Sec.~\ref{sec:femu:forward} into a problem in the displacements of a midsurface alone. The sheets of Cases~1 and~2 are loaded in their plane and reduce to classical plane stress. The arterial wall of Case~3 is loaded normal to itself and is modeled as a membrane. The last paragraph checks the plane-stress reduction against the mixed formulation.

\paragraph*{Kinematics}
The reference configuration is the set of points $\bX+X_3\vect{N}(\bX)$ with $\bX$ on a midsurface $A_0$ and $\abs{X_3}<H_0/2$, where $H_0$ is the uniform thickness and $\vect{N}$ the unit normal of $A_0$, flat for a sheet and curved for the arterial wall. At each point of $A_0$, the orthonormal tangent vectors $\vect{E}_\alpha$, $\alpha\in\{1,2\}$, with $\vect{E}_1\times\vect{E}_2=\vect{N}$, span the tangent plane; for the fiber-reinforced wall, they are the circumferential and axial directions of Eq.~\eqref{eq:fiber_directions}, projected into the plane of each element. The midsurface deforms to $\bx=\bX+\bu(\bX)$, and the surface gradient on $A_0$ is $\Grads\vect{w}=(\partial\vect{w}/\partial\xi_\alpha)\otimes\vect{E}_\alpha$, with $\partial/\partial\xi_\alpha$ the derivative along $\vect{E}_\alpha$. The surface deformation gradient $\bFs=\Grads\bx$ maps the tangent plane of $A_0$ to that of the current surface, spanned by the vectors $\bFs\vect{E}_\alpha$. The length of their cross product, $J_{\mathrm{s}}=\norm{\bFs\vect{E}_1\times\bFs\vect{E}_2}$, is the ratio of current to reference area.

Two assumptions reduce the deformation through the thickness: the material lines along $\vect{N}$ remain normal to the deformed midsurface, which neglects the transverse shears, and they stretch uniformly by $\lambda_3$. With the current unit normal $\vect{n}$,
\begin{equation}\label{eq:thin_F}
  \bF=\bFs+\lambda_3\,\vect{n}\otimes\vect{N}, \qquad
  \vect{n}=\frac{\bFs\vect{E}_1\times\bFs\vect{E}_2}{J_{\mathrm{s}}}, \qquad
  \lambda_3=\frac{1}{J_{\mathrm{s}}}.
\end{equation}
The last relation is the incompressibility constraint $J=\lambda_3J_{\mathrm{s}}=1$, solved for the thickness stretch: it holds identically and pointwise, and Eqs.~\eqref{eq:bvp_constraint} and~\eqref{eq:weakform_p} are no longer needed. Since $\vect{n}$ is normal to $\bFs\vect{E}_1$ and $\bFs\vect{E}_2$, the right Cauchy--Green tensor splits into $\bC=\bCs+\lambda_3^2\,\vect{N}\otimes\vect{N}$, with the surface tensor $\bCs=\bFs^\top\bFs$. Its trace and determinant on the tangent plane are $\tr\bCs=\norm{\bFs\vect{E}_1}^2+\norm{\bFs\vect{E}_2}^2$ and $\det\bCs=J_{\mathrm{s}}^2$, so the invariants read
\begin{equation}\label{eq:thin_invariants}
  \Ione=\tr\bCs+\frac{1}{\det\bCs}, \qquad
  \Itwo=\det\bCs+\frac{\tr\bCs}{\det\bCs}, \qquad
  \Ifour=\afib_{+}\cdot\bCs\afib_{+}=\norm{\bFs\afib_{+}}^2,
\end{equation}
and likewise $\Isix$, since both fiber directions lie in the tangent plane, $\afib_{\pm}\cdot\vect{N}=0$.

\paragraph*{Stress}
We impose the plane-stress condition $\vect{n}\cdot\bP\vect{N}=0$. The kinematics make the stress uniform through the thickness up to terms of order $H_0/R$, with $R$ the smallest radius of curvature of $A_0$. On a sheet, whose two faces are traction-free, the condition holds exactly. On the arterial wall, loaded by a cavity pressure $\pcav$, it neglects a normal stress whose Cauchy measure lies between $-\pcav$ on the loaded face and zero on the other, smaller than the in-plane stresses by that same order. The hydrostatic pressure of Eq.~\eqref{eq:augmented}, no longer determined by the constraint, follows from the condition: with $J=1$ and $\vect{n}\cdot\bF^{-\top}\vect{N}=1/\lambda_3$, Eq.~\eqref{eq:stress} gives
\begin{equation}\label{eq:thin_p}
  p=\lambda_3\,\vect{n}\cdot\bP_{\mathrm{iso}}\vect{N}.
\end{equation}
The transverse shear stresses vanish identically, for an isotropic material and for fibers in the tangent plane alike.\footnote{For $\afib\cdot\vect{N}=0$, the fiber stress is proportional to $\partial\Ifour/\partial\bF$, Eq.~\eqref{eq:fiber_stress}, which combines $(\bFs\afib)\otimes\afib$, whose first vector lies in the current tangent plane, normal to $\vect{n}$, and whose second lies in the reference tangent plane, normal to $\vect{N}$, with a multiple of $\bF^{-\top}$, which maps $\vect{N}$ to $\lambda_3^{-1}\vect{n}$ and the reference tangent plane to the current one. Neither carries a transverse shear.} The energy becomes a function of $\bFs$ alone, $\isoc{\Wenergy}_{\mathrm s}(\bFs)=\isoc{\Wenergy}(\isoc{\bF})$ with $\bF$ from Eq.~\eqref{eq:thin_F}. Its derivative follows by the chain rule through $\lambda_3$ and $\vect{n}$; the term carrying the variation of $\vect{n}$ drops, since that variation is normal to $\vect{n}$ and the transverse shear stresses vanish, which leaves
\begin{equation}\label{eq:thin_P}
  \bPs=\frac{\partial\isoc{\Wenergy}_{\mathrm s}}{\partial\bFs}=\bP\,(\bI-\vect{N}\otimes\vect{N}),
\end{equation}
the restriction of the stress of Eq.~\eqref{eq:stress} to the reference tangent plane. It derives from an energy, so the reduced problem is hyperelastic, and the hydrostatic pressure follows pointwise from the displacement. The thin body stores the energy $H_0\isoc{\Wenergy}_{\mathrm s}$ per unit reference area and carries the force $H_0\bPs$ per unit reference length. The thickness enters only through these two products, and the identification of Sec.~\ref{sec:femu:objective} therefore fixes $H_0\isoc{\Wenergy}_{\mathrm s}$ rather than $\isoc{\Wenergy}_{\mathrm s}$ itself: an error in $H_0$ scales every identified modulus by $1/H_0$.

\paragraph*{Flat sheet}
On a sheet, $A_0$ is flat and the loads act in its plane, so the midsurface stays in that plane and the displacement has the two components $\bu=(u_1,u_2)$, which depend on $(X_1,X_2)$ only. Then $\vect{n}=\vect{N}=\vect{e}_3$ and the surface deformation gradient reduces to $\bF_2=\bI_2+\Grad\bu\in\mathbb{R}^{2\times2}$, with the two-dimensional identity $\bI_2$ and $\Grad$ acting on $(X_1,X_2)$, so that $J_{\mathrm{s}}=\det\bF_2$. The neglected transverse shears are $F_{31}$ and $F_{32}$, proportional to the distance $X_3$ from the midplane and thus small in a thin sheet. Equations~\eqref{eq:thin_F}, \eqref{eq:thin_p} and~\eqref{eq:thin_P} become
\begin{equation}\label{eq:thin_flat}
  \bF=\begin{bmatrix}\bF_2&\vect{0}\\ \vect{0}^\top&F_{33}\end{bmatrix}, \quad
  F_{33}=\frac{1}{\det\bF_2}, \quad
  p=F_{33}\,(\bP_{\mathrm{iso}})_{33}, \quad
  \bP_2=(\bP_{\mathrm{iso}})_2-p\,\bF_2^{-\top},
\end{equation}
where $(\cdot)_2$ denotes the in-plane block and $\bP_2$ the reduced stress $\bPs$ written in it. Here the plane-stress condition is exact: $P_{13}=P_{23}=P_{33}=0$ throughout the thickness.

\paragraph*{Weak form}
The internal virtual work is $\int_{A_0}H_0\,\bPs(\bFs;\btheta):\Grads\vect{v}\dd A$ in either reduction; the loads differ. On a sheet, $\vect{b}$ and $\hat{\vect{t}}$ act in the plane and are uniform through the thickness, and the boundary of the midplane splits into $\partial A_0^{\bu}$ and $\partial A_0^{\vect{t}}$ as the lateral surface of $\Omega_0$ does. The weak form is then Eq.~\eqref{eq:weakform_u} with $\bPs$, the function spaces of Eq.~\eqref{eq:spaces} on $A_0$ with two components, and the integrals over $\Omega_0$ and $\partial\Omega_0^{\vect{t}}$ replaced by $H_0$ times integrals over $A_0$ and $\partial A_0^{\vect{t}}$. On a membrane, all three displacement components are prescribed on $\partial A_0$. The cavity pressure acts on the deformed arterial wall, normal to it, and the membrane carries it at its midsurface as the force $\pcav J_{\mathrm{s}}\vect{n}=\pcav\,\bFs\vect{E}_1\times\bFs\vect{E}_2$ per unit reference area, a follower load that grows with the current area. With the function spaces of Eq.~\eqref{eq:spaces} defined on $A_0$ and $\partial\Omega_0^{\bu}$ replaced by $\partial A_0$, the weak form reads: for every $t\in\mathcal{T}$, find $\bu\in\mathcal{S}$ such that
\begin{equation}\label{eq:thin_weakform}
  r_u(\bu,\vect{v};\btheta)=\int_{A_0}H_0\,\bPs(\bFs;\btheta):\Grads\vect{v}\dd A
  -\int_{A_0}\pcav J_{\mathrm{s}}\,\vect{n}\cdot\vect{v}\dd A=0
  \qquad\forall\,\vect{v}\in\mathcal{V}.
\end{equation}
On a sheet the loads are dead, hence conservative, and the load potential $L$ is the work of $\vect{b}$ and $\hat{\vect{t}}$. On a membrane, with the displacements prescribed on all of $\partial A_0$, the cavity pressure is conservative as well \citep{Schweizerhof1984}, with $L=\pcav V_{\mathrm{c}}$, where $V_{\mathrm{c}}$ is the volume the deformed midsurface encloses with any fixed surface spanning its rim, a choice that shifts $L$ by a constant only.
The virtual work of the load is then the first variation of $L$, so the weak form states that the potential
\begin{equation}\label{eq:thin_potential}
  \Pi(\bu;\btheta)=\int_{A_0}H_0\,\isoc{\Wenergy}_{\mathrm s}(\bFs;\btheta)\dd A-L(\bu)
\end{equation}
is stationary. The tangent of either reduction is symmetric, on a membrane including the stiffness of the pressure load, where the mixed formulation has the indefinite tangent of a saddle point.

\paragraph*{Discretization}
With no pressure unknowns, $\vect{y}=\vect{d}_\mathcal{F}$ and $\bR=\bR_{u,\mathcal{F}}$ in Eq.~\eqref{eq:discrete_residual}. On a sheet, we discretize the midplane by bilinear quadrilaterals with the nodal displacements $\vect{d}\in\mathbb{R}^{2n_a}$; the nodal residuals $\bR_u$ follow as in Eq.~\eqref{eq:global_residual}, with the factor $H_0$ and without the pressure residuals, and $\mathcal{D},\mathcal{F}\subset\{1,\ldots,2n_a\}$. On a membrane, we discretize $A_0$ by linear triangles, each of them flat, so that $\vect{E}_\alpha$, $\vect{N}$ and $\bFs$ are constant per element; the nodes lie at $\bX_a\in\mathbb{R}^3$ and carry the displacements $\vect{d}_a\in\mathbb{R}^3$. Automatic differentiation of the discrete potential $\Pi_h(\vect{d};\btheta)$ gives the nodal residuals $\bR_u=\partial\Pi_h/\partial\vect{d}$ and the tangent $\tens{K}=\partial^2\Pi_h/(\partial\vect{d}_\mathcal{F}\,\partial\vect{d}_\mathcal{F})$.

\paragraph*{Observations and reaction forces}
Both reductions compute the displacements of the midsurface, and the losses of Sec.~\ref{sec:femu:objective} compare them with the measured ones. On a sheet, planar DIC measures the two in-plane components, which Eq.~\eqref{eq:thin_F} makes independent of the position through the thickness, so $\Qu\in\mathbb{R}^{2n_u\times 2n_a}$ carries two instead of three components per node and measurement point. On a membrane, stereo DIC measures all three, so $\Qu$ keeps three components per node and measurement point. On a sheet the integrals carry the factor $H_0$, so the residuals $\bR_{u,\mathcal{D}}$ at the prescribed degrees of freedom are forces, and $\QF\in\mathbb{R}^{n_f\times|\mathcal{D}|}$ keeps its form with the in-plane directions $\vect{s}_k$. On a membrane, $\partial\Pi_h/\partial\vect{d}$ at those degrees of freedom differs from the reaction forces by terms that depend only on the positions of the nodes on $\partial A_0$; they would have to be subtracted before a reaction is compared with a measured force. Where no force is compared, $n_f=0$, so $\vect{g}_f=\vect{0}$ and $\vect{h}=\vect{g}_y$ in Eq.~\eqref{eq:chain}. Section~\ref{sec:femu:sensitivities} applies to both reductions with $\vect{y}=\vect{d}_\mathcal{F}$ (Appendix~\ref{app:sensitivities}).

\paragraph*{Check against the mixed formulation}
The specimens of Cases~1 and~2 can also be computed with the mixed formulation of Sec.~\ref{sec:femu:forward}, on the same geometry and with the same prescribed displacements, observation operator and time instants. We discretize one symmetric half of the thickness with a single layer of trilinear hexahedra, with $u_3=0$ on the midplane and the reaction force doubled. One element through the thickness resolves no more than a linear variation of the displacement and an element-wise constant thickness stretch, so the mixed model is close to the reduction and differs from it chiefly in enforcing incompressibility once per element instead of pointwise, at \num{16750} instead of \num{4838} degrees of freedom on TT2.

Evaluating the identified models in the mixed formulation leaves the reported quantities nearly unchanged. On TT1 and TT2, over all eleven time instants and all three identified models, the reaction forces of the two formulations differ by at most \qty{0.4}{\percent} and in the median by less than \qty{0.2}{\percent}; the reaction errors reported in Fig.~\ref{fig:case1_matrix} change by at most \num{0.25} percentage points and the median displacement errors by at most \num{0.01} percentage points. On the plate of Case~2 the reaction differs by \qty{0.04}{\percent} at most, and the stretch profile along the path, the most mesh-sensitive quantity reported here, by \qty{0.04}{\percent} in relative $L^2$. For a Neo-Hookean material the reactions agree to \qty{0.2}{\percent} as well. What this does not test is the identification itself, which was not repeated in the mixed formulation.

\bmsection{Evaluation from the forward solve\label{app:sensitivities}}
Besides $\vect{g}_u$, $\vect{g}_f$, and $\tens{K}$, the gradient of Eq.~\eqref{eq:adjoint} requires the vector $\vect{h}$, which contains the derivative $\partial\bR_{u,\mathcal{D}}/\partial\vect{y}$ of the reaction forces, the solution $\vect{\xi}$ of the adjoint system, and the derivatives $\partial\bR_{u,\mathcal{D}}/\partial\btheta$ and $\partial\bR/\partial\btheta$. This appendix shows how each of them follows from the converged Newton--Raphson iterate. As in Sec.~\ref{sec:femu:sensitivities}, we consider one time instant and drop the superscript $m$.

\paragraph*{Quantities of the forward solve} The Newton--Raphson method, Eq.~\eqref{eq:newton}, evaluates the residuals and the tangent at every iterate, including the final converged one. At the converged iterate, the FE solution, it has thus evaluated all nodal residuals $\bR_u$, among them the reactions $\bR_{u,\mathcal{D}}$ that enter $\vect{g}_f$, Eq.~\eqref{eq:loss_partials}. The tangent is assembled by automatic differentiation of all residuals with respect to all nodal displacements and element pressures. With the rows ordered as $(\bR,\bR_{u,\mathcal{D}})$ and the columns as $(\vect{y},\vect{d}_\mathcal{D})$, this full tangent reads
\begin{equation}\label{eq:full_tangent}
  \frac{\partial(\bR,\bR_{u,\mathcal{D}})}{\partial(\vect{y},\vect{d}_\mathcal{D})}
  =\begin{bmatrix}
     \tens{K} & \partial\bR/\partial\vect{d}_\mathcal{D}\\
     \partial\bR_{u,\mathcal{D}}/\partial\vect{y} & \partial\bR_{u,\mathcal{D}}/\partial\vect{d}_\mathcal{D}
   \end{bmatrix}.
\end{equation}
Its upper left block is the tangent $\tens{K}$ of Eq.~\eqref{eq:tangent}. The upper right block, the coupling block $\partial\bR/\partial\vect{d}_\mathcal{D}$, is the derivative of the equations of Eq.~\eqref{eq:discrete_residual} with respect to the prescribed displacements. The lower block row, the derivative of the reactions, is not needed for the solve and is not kept: in the implementation, its rows are overwritten when the Dirichlet conditions are imposed.

\paragraph*{The vector $\vect{h}$} The derivative $\partial\bR_{u,\mathcal{D}}/\partial\vect{y}$ in $\vect{h}$, Eq.~\eqref{eq:chain}, lies in the lower block row of Eq.~\eqref{eq:full_tangent} and is therefore not available. For the loads considered here --- dead loads in Cases~1 and~2, the conservative cavity pressure in Case~3 (Appendix~\ref{app:thin}) --- the full tangent is symmetric, so its lower left block is the transpose of its upper right block, $(\partial\bR_{u,\mathcal{D}}/\partial\vect{y})^\top=\partial\bR/\partial\vect{d}_\mathcal{D}$. The vector $\vect{h}$ thus follows from one matrix--vector product with the coupling block,
\begin{equation}\label{eq:adjoint_rhs}
  \vect{h}=\vect{g}_y+\Bigl(\frac{\partial\bR_{u,\mathcal{D}}}{\partial\vect{y}}\Bigr)^{\!\top}\vect{g}_f
  =\vect{g}_y+\frac{\partial\bR}{\partial\vect{d}_\mathcal{D}}\,\vect{g}_f .
\end{equation}

\paragraph*{Adjoint solve} The adjoint system $\tens{K}^\top\vect{\xi}=\vect{h}$, Eq.~\eqref{eq:adjoint}, is solved with the tangent $\tens{K}$ assembled at the converged iterate. Of its LU factorization, only the numerical part is computed anew; the symbolic part, which depends on the sparsity pattern alone, is shared with the Newton--Raphson iterations.

\paragraph*{Gradient by one vector--Jacobian product} The gradient in Eq.~\eqref{eq:adjoint} contains the parameter Jacobians of the reaction forces and of the equations $\bR=(\bR_{u,\mathcal{F}},\bR_p)$. The pressure equations $\bR_p(\vect{d})$ do not depend on $\btheta$, Eq.~\eqref{eq:discrete_residual}, so that, with $\vect{\xi}=(\vect{\xi}_\mathcal{F},\vect{\xi}_p)$, only $\bR_{u,\mathcal{F}}$ contributes to the second term. Both terms then contain rows of the same Jacobian $\partial\bR_u/\partial\btheta$ of all nodal residuals $\bR_u=(\bR_{u,\mathcal{F}},\bR_{u,\mathcal{D}})$ and combine into one product,
\begin{equation}\label{eq:vjp}
  \frac{\diff\ell}{\diff\btheta}
  =\Bigl(\frac{\partial\bR_{u,\mathcal{D}}}{\partial\btheta}\Bigr)^{\!\top}\vect{g}_f
  -\Bigl(\frac{\partial\bR_{u,\mathcal{F}}}{\partial\btheta}\Bigr)^{\!\top}\vect{\xi}_\mathcal{F}
  =\Bigl(\frac{\partial\bR_u}{\partial\btheta}\Bigr)^{\!\top}\vect{c}, \qquad
  \vect{c}=(-\vect{\xi}_\mathcal{F},\vect{g}_f)\in\mathbb{R}^{3n_a},
\end{equation}
with $\vect{c}$ partitioned like $\bR_u$. This vector--Jacobian product is evaluated by reverse-mode automatic differentiation of the residual assembly at the converged iterate, without forming $\partial\bR_u/\partial\btheta$ explicitly.

\paragraph*{Reduced formulations} In both reductions of Appendix~\ref{app:thin}, the pressure rows and columns are absent and $\vect{y}=\vect{d}_\mathcal{F}$; their full tangents are symmetric as well, and the above applies unchanged. On a membrane, the rows at the prescribed degrees of freedom are not the reaction forces, but they differ from them by terms that depend only on the positions of the boundary nodes, so their derivative with respect to $\vect{y}$ vanishes and the identity above survives.

\begin{figure*}[h]   
  \centering
  \includegraphics{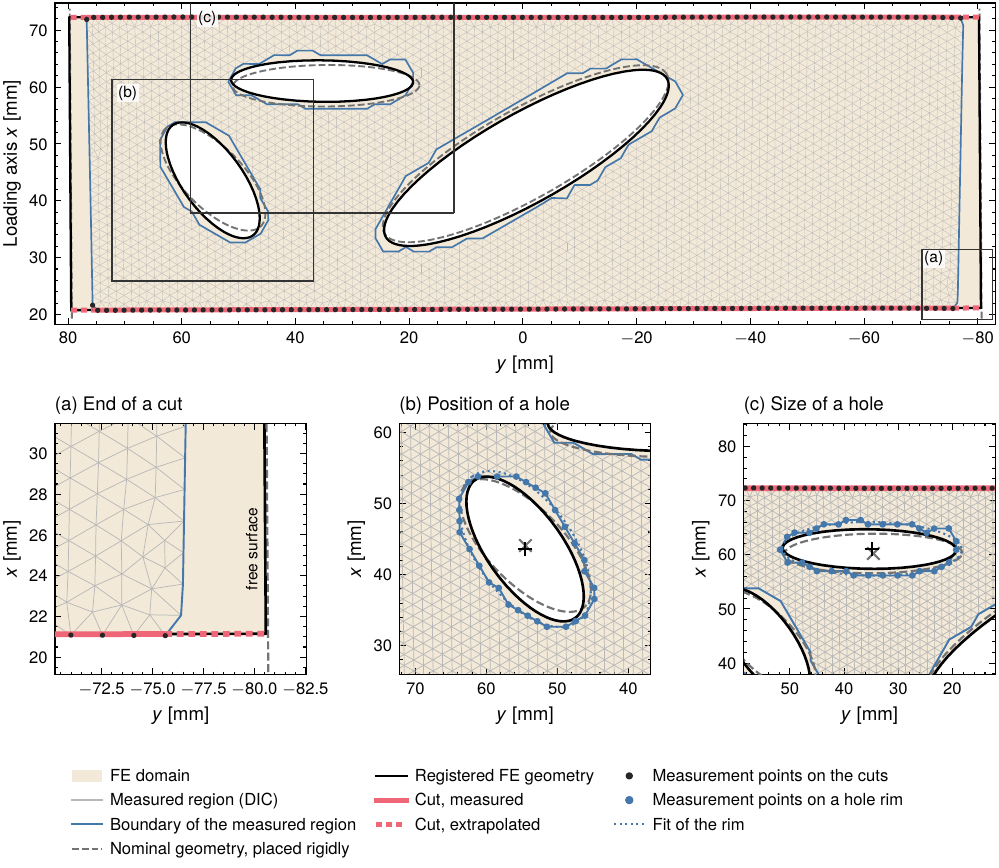}   
  \caption{Registered FE geometry of TT2. (a)~End of a cut, extrapolated over \qty{4.9}{\mm}. (b)~\emph{Position} of a hole: nominal center ($\times$) \qty{0.5}{\mm} from the measured one ($+$). (c)~\emph{Size} of a hole: slot tips shortened by \qty{2.4}{\percent}; this slot is also the hole furthest from its nominal position (\qty{0.9}{\mm}). The mesh follows the registered holes with a node every \qty{2.5}{\mm}.}
  \label{fig:registration}
\end{figure*}

\bmsection{Registered geometry of TT2\label{app:registration}}
Fig.~\ref{fig:registration} shows the registered geometry of TT2, as Fig.~\ref{fig:fe_geometry} does for TT1.

\bmsection{Hyperparameters\label{app:hyperparameters}}

Tab.~\ref{tab:case1_hyper} collects the hyperparameters of the Case~1 identifications.

\begin{table*}[h]
  \centering
  \caption{Hyperparameters of the Case~1 identifications
    (Sec.~\ref{sec:apps:case1:results}). The input grid interval is shared by
    all trainings; the grid intervals of the hidden layer follow from the
    initialization (Appendix~\ref{app:grids}). The cyclic learning rate is
    triangular between the two values, with a half cycle of \num{1000} epochs.
    The homogeneous fits are full batch, one Adam update per epoch; on TT1,
    each epoch sweeps the $n_T$ time instants
    (Sec.~\ref{sec:femu:training}).}
  \label{tab:case1_hyper}
  \begin{tabular}{lccc}
    \toprule
    & UT & UT \& PS & TT1 \\
    \midrule
    Formulation & -- & -- & plane stress (App.~\ref{app:thin}) \\
    Element & -- & -- & QUAD4 \\
    Element size $h$ / elements & -- & -- & \qty{2.5}{\mm} / \num{2616} \\
    Thickness $H_0$ & -- & -- & \qty{2.5}{\mm} \\
    \midrule
    Width $(n_0, n_1, n_2)$ & \multicolumn{3}{c}{$(2, 3, 1)$} \\
    Spline degree $d$ & \multicolumn{3}{c}{3} \\
    Input grid interval & \multicolumn{3}{c}{$[2.9, 14.0]$, 8 subintervals} \\
    Initialization & \multicolumn{3}{c}{Neo-Hookean, $\mu = \qty{0.67}{\mega\pascal}$ (UT, $\lambda \leq 1.1$)} \\
    Random initializations & \multicolumn{3}{c}{10} \\
    Optimizer & \multicolumn{3}{c}{Adam} \\
    \midrule
    Time instants $n_T$ & -- & -- & 11 \\
    Epochs & \num{20000} & \num{20000} & 300 \\
    Learning rate & \numrange{e-4}{e-2}, cyclic & \numrange{e-4}{e-2}, cyclic & \num{e-2} \\
    Force weight $\Lambda$ & -- & -- & 100 \\
    Penalty weight $\gamma$ & \num{e-4} & \num{e-3} & 0 \\
    \midrule
    Re-fit epochs & \num{20000} & \num{20000} & 300 \\
    Re-fit learning rate & \num{e-5} & \num{e-5} & \num{3e-4} \\
    \bottomrule
  \end{tabular}
\end{table*}

Tab.~\ref{tab:case2_hyper} collects the hyperparameters of the Case~2 identifications.

\begin{table*}[h]
  \centering
  \caption{Hyperparameters of the Case~2 identifications
    (Sec.~\ref{sec:apps:case2:results}), in the layout of
    Tab.~\ref{tab:case1_hyper}. The cyclic learning rate is triangular
    between the two values, with a half cycle of \num{1000} epochs for the
    homogeneous fits and of 50 on the plate. The homogeneous fits are
    full batch, one Adam update per epoch; the plate has a single time
    instant, so an epoch is likewise a single update.}
  \label{tab:case2_hyper}
  \begin{tabular}{lccc}
    \toprule
    & UT & UT \& PS & TT \\
    \midrule
    Formulation & -- & -- & plane stress (App.~\ref{app:thin}) \\
    Element & -- & -- & QUAD4 \\
    Element size $h$ / elements & -- & -- & \qty{1}{\mm} / \num{8829} \\
    Thickness $H_0$ & -- & -- & \qty{1.75}{\mm} \\
    \midrule
    Width $(n_0, n_1, n_2)$ & \multicolumn{3}{c}{$(2, 3, 1)$} \\
    Spline degree $d$ & \multicolumn{3}{c}{3} \\
    Input grid interval & \multicolumn{3}{c}{$[2.9, 10.0]$, 8 subintervals} \\
    Initialization & \multicolumn{3}{c}{Neo-Hookean, $\mu = \qty{0.32}{\mega\pascal}$ (UT, $\lambda \leq 1.1$)} \\
    Random initializations & \multicolumn{3}{c}{2} \\
    Optimizer & \multicolumn{3}{c}{Adam} \\
    \midrule
    Time instants $n_T$ & -- & -- & 1 \\
    Epochs & \num{20000} & \num{20000} & 800 \\
    Learning rate & \numrange{e-4}{e-2}, cyclic & \numrange{e-4}{e-2}, cyclic & \numrange{e-4}{e-2}, cyclic \\
    Force weight $\Lambda$ & -- & -- & \num{e-3} \\
    Penalty weight $\gamma$ & \num{e-4} & \num{e-3} & 0 \\
    \midrule
    Re-fit epochs & \num{20000} & \num{20000} & 100 \\
    Re-fit learning rate & \num{e-5} & \num{e-5} & \num{3e-4} \\
    \bottomrule
  \end{tabular}
\end{table*}

Tab.~\ref{tab:case3_hyper} collects the hyperparameters of the Case~3 identifications.

\begin{table*}[h]
  \centering
  \caption{Hyperparameters of the Case~3 identifications
    (Sec.~\ref{sec:apps:case3:results}), in the layout of
    Tab.~\ref{tab:case1_hyper}. Case~3 has a single specimen, so the columns
    are the two training ladders. The isotropic network has no hidden layer,
    so the isotropic part of the energy is a sum of one univariate function
    per invariant; the width and grid rows of $\KANfib$,
    Eq.~\eqref{eq:energy_fiber}, are its own. The grid intervals follow the
    FE solutions of the training instants rather than the measured states
    (Appendix~\ref{app:grids}). Each ladder is initialized from its own HGO
    reference, fitted by FEMU on the same instants, with both
    initializations restricted to stretches $\lambda \leq 1.1$ as in
    Cases~1 and~2: the base scales of the isotropic network to the
    Neo-Hookean energy at $\mu_{\mathrm{HGO}}$, the fiber network to the
    HGO fiber term. The cyclic learning rate is triangular between the two
    values, with a half cycle of 15 epochs; each epoch sweeps the $n_T$
    instants.}
  \label{tab:case3_hyper}
  \begin{tabular}{lcc}
    \toprule
    & \qty{20}{\kilo\pascal} ladder & \qty{50}{\kilo\pascal} ladder \\
    \midrule
    Formulation & \multicolumn{2}{c}{membrane (App.~\ref{app:thin})} \\
    Element & \multicolumn{2}{c}{TRI3} \\
    Mesh / elements & \multicolumn{2}{c}{DIC triangulation, \num{1203}} \\
    Thickness $H_0$ & \multicolumn{2}{c}{\qty{0.85}{\mm}} \\
    \midrule
    Width $(n_0, n_1)$ & \multicolumn{2}{c}{$(2, 1)$, no hidden layer} \\
    Spline degree $d$ & \multicolumn{2}{c}{3} \\
    Input grid interval & $[2.9, 7.6]$, 8 subintervals & $[2.9, 8.8]$, 8 subintervals \\
    $\KANfib$ width $(n_0, n_1)$ & \multicolumn{2}{c}{$(1, 1)$} \\
    $\KANfib$ grid interval & $[0, 3.7]$, 8 subintervals & $[0, 4.9]$, 8 subintervals \\
    Fiber angle $\fibangle$ & \multicolumn{2}{c}{\ang{41}, fixed} \\
    Initialization, isotropic & Neo-Hookean, $\mu = \qty{0.013}{\mega\pascal}$ ($\lambda \leq 1.1$)
                              & Neo-Hookean, $\mu = \qty{0.014}{\mega\pascal}$ ($\lambda \leq 1.1$) \\
    Initialization, $\KANfib$ & HGO, $k_1 = \qty{0.011}{\mega\pascal}$, $k_2 = 1.43$ ($\lambda \leq 1.1$)
                              & HGO, $k_1 = \qty{0.010}{\mega\pascal}$, $k_2 = 1.54$ ($\lambda \leq 1.1$) \\
    Random initializations & \multicolumn{2}{c}{5} \\
    Optimizer & \multicolumn{2}{c}{Adam} \\
    \midrule
    Time instants $n_T$ & 9 & 13 \\
    Epochs & \multicolumn{2}{c}{600} \\
    Learning rate & \multicolumn{2}{c}{\numrange{e-4}{e-2}, cyclic} \\
    Force weight $\Lambda$ & \multicolumn{2}{c}{-- (no force measured)} \\
    Penalty weight $\gamma$ & \multicolumn{2}{c}{\num{e-3}} \\
    \midrule
    Re-fit epochs & \multicolumn{2}{c}{300} \\
    Re-fit learning rate & \multicolumn{2}{c}{\num{3e-4}} \\
    \bottomrule
  \end{tabular}
\end{table*}

\clearpage

\bmsection{Full-field results on the withheld specimen TT2\label{app:tt2}}

\begin{figure*}[h]
  \centering
  \includegraphics{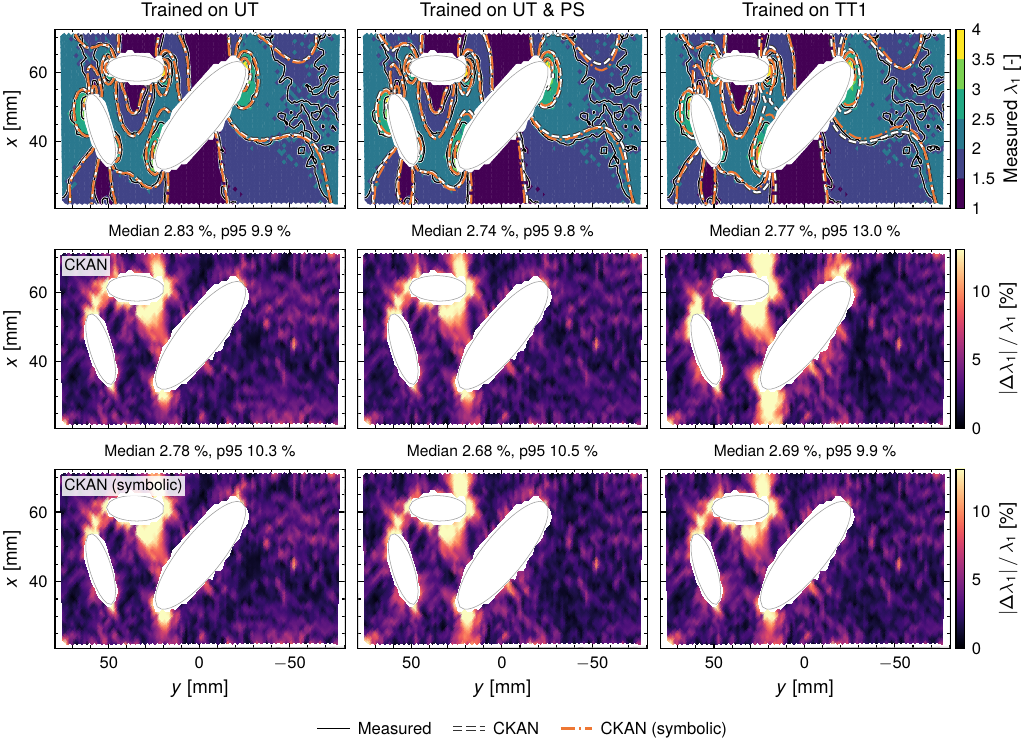}
  \caption{Maximum principal stretch on TT2 at the final time instant, one
    column per training set. Top: measured $\lambda_1$ at the DIC-triangle
    centroids ($\bF$ from the dataset, filled, black contours) with the CKAN
    (blue dashed) and its symbolic form (orange dash-dotted), FE quadrature
    values interpolated cubically. Middle and bottom: pointwise relative
    error $|\Delta\lambda_1| / \lambda_1$ for the two forms; the subplot
    titles give the median and the 95th percentile.}
  \label{fig:case1_stretch_TT2}
\end{figure*}

\begin{figure*}[h]
  \centering
  \includegraphics{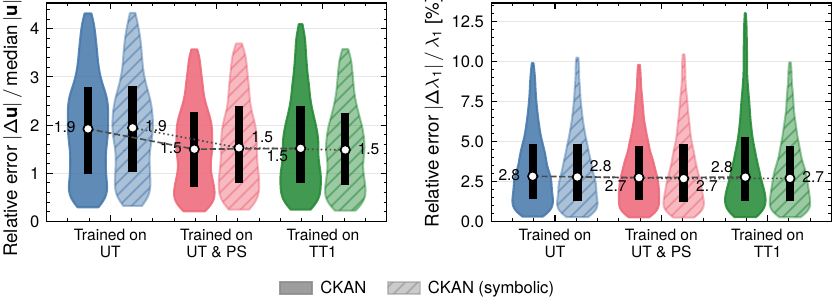}
  \caption{Distributions of the pointwise relative errors on TT2 at the
    final time instant, for the CKAN and its symbolic form trained on each
    training set: $|\Delta\bu|$ relative to the median $|\bu|$ of the step
    (left) and the error of the maximum principal stretch,
    $|\Delta\lambda_1| / \lambda_1$ (right). The violins span the central
    \qty{90}{\percent} of the data; the black bar marks the central
    \qty{50}{\percent} and the white dot the median of the full data, with
    its value printed beside it. The medians equal those in the subplot
    titles of Figs.~\ref{fig:case1_field_u_TT2}
    and~\ref{fig:case1_stretch_TT2}.}
  \label{fig:case1_errors_TT2}
\end{figure*}

Figs.~\ref{fig:case1_stretch_TT2} and~\ref{fig:case1_errors_TT2} supplement the
displacement field of Fig.~\ref{fig:case1_field_u_TT2} with the maximum
principal stretch on TT2 and with the distributions of both errors.

\end{document}